\documentclass[twocolumn]{emulateapj}
\usepackage{graphicx}
\usepackage{amssymb}
\usepackage{amsmath}
\usepackage{subfigure}
\usepackage{multirow}
\usepackage{threeparttable}
\usepackage{array}
\usepackage{natbib}
\usepackage{xcolor}
\usepackage{subfigure}

\usepackage{hyperref}
\hypersetup{
    colorlinks=true,
    linkcolor=red,
    filecolor=magenta,      
    urlcolor=cyan,
}

\def\MgX{{\rm Mg} {\sc x}}

\def\NeVIII{{\rm Ne} {\sc viii}}

\def\OII{{\rm O} {\sc ii}}
\def\OIII{{\rm O} {\sc iii}}
\def\OIV{{\rm O} {\sc iv}}

\def\OVI{{\rm O} {\sc vi}}
\def\OVII{{\rm O} {\sc vii}}
\def\OVIII{{\rm O} {\sc viii}}

\def\SiIII{{\rm Si} {\sc iii}}
\def\SiIV{{\rm Si} {\sc iv}}
\def\CII{{\rm C} {\sc ii}}
\def\CIII{{\rm C} {\sc iii}}

\def\HI{{\rm H} {\sc i}}

\begin{document}
\title{The Mass and Absorption Columns of Galactic Gaseous Halos}
\author{Zhijie Qu and Joel N. Bregman}
\affil{Department of Astronomy, University of Michigan, Ann Arbor, MI 48104, USA}
\email{quzhijie@umich.edu}
\email{jbregman@umich.edu}

\begin{abstract}
The galactic gaseous halo is a gas reservoir for the interstellar medium in the galaxy disk, supplying materials for star formation. We developed a gaseous halo model connecting the galaxy disk and the gaseous halo by assuming that the star formation rate on the disk is balanced by the radiative cooling rate of the gaseous halo, including stellar feedback. In addition to a single-temperature gaseous halo in collisional ionization equilibrium, we also consider the photoionization effect and a steady-state cooling model. Photoionization is important for modifying the ion distribution in low-mass galaxies and outskirts of massive galaxies due to the low densities. The multi-phase cooling model dominates the region within the cooling radius, where $t_{\rm cooling}=t_{\rm Hubble}$. Our model reproduces most of the observed high ionization state ions for a wide range of galaxy masses (i.e., {\OVI}, {\OVII}, {\NeVIII}, {\MgX}, and {\OVIII}). We find that the {\OVI} column density has a narrow range around $\approx10^{14}\rm~cm^{-2}$ for halo masses from $M_\star\approx3\times10^{10}~M_\odot$ to $6\times10^{12}~M_\odot$, which is consistent with some but not all observational studies. For galaxies with halo masses $\lesssim3\times10^{11}~M_\odot$, photoionization produces most of the {\OVI}, while for more massive galaxies, the {\OVI} is from the medium that is cooling from higher temperatures. Fitting the Galactic (Milky-Way) {\OVII} and {\OVIII} suggests a gaseous halo model where the metallicity is $\approx 0.55~Z_\odot$ and the gaseous halo has a maximum temperature of $\approx 1.9\times10^6\rm~K$. This gaseous halo model does not close the census of baryonic material within $R_{200}$.

\end{abstract}

\keywords{galaxies: halos -- galaxies: CGM -- galaxies: X-ray -- quasars: absorption lines}
\maketitle

\section{Introduction}
Recently, more and more observational evidence has been found to show the importance of gaseous components (the circumgalactic medium; CGM) in galaxy halos (\citealt{Anderson:2010aa, Menard:2010aa, Werk:2014aa}; see the review of \citealt{Tumlinson:2017aa}). These gaseous components surrounding the galaxy disk are formed during the galaxy formation and are modified by various feedback processes, such as stellar feedback and active galactic nucleus (AGN) feedback \citep{White:1991aa}. The existence of a CGM also modifies the evolution of the galaxy by providing fresh materials for star formation \citep{Keres:2005aa, Sancisi:2008aa, Keres:2009aa}, and by heating materials accreted from the intergalactic medium (IGM) through the gravitational potential release and the accretion shock \citep{Mo:2010aa}. 

The existence of gaseous halos is also helpful to explain various observational issues, such as the missing baryon problem (the baryonic fraction is significantly lower than the cosmic baryonic fraction of 0.16; \citealt{Dai:2010aa, McGaugh:2015aa, Planck-Collaboration:2016aa}). One solution is that the missing baryons stay in the galaxy but in an invisible phase (low density and high temperature), which could be the hot gaseous halo \citep{Fukugita:2006aa, Bregman:2007ab}. Theoretically, simulations found that the cool gas ($10^4~\rm K$) in the early Universe ($z > 4$) is heated, becoming a warm-hot intergalactic medium ($10^5 - 10^7\rm~K$) during galaxy formation, which accounts for more than $30\%$ of total baryons \citep{Weinberg:1997aa, Cen:1999aa}.

The final temperature of these heating processes is about the virial temperature of the galaxy, which is determined by the galaxy halo mass. Massive galaxies have higher virial temperatures than low-mass galaxies, and the virial temperature of low-mass galaxies ($M_{\rm h} \sim 10^{11}~M_\odot$) is around $10^{5.5}\rm~K$, which is also the peak temperature of the radiative cooling curve and can lead to rapid cooling with a cooling timescale of $< 1~\rm Gyr$. Therefore, whether the gaseous halo exists is a result of the competition between various heating processes and the radiative cooling, and this competition results in a multi-phase medium in the gaseous halo \citep{Oppenheimer:2016aa}.

Multi-wavelength observations of both emission and absorption reveal different phase mediums in the gaseous halo. The hot components in the gaseous halo can be detected in emission by direct X-ray imaging \citep{Anderson:2010aa, Bogdan:2013aa, Goulding:2016aa, Li:2016aa}, and in absorption or emission from high ionization state ions (e.g., {\OVII}, {\NeVIII} and {\MgX}; \citealt{Nicastro:2002aa, Savage:2005aa, Miller:2015aa, Qu:2016aa}). These studies found that the mass of the hot gaseous halo is comparable to the stellar mass of the galaxy, and about half of total baryons are still missing. Some studies show that the hot gas may account for all missing baryons in the Milky Way (MW; \citealt{Gupta:2012aa, Nicastro:2016aa}), however, they overestimate the emission measurement by more than one order of magnitude \citep{Bregman:2018aa, Li:2018aa}. Ultraviolet (UV) absorption line studies on low and intermediate ionization state ions show the existence of cool clouds in the halos, but the mass is model-dependent with a variation from $6\%$ to $40\%$ of the total baryon mass \citep{Werk:2014aa, Stern:2016aa}.

With a large amount of gas in the halo, radiative cooling may lead to a significant cooling flow onto the galaxy disk, which will be transformed into the stellar content of the disk through star formation \citep{Sancisi:2008aa}. This astrophysical connection between the radiative cooling and the star formation suggests that the cooling rate and the star formation rate (SFR) should be comparable with each other. Although the net cooling rate is also modified by heating from galactic feedback or accretion from the IGM, observations showed that the cooling rate is approximately the SFR for star-forming galaxies (with large scatter; \citealt{Li:2014aa}).

In this paper, our starting point is that the SFR is balanced by the radiative cooling rate of the gaseous halo within the cooling radius. Then, we employ a set of assumptions for the gaseous halo -- the density profile, the temperature distribution, hydrostatic and ionization equilibrium, and build up a halo model to connect the properties of the gaseous halo (i.e., mass and ion column density) to other galaxy properties (i.e., stellar mass and star formation rate). The details of the model assumptions are described in Section 2. In Section 3, we present the mass and column density of the gaseous halos, and their dependence on model parameters (e.g., stellar mass, SFR, or metallicity). The comparison with observations and implications are discussed in Section 4; our results are summarized in Section 5.

\section{Model}
We consider a spherical volume-filling gaseous halo model to connect the galaxy properties with the gaseous halo properties. In this section, the employed assumptions will be described and discussed.

\subsection{General Picture}
During the formation of the galaxy, the accreted material is heated by the released gravitational potential through the accretion shock. Without radiative energy losses, the final temperature of a gravitationally self-bound system is the virial temperature that is determined by the total mass. However, a realistic gaseous halo suffers from radiative cooling, which is crucial for the formation of galaxy disk.

Once the galaxy disk is formed, star formation leads to stellar feedback, injecting gas, dust, and energy into the gaseous halo. Stellar feedback affects the galaxy in several ways: stellar winds of massive stars; mass-loss of asymptotic giant branch stars; and supernovae from either massive stars or degenerate stars \citep{Zaritsky:1994aa, Willson:2000aa, Scannapieco:2008aa}. Another main feedback channel is the central supermassive black halo that is in an active galactic nuclei (AGN) phase, which injects ionizing photons and high-energy particles \citep{Fabian:2012aa}. These feedback processes can offset radiative cooling, or reheat the cooled gas \citep{Li:2015aa}. Although these processes are poorly resolved and implemented with different subgrid models in cosmological simulations, their effects on the galaxy evolution have been confirmed showing that no single feedback channel can dominate across all galaxy masses \citep{Vogelsberger:2014aa, Schaye:2015aa, Hopkins:2017aa}. However, the relative contributions for different processes are still controversial \citep{Nelson:2017aa, Suresh:2017aa}.

Besides the feedback from the galaxy disk, accretion from the IGM also provides additional energy to the gaseous halo as material falls deeper into the gravitational potential well. Then, the energy conservation of the gaseous halo leads to
\begin{equation}
L_{\rm net, cl} = L_{\rm rad} - L_{\rm net, acc} - L_{\rm fb},
\end{equation}
where symbols denote the net cooling, the radiative cooling, the net accretion heating and the feedback heating. For simplicity, we ignore the heating from the accretion of the IGM gas in our models, since the actual value of accretion heating depends on several uncertain factors -- the accretion rate from IGM, the accretion shock process and the structure around the virial radius. However, an estimation shows that the contribution from accretion heating is not significant when the hot gaseous halo already exists. Assuming the accreted material is virialized at the virial radius, the released gravitational potential energy is $2k_{\rm B}T_{\rm vir}$, which is slightly larger than the internal energy of the virialized halo of $3/2k_{\rm B}T_{\rm vir}$. Additionally, the energy to ionize electrons from atoms will increase the internal energy by several tens of $\rm eV$ per atom, which is equivalent to a temperature around $10^5-10^6\rm~K$ (depending on the ionization state that is proportional to the virial temperature). Therefore, the energy used to ionize atoms cannot be transformed into internal energy, which decreases the net heating from the IGM accretion. Finally, we consider the net cooling rate that is only related to the radiative cooling and the heating due to galactic feedback.

The net cooling flux is related to the accretion flow since the cooled gas cannot be buoyant in the halo due to the gravitational potential. Once the gas from the halo is accreted onto the disk, it will interact with the disk interstellar medium or outflows launched from the disk, which leads to the disruption of the cool gas and the condensation of the hot gas \citep{Marinacci:2010aa, Scannapieco:2015aa}. Additionally, various processes are involved in this interaction, such as the disk dynamics and the thermal conduction, which lead to complex situations in different galaxies \citep{Oosterloo:2007aa, Armillotta:2016aa, Zheng:2017aa}. These phenomenon are beyond the scope of this paper, therefore, we assume that the accreted cold gas could be mixed with the existed ISM instantly to avoid detailed interactions between disk and halo gases.

Studies of the MW molecular clouds showed that the star formation timescale is comparable to the dynamical timescale of the cloud ($\sim 1-10\rm~Myr$), and the star formation efficiency is less than $2\%$ \citep{Larson:1981aa, Myers:1986aa, Leroy:2008aa}. Considering the SFR of the MW as $\approx 1 ~M_\odot~\rm yr^{-1}$ \citep{Robitaille:2010aa}, around $100~M_\odot~\rm yr^{-1}$ gas should be transformed into the star-forming molecular clouds since the lifetime of molecular clouds is short ($\lesssim 20\rm~Myr$ ; \citealt{Larson:1981aa, Larson:1994aa}). The total atomic gas mass in the MW is around $7\times 10^{9}~M_\odot$ \citep{Nakanishi:2016aa}, which means that the atomic gas will be refreshed in around $70~\rm Myr$. Therefore, the timescale is around several $10^7\rm~yr$ to form stars using accreted cool gas from the gaseous halo. This timescale is comparable with the timescale of current measurement methods (i.e., UV/IR) of SFR for external galaxies, which measure the average SFR over $10^7$ to $10^8\rm~yr$ \citep{Madau:2014aa}. In this sense, the measured net cooling flow mass has a physical connection with the measured SFR. 

The cold mode accretion provides an additional gas origin besides the hot mode accretion (the radiative cooling and accretion of the virialized halo), which requires the density of at least one order of magnitude higher than $n_{200}$ ($10^{-3} - 10^{-4}\rm~cm^{-3}$) and the low temperature ($10^{4-5}\rm~K$) for the gas to remain cool during the accretion \citep{Keres:2005aa}. The cold mode accretion leads to cool gas filaments in the halo,  directly connecting the disk and the IGM and transporting gases into the disk \citep{Keres:2009aa}. However, the existence of a hot ambient halo ($T \approx T_{\rm vir}$) near hydrostatic equilibrium could destroy these cold gas filaments by the mixing and interaction, which makes the contribution from the cold mode accretion less than one-third of the hot mode in the low redshift universe ($z<2$; \citealt{Nelson:2013aa}). Therefore, involving cold mode accretion will not break the balance between the cooling flow and the star formation, so we adopt the assumption that the net cooling rate is equal to the SFR.

Feedback processes must the included, as they will offset some of the radiative cooling. For a star-forming galaxy without a merger, the gas for star formation is originally from the gaseous halo, and the accretion from a gaseous halo is modified by the strength of stellar feedback (i.e., proportional to the SFR) when the redshift is low. Therefore, for a galaxy dominated by stellar feedback (with a dim AGN or without an AGN), the stellar feedback strength is proportional to the radiative cooling rate, which can be modeled as $\dot{M}_{\rm stellar, h} = \alpha \dot{M}_{\rm rad, cl}$. Then, a simple relationship between the SFR and the radiative cooling rate is
\begin{equation}
{\rm SFR} = \gamma \dot{M}_{\rm rad, cl},
\end{equation}
where $\gamma = 1-\alpha$ is smaller than unity to account the heating by stellar feedback, and $\dot{M}_{\rm rad, cl}$ is the total radiative cooling rate of the gaseous halo. For simplicity, we assume that $\gamma$ is unity for the following calculation; the effect of variations in this $\gamma$ factor is discussed in Section 4.

This relationship will be broken by several physical processes, such as feedback from an AGN or a starburst event. For AGN feedback, there is no direct connection with the SFR, therefore, there is no direct relationship between AGN feedback heating and the radiative cooling. For merging galaxies that trigger starburst events, the connection between the SFR and the radiative cooling rate is not valid either, since the interaction between gases in the two galaxies triggers the star formation, which is not related to the gaseous halo cooling. Therefore, Equation (2) is only applicable for stably-evolving star-forming galaxies  without powerful AGNs.

Therefore, in our model, the SFR and radiative cooling from the gaseous halo are tightly connected. This model is most applicable to field galaxies, rather than group or cluster galaxies, which can be greatly affected by the intragroup or intracluster medium \citep{Balogh:1998aa}. With these constraints, we adopt the conditions where the SFR is equal to the radiative cooling rate of the gaseous halo. The radiative cooling rate is limited within the cooling radius, where the cooling timescale is equal to the Universe age ($13.8\rm~Gyr$; \citealt{Planck-Collaboration:2016aa}) or the cosmic epoch at a given redshift. In the following calculation, we use $H_0 = 67.8 \rm~km~s^{-1}~Mpc^{-1}$, $\Omega_{\rm m} = 0.308$, and $\Omega_{\rm b} = 0.0483$ \citep{Planck-Collaboration:2016aa}.

\subsection{Galaxy and Gaseous Halo Properties}

To construct sample galaxies, we adopt several empirical relationships. For a given stellar mass, we obtain the halo mass based on the stellar mass-halo mass (SMHM) relationship \citep{Behroozi:2013aa, Kravtsov:2014aa}. These two SMHM relationships sdiverge when $M_{\rm h} > 10^{11.5}~M_\odot$, and \citet{Kravtsov:2014aa} has a higher stellar mass than Behroozi's relationship. At the halo mass of $10^{13.5}~M_\odot$, the stellar mass difference is around $0.5\rm~ dex$. We choose the Kravtsov's SMHM relationship, since it describes the case that is more similar to the MW, where a $\approx 2\times 10^{12} ~M_\odot$ halo hosts a $5-8\times 10^{10}~M_\odot$ galaxy disk. Once the halo mass is determined, the virial radius and the virial temperature are calculated as:
\begin{eqnarray}
R_{\rm vir} &=& R_{200} = \frac{M_{\rm h}}{4\pi \Delta_{\rm vir} \rho_{\rm crit}/3}, \notag \\
V_{\rm c}^2 &=& \frac{G M_{\rm h}}{R_{\rm vir}} = 100 H_0^2 R_{\rm vir}^2, \notag \\
T_{\rm vir} &=& \frac{\mu m_{\rm p}}{2k_{\rm B}} V_{\rm c}^2,
\end{eqnarray}
where $\Delta_{\rm vir} = 200$ is the collapse factor, and $\rho_{\rm crit} = 3H_0^2/8\pi G$ is the cosmic critical density. The quantities $R_{\rm vir}$ and $T_{\rm vir}$ are input parameters of our models and can be varied by introducing additional factors (as the model of the MW in Section 4.5). Therefore, the choice of the SMHM relationship does not affect our results significantly.

The star formation rate can be inferred using the star formation-stellar mass plane \citep{Renzini:2015aa, Morselli:2016aa}:
\begin{equation}
\log({\rm SFR}) = (0.72 \pm 0.02) \log M_\star - 7.12,
\end{equation}
in the stellar mass range of $M_\star = 10^{8.5}-10^{11.25} M_\odot$. Therefore, we set the range of halo mass to $10^{10.5} - 10^{13.5}~M_\odot$. The star formation also has a dependence on the redshift \citep{Pannella:2009aa}:
\begin{equation}
\overline{\rm SFR} \approx 270 \frac{M_\star}{10^{11}~M_\odot}\left(\frac{t}{3.4\times 10^9\rm~yr}\right)^{-2.5}\frac{M_\odot}{\rm yr},
\end{equation}
where $t$ is the cosmic epoch. This relationship can be rewritten as a dependence on the redshift directly ${\rm sSFR} \propto (1+z)^3$ at $z<2$ \citep{Lilly:2013aa}.

The structure of the gaseous halo is also fixed to reduce the degree of freedom, and we adopt the $\beta$-model for gaseous halos for all galaxies with different masses, which has the density profile:
\begin{equation}
\rho(r) = n_0 \left(1+\frac{r^2}{r_{\rm c}^2}\right)^{-3\beta/2},
\end{equation}
where $n_0$ is the normalization parameter, and $r_{\rm c}$ is the core radius. Normally, core radii for galaxies are small, and cannot be modeled for isolated galaxies \citep{Li:2016aa}. Then, we rewrite the profile as 
\begin{equation}
\rho(r) = \frac{n_0r_{\rm c}^{3\beta}}{r^{3\beta}},
\end{equation}
which is valid for $r >> r_{\rm c}$, and then the degeneracy of $n_0r_{\rm c}^{3\beta}$ will not be broken. X-ray imaging studies on nearby massive galaxies showed that the $\beta$ factor is around $0.5$ within the radius $\leq 50\rm~kpc$ \citep{Anderson:2016aa}. Recently, the Circum-Galactic Medium of MASsive Spirals (CGM-MASS) project shows that $\beta$ is a constant of $\approx 0.4$ extended to around the half of virial radius ($\approx 200\rm~kpc$ for massive star-forming spiral galaxies; \citealt{Li:2017aa}). Therefore, we adopted $\beta$ as a constant over all of the radius range for one gaseous halo, but $\beta$ can be varied for different models. 

Since the total mass of the $\beta$-model is not convergent with increasing radius, we need to set the radius range for this model. In the inner region, other physical processes occur (e.g., the interaction with disk gases), therefore, the hydrostatic assumption is broken, and the $\beta$-model may not be applicable. This radius is set by the competition between the free-fall timescale and the radiative cooling timescale, which is around $5-10\rm~kpc$ using the radiative cooling timescale of the MW from \citet{Miller:2015aa}. Massive galaxies have larger inner radii that can be larger than $10\rm~ kpc$, but our model does not show significant dependence on the innermost radius. From $5~\rm kpc$ to $10~\rm kpc$, the mass of the gaseous halo is increased by up to $15\%$, which is only for the most massive galaxies ($M_{\rm h}>10^{13}~M_\odot$) due to their small cooling radii. For $L^*$ galaxies, this change is smaller than $10\%$, therefore, we fix the innermost radius as $5\rm~kpc$ for all galaxies. For the outer region, the maximum radius is set to the virial radius for a given halo mass, which means that the density goes to zero at the virial radius. However, it is shown that the massive system (galaxy cluster) could have detectable gas reaching $R_{200}$, which implies that the gaseous component could extend beyond the virial radius \citep{Baldi:2012aa}. Therefore, this assumption may not be correct, however, there are no other means to set an unbiased boundary condition.

The normalization parameter $n_0$ is calculated based on the assumption that the SFR is equal to the radiative cooling rate:
\begin{equation}
{\rm SFR} = \int_{5\rm kpc}^{{\rm min}\{R_{200}, R_{\rm cl}\}} \frac{\overline{\Lambda}(r) n^2(r)}{\overline{\epsilon}(r)} \mu m_{\rm p}4\pi r^2 {\rm d}r,
\end{equation}
where $\overline{\Lambda}(r)$ is the average radiative cooling emissivity, while the $\overline{\epsilon}(r)$ is the average internal energy at a given radius, defining as
\begin{eqnarray}
\overline{\Lambda} &=& \frac{\int_{T_{\rm min}}^{T_{\rm max}} M(T) n(T) \Lambda(T) {\rm d}T}{\int_{T_{\rm min}}^{T_{\rm max}} M(T) n(T) {\rm d}T},\\
\overline{\epsilon} &=& \int_{T_{\rm min}}^{T_{\rm max}} M(T) \frac{3}{2} k_{\rm B}T {\rm d}T.
\end{eqnarray}
$M(T)$ is the mass distribution depending on the temperature, which has the normalization of $\int_{T_{\rm min}}^{T_{\rm max}} M(T) {\rm d}T = 1$. Here, we also assume that the average mass of particles ($\mu$) is $0.59$ for the temperature range considered ($T>10^{4.5}\rm~K$), since this value is dominated by the ionization state of the hydrogen, which is almost completely ionized in this temperature range. The choice on the radiative cooling model will be discussed in the following section.

The radial dependence of temperature is still observationally poorly constrained for isolated galaxies. X-ray studies on galaxy clusters showed that the temperature variation is less than one order of magnitude within $R_{500}$ \citep{Baldi:2012aa}. For isolated star-forming galaxies, \citet{Anderson:2016aa} showed that NGC 1961 also has a small variation, but only out to $\sim 50\rm~kpc$. Here, we assume that there is no radius-dependence of the temperature.

\subsection{Cooling Emissivity}
The radiative cooling rate is directly affected by the emissivity, which has a dependence on the temperature, the density and the metallicity. For the temperature range of a gaseous halo ($\sim 10^{4.5} - 10^7\rm~K$), the radiative cooling is dominated by lines of various ions. Therefore, for a given temperature and density, the ionization state of different ions can be determined and the cooling rate is calculated involving the metallicity. Here, we assume that the gaseous halo is in ionization equilibrium, and consider two ionization processes -- collisional ionization and the modification due to photoionization.

For collisional ionization equilibrium (CIE), we adopt the emissivity calculated using CHIANTI (version 8.0.6; \citealt{Del-Zanna:2015aa}). In this calculation, the metallicity is set to $0.1~Z_\odot$, $0.3~Z_\odot$, $1.0~Z_\odot$ and $2.0~Z_\odot$, and the solar metallicity of $Z_\odot = 0.0142$ is adopted from \citet{Asplund:2009aa}.

The photoionization due to the ultraviolet background (UVB) can modify the ionization distribution of different elements \citep{Wiersma:2009aa}, and the photoionization model is employed to model the low and intermediate ionization ions in intervening systems \citep{Savage:2014aa, Werk:2014aa}. Also, the high ionization state ions might be photoionized at low densities of $\lesssim 10^{-5}\rm~cm^{-3}$, which is the expected density in the outskirts of gaseous halos \citep{Hussain:2015aa, Hussain:2017aa}. Therefore, we include the photoionization from the UVB to compare with the pure collisional ionization mode.

Galaxies also provide a part of the ionizing flux to photoionize the CGM or nearby IGM, which is known as the escaping ionizing flux. The escape fraction is believed to be large ($\gtrsim 10\%$) in the early Universe ($z>6$) to contribute to the re-ionization \citep{Mitra:2013aa}, while studies of the low redshift IGM ($z<2$) found the escape fraction is several percent \citep{Khaire:2015ab}. The small escape fraction has the implication that those ionizing photons mainly affect the innermost $\sim 50\rm~kpc$ region of the gaseous halo, thus, we ignore ionizing photons from the galaxy disk \citep{Suresh:2017aa}.

\begin{figure}
\begin{center}
\includegraphics[width=0.48\textwidth]{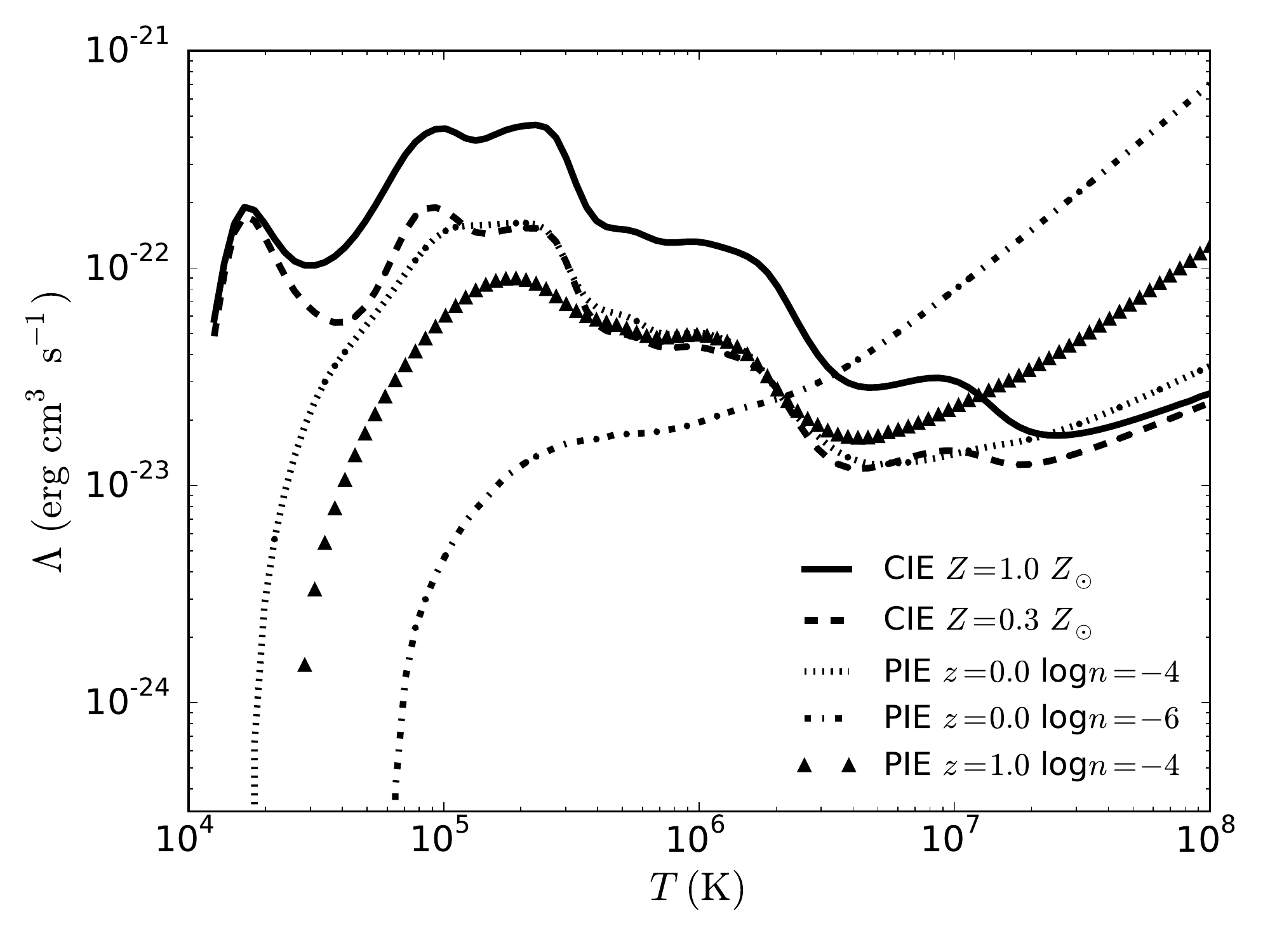}
\end{center}
\caption{Comparison between the cooling curve for pure collisional ionization (CIE) and with the modification from photoionization (PIE). The cooling curves in CIE have metallicities of $1~Z_\odot$ (the solid line) and $0.3~Z_\odot$ (the dashed line). The three PIE cooling curves all have the same metallicity of $0.3~Z_\odot$. The dotted line has a density of $10^{-4}~\rm cm^{-3}$ (typical of the density of the inner gaseous halo) and at $z=0$, while the dash-dotted line has a lower density of $10^{-6}\rm~cm^{-3}$ (typical of the density in the halo  outskirts) at the same redshift. The up-triangle shows the cooling curve at $z=1$ with a density of $10^{-4}\rm~cm^{-3}$.}
\label{cooling}
\end{figure}

For the photoionization equilibrium (PIE), we adopt the calculation from \citet{Oppenheimer:2013aa}, who tabulated results for different redshifts, densities, and temperatures. Several UVB models have been provided, and we choose the UVB form \citet{Haardt:2012aa} in our models. In this database, authors also include the cosmic microwave background (CMB), with the dependence on the redshift. The existence of the CMB provides a large number of low-energy photons, which can be heated by inverse Compton scattering, thereby cooling the high-temperature electrons. 

In Fig. \ref{cooling}, we show the comparison between CIE and PIE cooling curves. High energy photons from the UVB photoionize low ionization state ions to higher states, which suppresses the cooling in the low-temperature region. Due to the lack of {\HI}, the first peak around $2\times 10^4\rm~K$ is missing. The photoionization also changes the ionization fraction of metals and contributions to the radiative cooling, suppressing low ionization state cooling (e.g., {\CII} and {\OII}) and increasing high ionization cooling (e.g., {\OVI}). Therefore, the cooling emissivity is lower in the low-temperature regime for the PIE model than the CIE model. Inverse Compton cooling due to CMB dominates the high temperature and low-density gas. The emissivity of inverse Compton scattering is proportional to $nT$, while the free-free emission has the dependence $n^2T^{1/2}$. Therefore, there is always a critical combination of temperatures and densities, above which the inverse Compton cooling is dominant. However, in the low redshift Universe ($z<2$), the number density of CMB photons is sufficiently low so that gases have a cooling timescale longer than the Hubble timescale. Therefore, the effect due to the CMB can be ignored for the low redshift ($z< 2$) Universe. The effect of the radiative cooling model will be described in details in Section 3.

\subsection{Temperature Dependence of the Mass Distribution}
Multi-phase gas in gaseous halos have been detected by various observations \citep{Nicastro:2002aa, Danforth:2008aa, Anderson:2013aa, Werk:2013aa, Savage:2014aa, Qu:2016aa}. Unfortunately, obtaining an accurate distribution of the multi-phase medium by mass remains a challenge both observationally or theoretically (e.g., the divergence on {\OVI} abundance; \citealt{Oppenheimer:2016aa, Suresh:2017aa}). Therefore, for the simplest model, we assume that the gaseous halo is a single phase medium at the virial temperature.

\begin{figure}
\begin{center}
\includegraphics[width=0.48\textwidth]{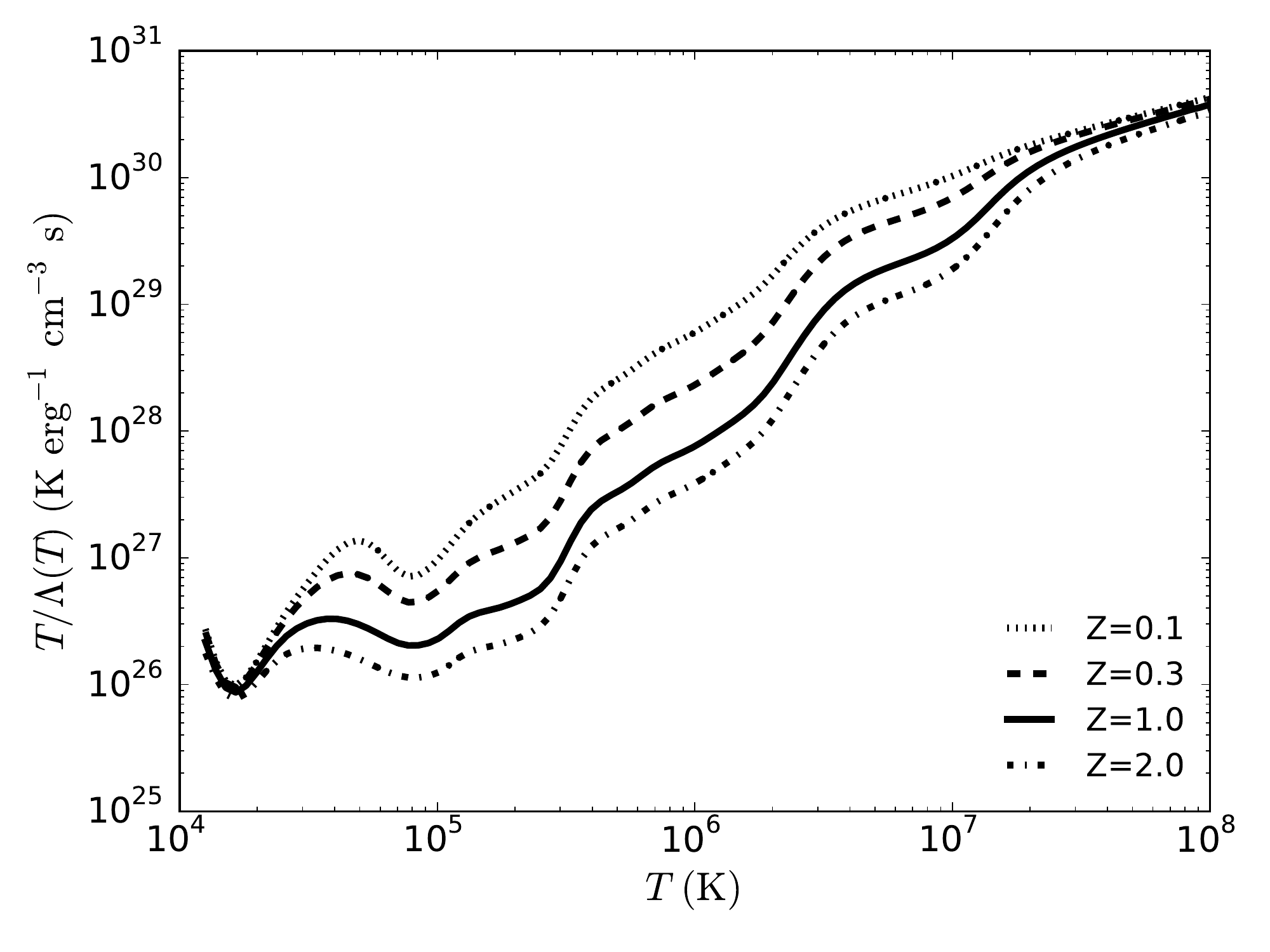}
\end{center}
\caption{The unnormalized mass distribution for cooling gas as a function of temperature. {\tt CIE} cooling curves with metallicities  of $0.1~Z_\odot$, $0.3~Z_\odot$, $1.0~Z_\odot$ and $2.0~Z_\odot$ are shown in dotted, dashed, solid and dot-dashed lines, respectively. With temperature limits, which are related to the galaxy mass, this function can be normalized as $\int_{T_{\rm min}}^{T_{\rm max}} M(T) {\rm d}T = 1$ to obtain the mass distribution $M(T)$.}
\label{MT}
\end{figure}

We also consider a stable cooling model, which is a time-independent solution. In this model, we assume the mass cooling rate is the same at all temperatures:
\begin{equation}
L(T) = \Lambda(T) n^2(T) \frac{M(T)}{\mu m_{\rm p}n(T)} = const.,
\end{equation}
where $L(T)$ is the luminosity at a given temperature $T$, and $M(T)$ is the mass distribution dependence on the temperature. Another assumption is the pressure balance, which implies $n(T) \propto 1/T$. Thus, the mass distribution is 
\begin{equation}
M(T) = \frac{T}{\Lambda(T)}/\int_{T_{\rm min}}^{T_{\rm max}} \frac{T}{\Lambda(T)} {\rm d}T,
\end{equation}
the temperature upper limit ($T_{\rm max}$) is set to the virial temperature, while the lower limit is fixed to $10^{4.5}\rm~K$, under which forbidden lines dominate the cooling, along with dust and molecules. An example of $M(T)$ without normalization is shown in Fig. \ref{MT}.

\begin{figure*}
\begin{center}
\subfigure{
\includegraphics[width=0.48\textwidth]{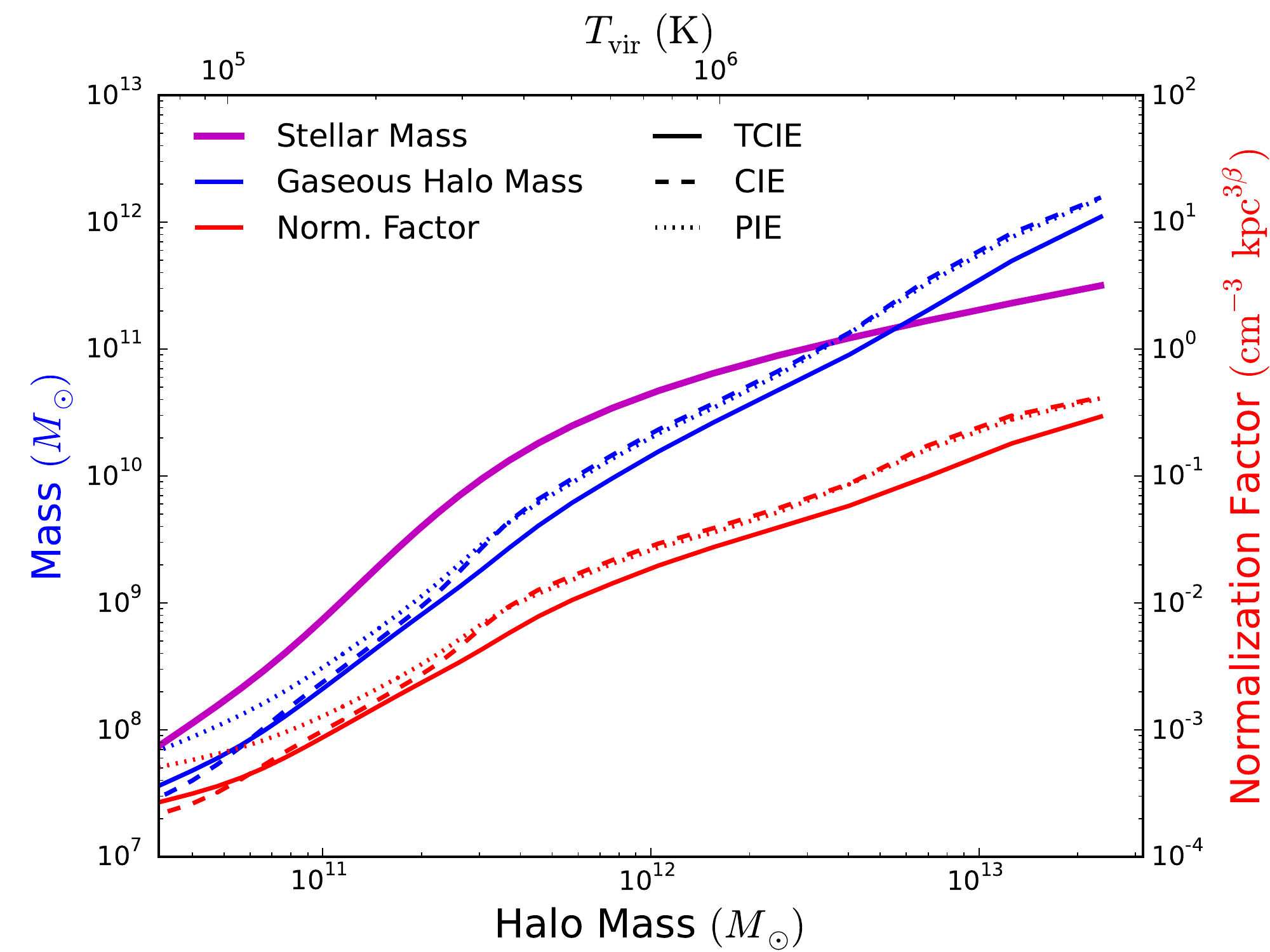}
\includegraphics[width=0.48\textwidth]{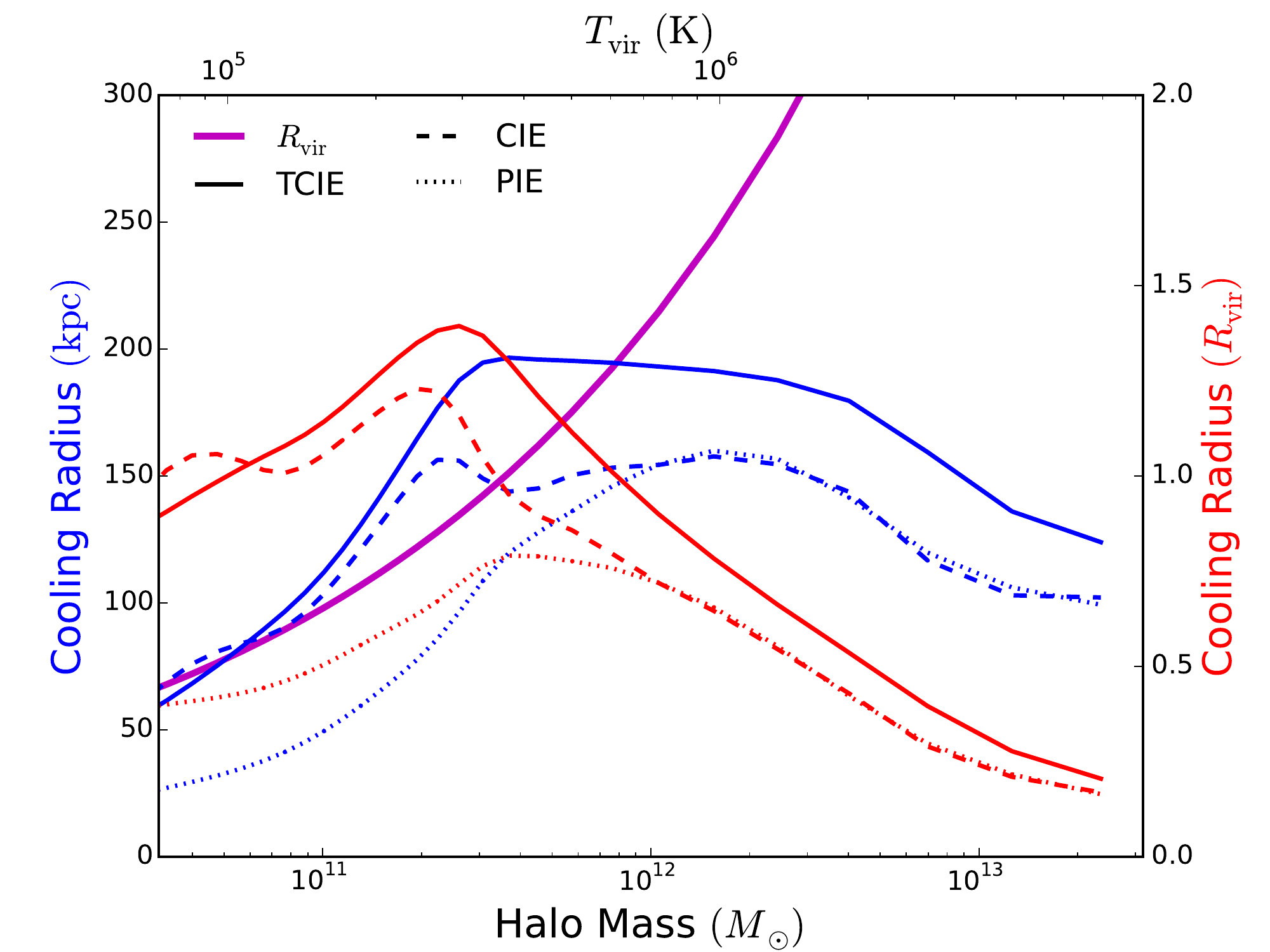}
}
\end{center}
\caption{{\it Left panel}: The gaseous halo mass and the normalization factor of different models at $z=0$. The blue lines are the gaseous halo masses, while the red lines are the normalization factors in the $\beta$-model. {\tt CIE}, {\tt PIE} and {\tt TCIE} models are shown in dash-dotted, dashed, and solid lines. The magenta line is the stellar mass from stellar mass-halo mass relationship \citep{Kravtsov:2014aa}. {\it Right panel}: The cooling radius, the radius within which the cooling time equals the local Hubble time, as a function of halo mass. The blue lines are the absolute cooling radius (left scale), while the red lines are the cooling radius in the unit of the virial radius (right scale). The range in the cooling radius only changes by a factor of four over the range in which the halo mass changes by three orders of magnitude. }
\label{fiducial}
\end{figure*}

Our model does not include all relevant physics that occurs in galaxy halos (e.g., thermal instabilities), but it allows us to explore a wide range of parameter space and to identify robust results. Detailed calculations show that the stable cooling model has applicability for the cooling in a temperature range of $10^4\rm~K$ to $10^{6.5}\rm~K$ in stellar feedback dominated galaxies \citep{Thompson:2016aa}. Their breaking of this cooling assumption in the high-temperature range is mainly because they consider the hot gas from the stellar feedback, which softens the assumed boundary condition that the high-temperature gas can be supplied infinitely. However, in the gaseous halo scenario, this condition could be satisfied when a hot and long radiative-cooling timescale gaseous halo exists.

\section{Results}
We calculate three models with different cooling models and temperature distributions -- {\tt CIE}: the single temperature collisional ionization model; {\tt {\tt PIE}}: the single temperature photoionization model; {\tt TCIE}: the collisional ionization model with the mass distribution described in Section 2.4. In this section, we show our main results for these models on the gaseous halo mass and the ion column density.

\subsection{Fiducial Galaxies}
There are four factors affecting the properties of the gaseous halo in our simplified models -- the metallicity ($Z$), the specific star formation rate ($\rm sSFR$ defined as ${\rm SFR}/M_{\star}$), the slope of the $\beta$-model ($\beta$) and the redshift ($z$). Based on these four dimensions, we have fiducial galaxies defining as $\log M_{\rm h} = 10.5-13.3$, $Z=0.3 ~Z_{\odot}$ (cosmic metallicity), $\rm sSFR=10^{-10}~yr^{-1}$ (star-forming), $\beta=0.5$ (hydrostatic equilibrium structure) and $z=0$. For each modeled gaseous halo, we calculate the gaseous halo mass enclosed in the radius range of $5\rm~ kpc$ to the virial radius, and the cooling radius. The calculation results are shown in the Fig. \ref{fiducial}.

For the fiducial case, all three models have masses of gaseous halos that are smaller than corresponding stellar masses around the (sub-)$L^*$ galaxies. The largest difference of $\approx 0.5\rm ~dex$ occurs at sub-$L^*$ galaxies ($M_{\rm h} \approx 4\times 10^{11}~M_\odot$). Overall, the {\tt CIE} model shows convergence with {\tt TCIE} in the low-mass region and converges with {\tt PIE} for massive galaxies. With the halo mass decreasing, the temperature range for {\tt TCIE} (with $T_{\rm min} = 3\times 10^{4}$) is also decreasing, which leads to the similarity with {\tt CIE}. Both of collisional ionization models have lower mass gaseous halos than the {\tt PIE} model because they have a higher radiative emissivity in the low temperature, and the photoionization due to the UVB can support a relatively more massive halo for low-mass galaxies.

In the massive galaxy range, the radiative cooling is reduced at high temperatures, which results in a massive gaseous halo, consistent with theoretical expectations \citep{Mo:2010aa}. These halos are supported by their buoyancy even for the {\tt PIE} model. The convergence between {\tt CIE} and {\tt PIE} is due to the higher density in massive galaxies -- in the inner region (inside of the cooling radius), the average density is higher than $10^{-4}\rm~cm^{-3}$. This high density corresponds to the low ionization parameter ($U=n_{\rm ph, ionizing}/n_{\rm H}$), indicating the weakening of the photoionization. As shown in Fig. \ref{cooling}, the CIE cooling is consistent with the PIE cooling with a density of $10^{-4}\rm~cm^{-3}$. {\tt TCIE} has a lower mass gaseous halo than {\tt CIE} or {\tt PIE}, due to its higher average emissivity since this model always has low-temperature gas with a higher radiative emissivity.

\begin{figure*}
\begin{center}
\includegraphics[width=0.98\textwidth]{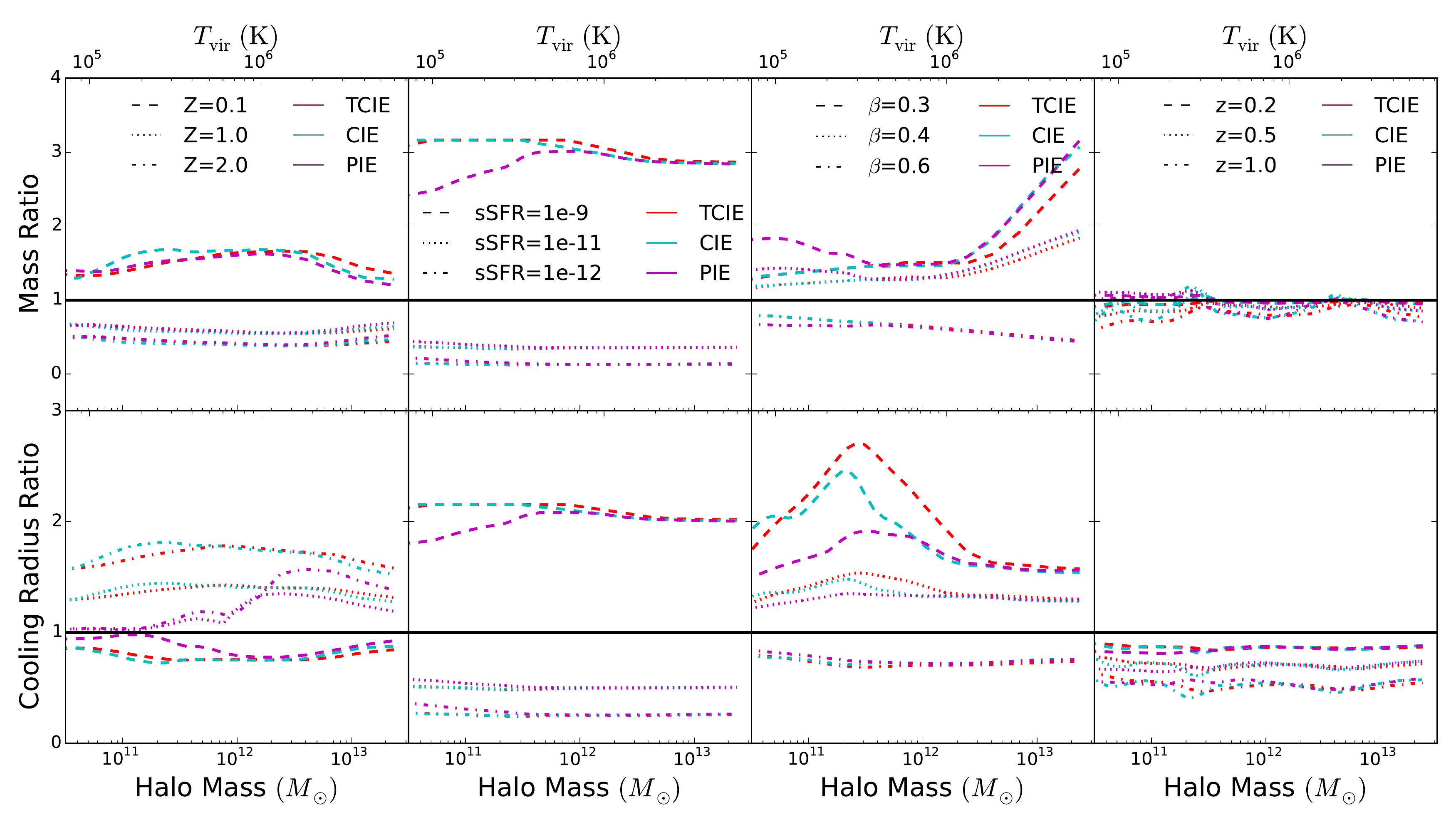}
\end{center}
\caption{Gaseous halo properties as a function of the halo mass for variations in the metallicity (left), the sSFR, the slope of the density profile, and the redshift (right). {\tt CIE}, {\tt PIE} and {\tt TCIE} models are shown in cyan, magenta and red, and the ratios are the values relative to the functional form of the fiducial galaxy has parameters of $Z=0.3~Z_\odot$, ${\rm sSFR} = 10^{-10}\rm~yr^{-1}$, $\beta=0.5$, and $z=0$ (in the solid black lines).}
\label{halo_prop}
\end{figure*}

Overall, the cooling radius varies only modestly over the halo mass range in each model. Specifically, the variation is less than one order of magnitude, and this variation corresponds to the changes in the average emissivity. With the higher emissivity, the cooling radius is larger, however, the changes in the cooling radius is smaller than the emissivity changes. Meanwhile, the cooling radius shows a similar convergence as the gaseous halo mass between {\tt CIE} and {\tt TCIE} for low-mass galaxies, and between {\tt CIE} and {\tt PIE} for massive galaxies.

\subsection{The Effect of Galaxy Properties}
By changing the four parameters ($Z$, $\rm sSFR$, $\beta$ and $z$), we show the effect of these parameters on the resulting hot halo and column densities. For each parameter, we have four choices: the metallicity -- $0.1~Z_\odot$, $0.3~Z_\odot$, $1~Z_\odot$, or $2~Z_\odot$; the specific SFR -- $10^{-9}\rm~yr^{-1}$, $10^{-10}~\rm yr^{-1}$, $10^{-11}~\rm yr^{-1}$, or $10^{-12}~\rm yr^{-1}$; the $\beta$ parameter -- $0.3$, $0.4$, $0.5$, or $0.6$; and the redshift -- $0$, $0.2$, $0.5$, or $1$. In Fig. \ref{halo_prop}, we show the change corresponding to these parameters as the ratio between varied models and the fiducial model. 

The high metallicity increases the cooling emissivity, which reduces the normalization parameter in the $\beta$-model, and subsequently the halo mass. With the variation of metallicity in the range $0.1-2.0~Z_\odot$, the change of the gaseous halo mass is less than a factor of $5$, whereas the change of metallicity is a factor of 20. This implies that lower metallicity gaseous halos have a lower total metal mass to account for the same cooling rate. The cooling radius of {\tt CIE} and {\tt TCIE} models has a positive dependence on the metallicity due to the increase of emissivity. However, inclusion of photoionization shows a similar cooling radius for different metallicities in the low-mass end, which indicates that the radiative cooling due to the low ionization metal ions is suppressed by the photoionization.

A $\rm sSFR$ of $10^{-11}~\rm yr^{-1}$ is used as a boundary between a star-forming galaxy and a quiescent galaxy, while normal star-forming galaxies have $\rm sSFR$ around $10^{-10}$ \citep{Renzini:2015aa}. By increasing the $\rm sSFR$, the total radiative cooling rate is increased, which means that a massive gaseous halo is needed. For {\tt CIE} and {\tt TCIE}, this effect is almost a constant over all mass regions, and {\tt PIE} shows a similar tendency in the high mass region ($>10^{12}~M_\odot$). However, for low-mass galaxies, {\tt PIE} models with different sSFR values show a significant convergence of the gaseous halo mass, which indicates the effect of a changing sSFR is not as large as {\tt CIE} or {\tt TCIE}. The reason for these phenomena is that there are two ways to increase the radiative cooling rate -- higher density or higher emissivity. In {\tt CIE} and {\tt TCIE} models, the emissivity cannot be increased when the temperature distribution is fixed. Therefore the only way to raise the cooling rate is by increasing the density, which makes the density proportional to the square root of the sSFR for all halo masses. In the {\tt PIE} model, the emissivity has a dependence on the density as shown in Fig. \ref{cooling}. Within the cooling radius, the density is higher than $10^{-4}\rm~cm^{-2}$, and the PIE cooling curve does not deviate from the CIE cooling curve significantly in the temperature range of $\approx 10^{5.5}-10^6 \rm~K$. For galactic gaseous halos with these temperatures ($M_{\rm h} > 10^{12}~M_\odot$), the emissivity shows a similar behavior as the CIE models, therefore, the change of the density (and the gaseous halo mass) is also similar to the CIE models. For low-mass galaxies with low-temperature halos, the PIE cooling deviates from the CIE cooling curve significantly since the cooling is suppressed for low ionization state ions. Raising the density decreases the photoionization effect, and increases the emissivity to the value of CIE models. Therefore, in the {\tt PIE} model, the high density not only increases the cooling rate by the squared dependence on the density itself but also increases the emissivity, which leads to a smaller change in the density to account for the high sSFR.

The variation of the $\rm sSFR$ is equivalent to changing the $\gamma$ factor with the $\rm sSFR$ unchanged. For an example, $\rm sSFR = 10^{-9}$ with $\gamma = 1$ is the same model as $\rm sSFR = 10^{-10}$ but with $\gamma = 0.1$. Therefore, our models show that the gaseous halo mass has a square root dependence on the inverse $\gamma$ factor.

\begin{figure*}
\begin{center}
\subfigure{
\includegraphics[width=0.48\textwidth]{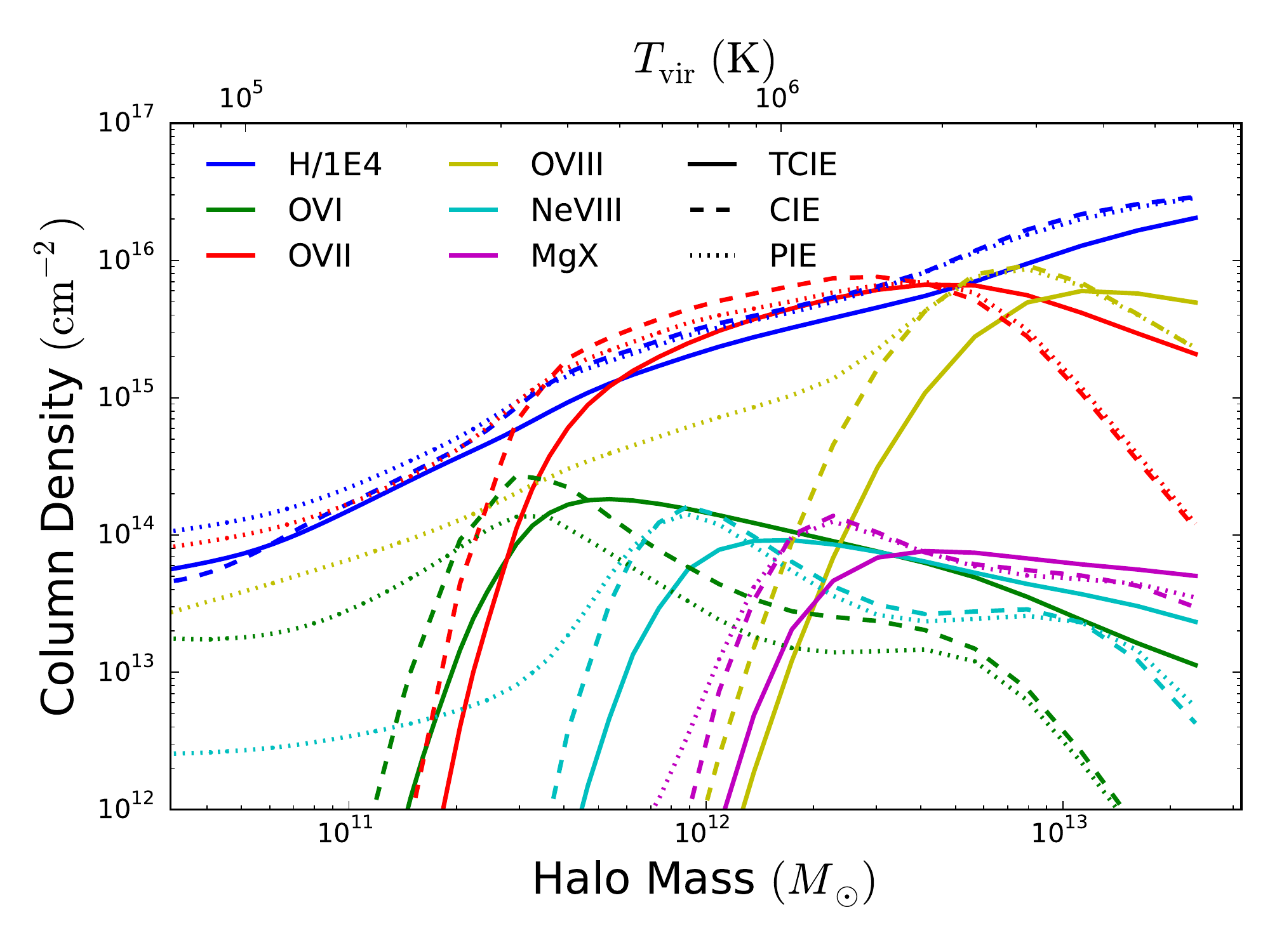}
\includegraphics[width=0.48\textwidth]{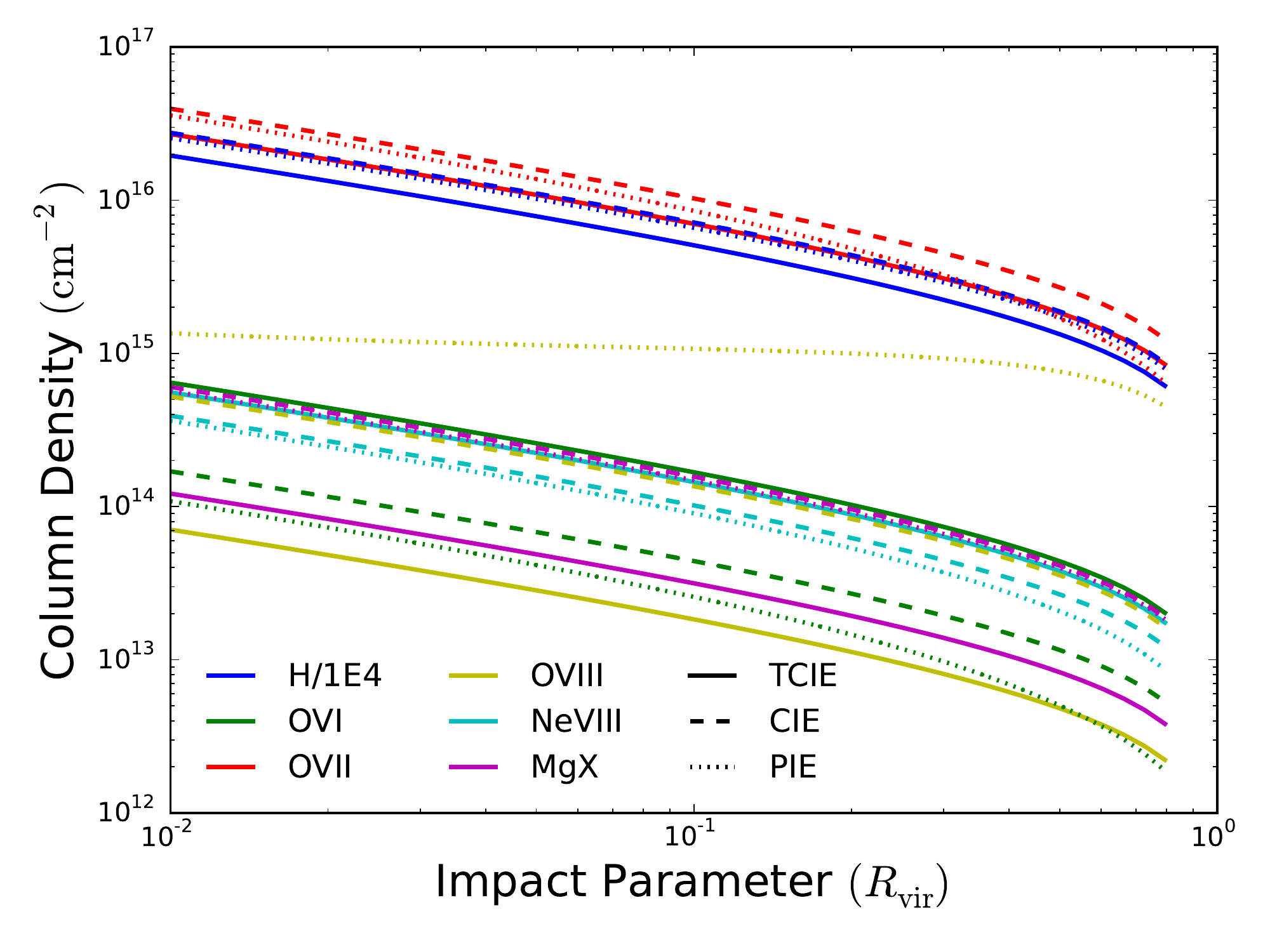}
}
\end{center}
\caption{Comparison of the ion column density between three models, {\tt CIE}, {\tt PIE} and {\tt TCIE} at $z=0$, which are shown in dashed, dotted and solid lines, respectively. Different ions are shown in different colors: blue -- hydrogen (reduced by a factor of $10^{4}$); green -- {\OVI}; red -- {\OVII}; yellow -- {\OVIII}; cyan -- {\NeVIII}; and magenta -- {\MgX}. {\it Left panel}: The column density dependence on the halo mass. The galaxy sample is the fiducial galaxy locus, and the impact parameter is fixed to $0.3~R_{\rm vir}$. {\it Right panel}: The column density dependence on the impact parameter for the galaxy with $M_\star = 7\times 10^{10}~M_\odot$, and ${\rm SFR} = 3~M_\odot\rm~yr^{-1}$, $Z=0.3~Z_\odot$, and $\beta =0.5$.}
\label{ions_models}
\end{figure*}

With a larger $\beta$, the gas is more concentrated in the central region (at the same gas mass), which leads to the higher emissivity. Since the mass is linearly dependent on the density, the larger $\beta$ results in a smaller gaseous halo mass. Due to the concentrated emission, the cooling radius is also decreases as $\beta$ is increases. The effect of larger $\beta$ also has a dependence on the halo mass -- with a more massive halo, the ratio of masses is larger, as shown in Fig. \ref{halo_prop}. This correlation occurs because the massive galaxy has a relatively small cooling radius compared to the virial radius, and the flat $\beta$-model can host more mass in the region out of the cooling radius.

A higher redshift can affect the gaseous halo through three means -- the higher gas density, the younger Universe age and the more intense cosmic background. First, due to the Universe expansion, the higher redshift Universe has a higher density, which leads to the smaller virial radius, and hence higher virial temperature. Second, the younger Universe age at the higher redshift determines a shorter cooling timescale, which reduces the cooling radius and the total cooling rate within this radius. Since {\tt CIE} and {\tt TCIE} models have no photoionization involved, these two models are only affected by these two factors. The gaseous halo mass has an anti-correlation with the cooling emissivity of the gas, but the emissivity is a result of a competition of two factors -- the increasing emissivity due to the higher density and the changing due to virial temperature. Together, these complex effects leads to a small variation of the gaseous halo mass, with small variations reflecting the shape of the cooling curve (i.e., the bump around $T_{\rm vir} \approx 10^{5.5}\rm~ K$, also the peak of the cooling curve). For the cooling radius, the effect is clear that the higher redshift leads to a smaller cooling radius. The cooling radius of a $z=0$ galaxy is about 1.5 times larger than the same mass galaxy at $z=1$ galaxies. 

The cosmic background includes two parts -- the UVB and the CMB. As stated in Section 2.3, the inverse Compton cooling is negligible in the low-redshift universe ($z<6$), but the UVB changes the ionization state distribution, leading to the reduced radiative cooling in the low-temperature region. Therefore, for the low-mass galaxies ($M_{\rm h} \lesssim 10^{11.3}~M_\odot$), a more massive gaseous halo is required to account for the same SFR with the UVB increasing at the higher redshift, which results in the mass ratio being slightly larger than 1. For the high-temperature end, the {\tt PIE} model converges with the {\tt CIE} model, which is expected.

\begin{figure*}
\begin{center}
\subfigure{
\includegraphics[width=0.48\textwidth]{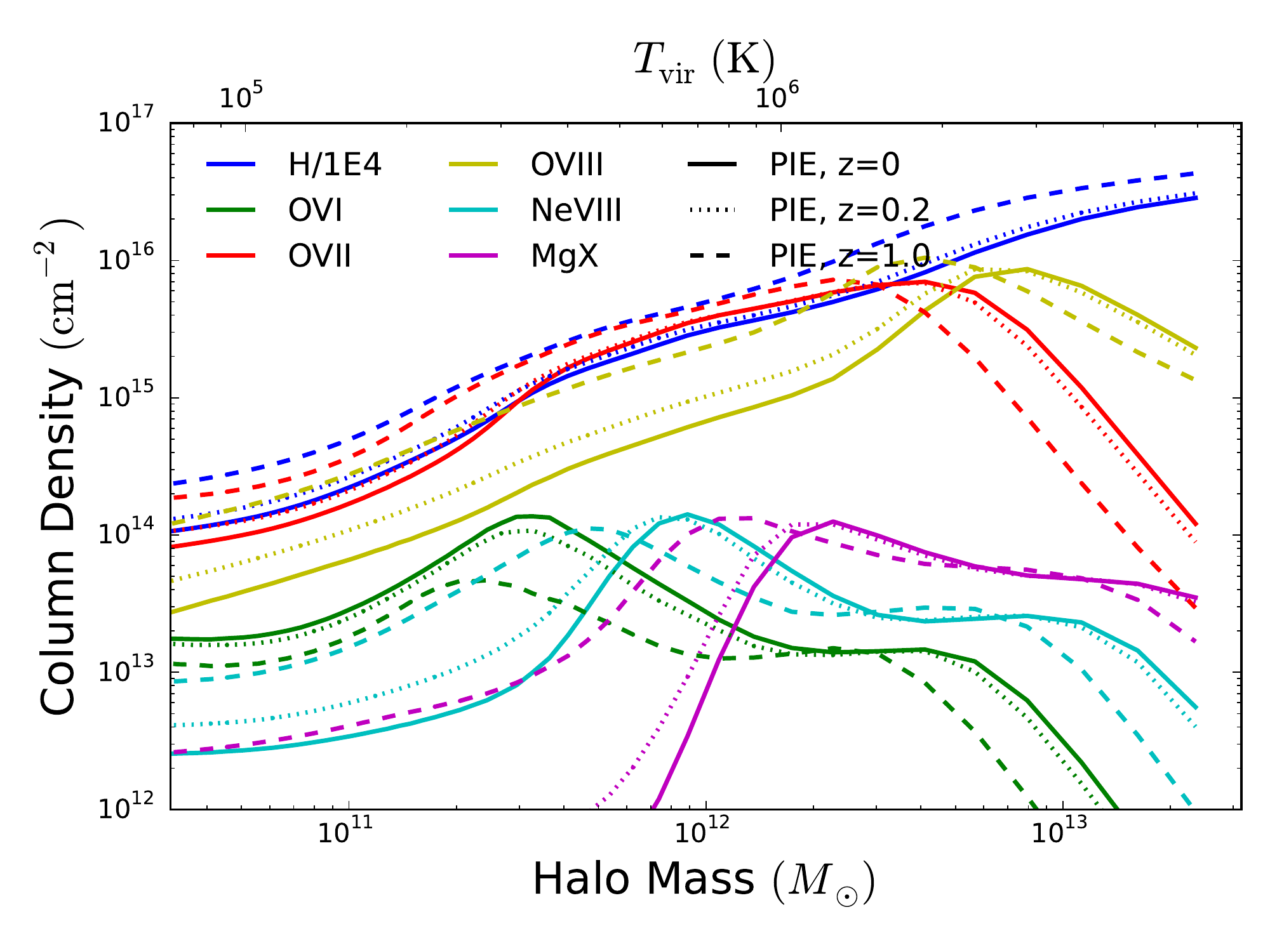}
\includegraphics[width=0.48\textwidth]{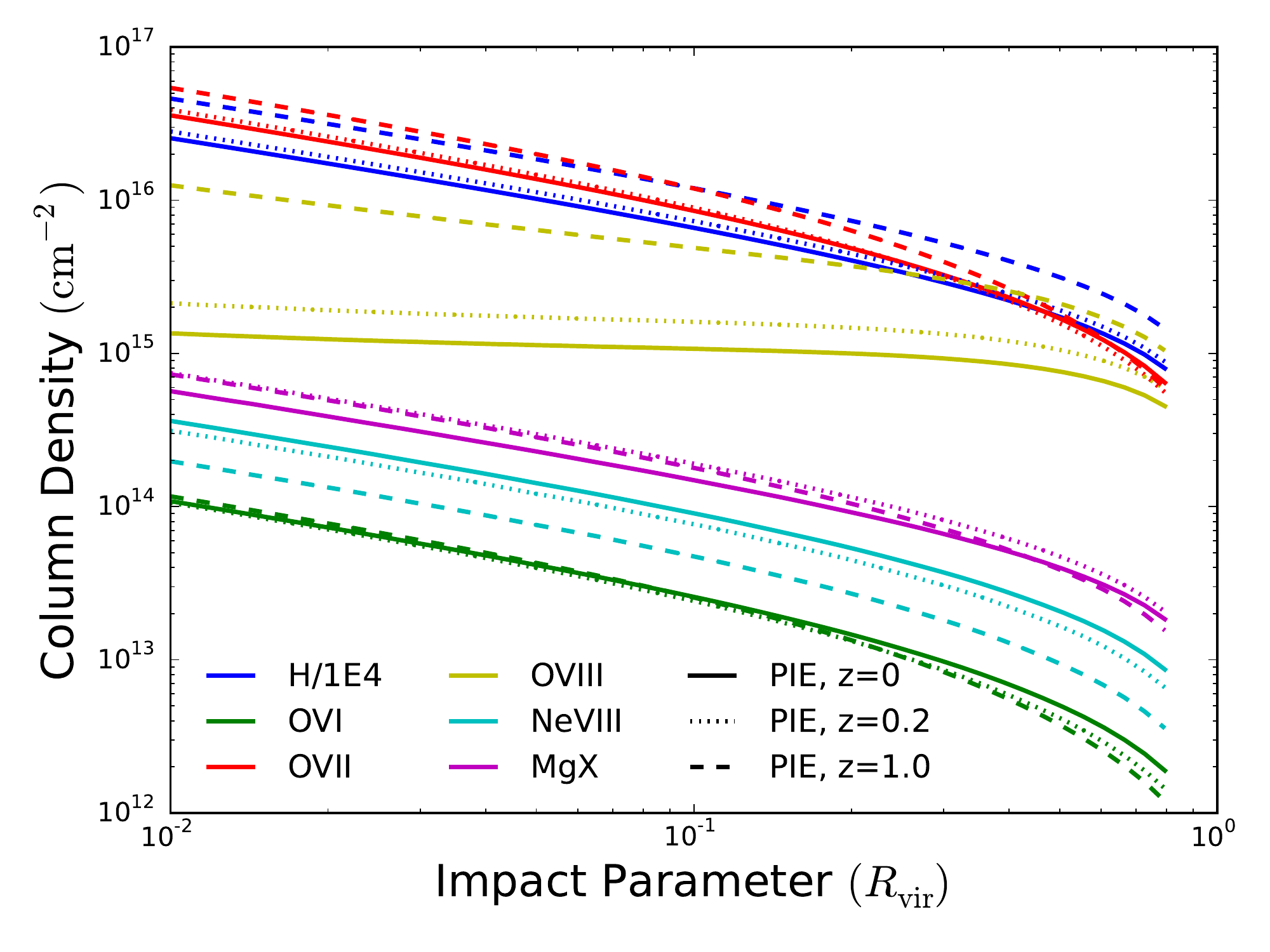}
}
\end{center}
\caption{Comparison of the ion column density of the {\tt PIE} model at different redshifts. Ions are encoded in the same colors as Fig. \ref{ions_models}, while solid, dotted, and dot-dashed lines are {\tt PIE} models with $z=0$, $z=0.2$, $z=1.0$, respectively. The galaxy sample configuration is also the same as Fig. \ref{ions_models}.  {\it Left panel}: The column density dependence on the halo mass. Since the virial temperature has the dependence on the redshift, the shown $T_{\rm vir}$ is at $z=0$. {\it Right panel}: The column density dependence on the impact parameter. The higher redshift leads to a higher virial temperature, and shifts the peaks of ions to lower-mass galaxies. Both the higher virial temperature and more intense UVB increase the higher ionization state ion column densities for higher redshift galaxies.}
\label{ions_z}
\end{figure*}

\subsection{The Ion Column Densities}
With the calculated density profile and the temperature distribution, we calculate the column density for ions of interests (mainly high ionization state ions), which are more common in the hot ambient medium. For {\tt CIE} and {\tt TCIE} models, we adopt the ionization distributions from \citet{Bryans:2006aa}, which only has a dependence on the temperature. The {\tt PIE} ionization fraction is adopted from \citet{Oppenheimer:2013aa}, which is tabulated based on the redshift, the metallicity, the density and the temperature.

For the {\tt TCIE} model, we calculate the average ionization fraction using
\begin{equation}
\overline{f}_{\rm i} = \int_{T_{\rm min}}^{T_{\rm max}} f_{\rm i}(T) M(T) {\rm d}T.
\end{equation}
Here, we assume that multi-phase medium has similar covering factors around 1, which implies that the multi-phase medium is well-mixed. This assumption should be good for high ionization state ions. \citet{Werk:2013aa} shows that intermediate ionization ions (i.e., {\CIII}, {\SiIII}, {\SiIV}) and high ionization ions (i.e., {\OVI}) have comparable covering factors around $0.8$, except that {\OVI} seems to be less in quiescent galaxies. However, this is probably caused by quiescent galaxies that are usually massive galaxies with higher virial temperatures \citep{Oppenheimer:2016aa}. %Overall, this assumption can be a useful assumption to transfer the mass distribution into the column density distribution.

In Fig. \ref{ions_models}, we compare the three models ({\tt CIE}, {\tt TCIE} and {\tt PIE}), showing ion column densities for star-forming galaxies (i.e., $Z=0.3~ Z_\odot$, $z=0$, $\rm sSFR = 10^{-10}\rm~yr^{-1}$ and $\beta= 0.5$). To show the dependence on the stellar mass, we fix the impact parameter to $0.3~R_{\rm vir}$, which is a typical impact parameter in the COS-Halos program and also leads to a similar column for ions observed in the MW from the Sun. Similar to the result of the gaseous halo mass, {\tt CIE} and {\tt PIE} shows the convergence of {\HI} column densities from $M_{\rm h} = 4\times 10^{11}~M_\odot$ and above, which indicates that the cooling emissivity is almost the same. However, other ions do not show the same similarity, which indicates that the ionization fractions are not similar. Only the most massive halo ($M_h>10^{13}~M_\odot$) has similar ionization fractions for {\tt CIE} and {\tt PIE} due to the relatively higher density. In low-mass galaxies (at the left side of the ionization peak for different ions), {\tt PIE} leads to extended tails for high ionization state ions (e.g., {\OVI} and {\OVII}) because of the low density.

Compared to {\tt CIE} or {\tt PIE} without cooling temperature distributions, {\tt TCIE} does not show the shape of the ionization fraction function directly, but it shows a flattened peak for high ionization state ions. For {\OVI} and {\OVII}, column density peaks are $\gtrsim 10^{14}\rm~cm^{-2}$ and $7\times 10^{15}\rm~cm^{-2}$, respectively. {\OVI} is higher than $ 10^{13.5}\rm ~cm^{-2}$ over a halo mass range of $2\times 10^{11} ~M_\odot$ to $4\times 10^{12}~M_\odot$. {\NeVIII} and {\MgX} show comparable flattened column density distributions in the range of $10^{13.5}-10^{14.0}\rm~cm^{-2}$, while {\NeVIII} occurs in lower mass galaxies compared to {\MgX}.

To show the ion column density dependence on the impact parameter, we choose a MW-like galaxy with $M_\star = 7\times 10^{10}~M_\odot$, ${\rm SFR} = 3~M_\odot\rm~yr^{-1}$, $Z=0.3$, $\beta = 0.5$ and $z=0$. In {\tt CIE} and {\tt TCIE} models, the ionization fraction does not have a dependence on the density, so the average ionization fraction has no dependence on the impact parameter. Therefore, all columns follow a general radial decrease of the $\beta$-model. For the {\tt PIE} model, the significant flattening of {\OVIII} in the small impact parameter region shows that the ionization fraction in the inner region is much smaller than the outer region, where the photoionization generates more {\OVIII} \citep{Oppenheimer:2013aa}. The turnover point is about the half of the virial radius, where about half of {\OVIII} is produced beyond this radius.

We consider the redshift dependence of the {\tt PIE} model in Fig. \ref{ions_z}, which is otherwise similar to Fig. \ref{ions_models}. At higher redshifts, the more intense UVB leads to a stronger tail of high ionization state ions in low-mass galaxies. Also, high ionization ions (e.g., {\OVIII}) show peaks at the lower halo mass for the high redshift galaxy, as the {\OVIII} column density peak moves from $M_{\rm h} \approx 7\times 10^{12}~M_\odot$ ($z=0$) to $M_{\rm h} \approx 4\times 10^{12}~M_\odot$ ($z=1$). This is mainly due to the increasing virial temperature at higher redshifts. For these higher virial temperatures, the {\OVIII} column density is no longer flat in the small impact parameter region at $z = 1$, which means that the ionization of {\OVIII} is no longer dominated by the photoionization. Although the high density in the inner region still reduces the ionization fraction of {\OVIII}, a significant amount of {\OVIII} is produced in the inner region through collisional ionization.

\begin{figure*}
\begin{center}
\subfigure{
\includegraphics[width=0.48\textwidth]{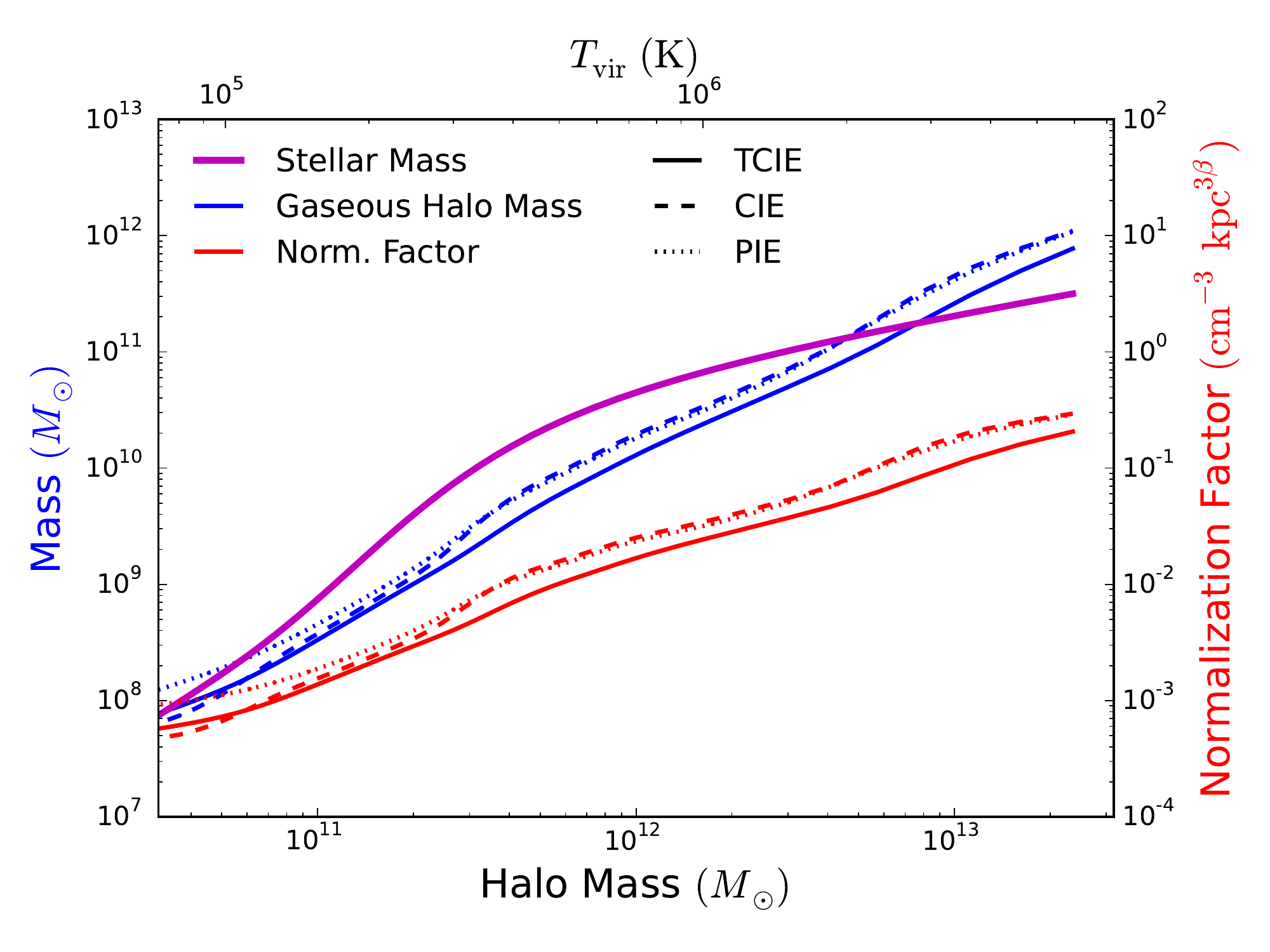}
\includegraphics[width=0.48\textwidth]{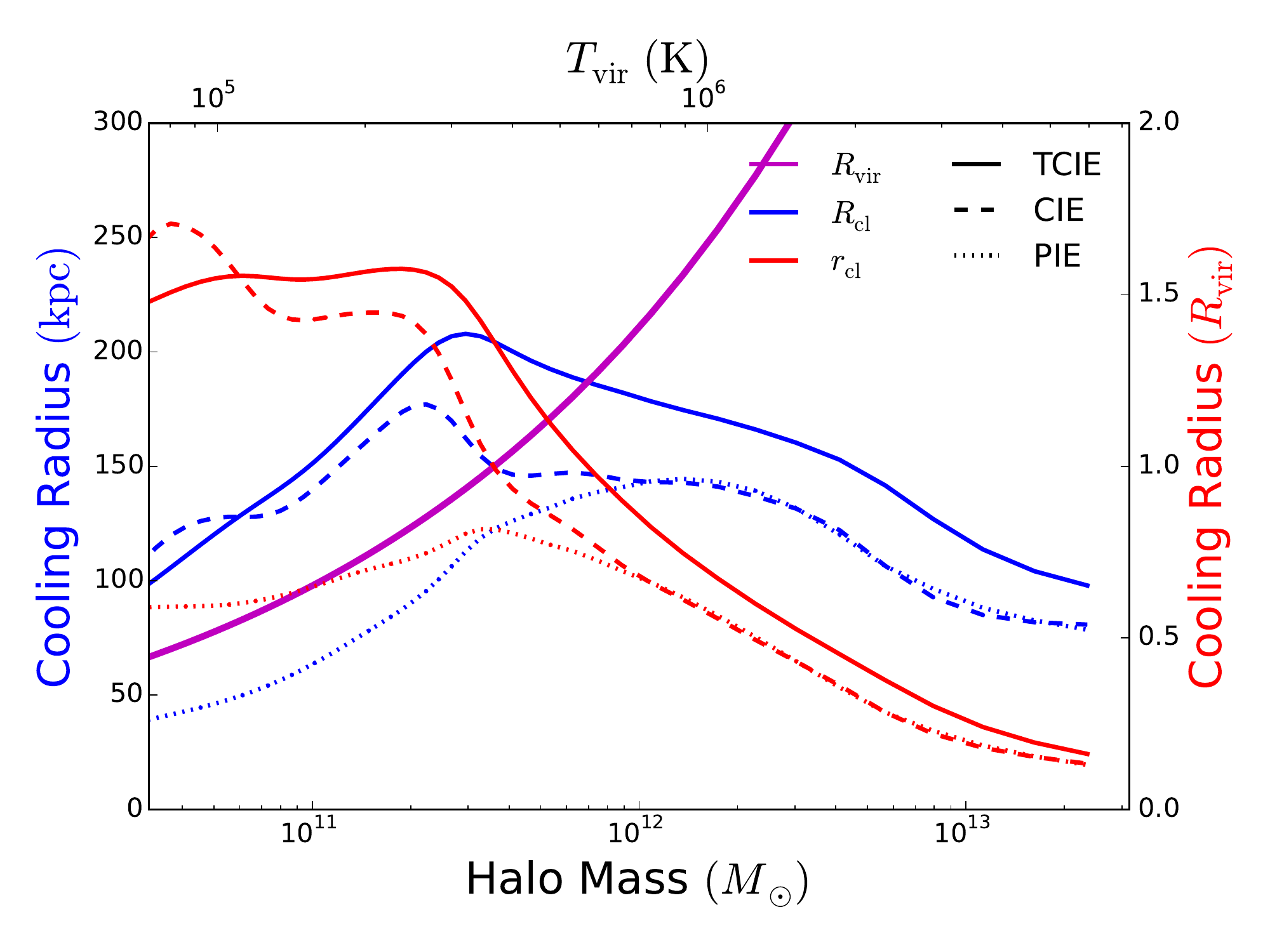}
}
\end{center}
\caption{The gaseous halo mass and the cooling radius of galaxies where the SFR is given by the mean fundamental plane SFR-stellar mass relationship of \citet{Morselli:2016aa}. {\it Left panel}: The gaseous halo mass and the normalization factor. {\it Right panel}: The cooling radius dependence on the halo mass. These two plots are encoded in the same way as Fig. \ref{fiducial}. The relatively higher sSFR for low-mass galaxies, leads to more massive gaseous halos and larger cooling radius, while massive galaxies have lower mass and smaller cooling radius compared to Fig. \ref{fiducial}.}
\label{typical}
\end{figure*}

\subsection{Galaxies with the SFR Main Sequence}
Using the relationship between the $\rm SFR$ and the stellar mass \citep{Morselli:2016aa}, we generate a set of galaxies with typical SFR, and calculate the ion column densities, showing the result in Fig. \ref{typical}. The $\rm sSFR$ has a weak dependence on the stellar mass as ${\rm sSFR} \propto M_\star^{-0.28}$, so the changes are modest compared to Fig. \ref{fiducial}, where the sSFR is constant for different galaxies. Due to the high sSFR of low-mass galaxies ($M_{\rm h}<10^{11}~M_\odot$), these galaxies have higher normalization factors in the $\beta$-model and gaseous halo masses. The gaseous halo mass at the high-mass end ($M_{\rm h} \approx 10^{13}~M_\odot$) is decreased by a factor of two, due to the small sSFR. The cooling radius has a moderately narrow range of $50\rm~kpc$ to $200\rm~ kpc$ as a function of the halo mass for a given model. Therefore, the increasing $R_{\rm vir}$ leads to a decrease of the relative cooling radius in units of $R_{\rm vir}$. Compared to Fig. \ref{fiducial}, the cooling radius in the low-mass range is raised until $M_{\rm h} \approx 10^{11.5}~M_\odot$, while it is suppressed for the massive galaxy, corresponding to the change of sSFR. For dwarf galaxies ($M_{\rm h}\lesssim 5\times 10^{11}~M_\odot$), the entire gaseous halo is radiatively cooling in {\tt CIE} or {\tt TCIE} models, while the {\tt PIE} model always shows the cooling radius smaller than the virial radius.

\begin{figure}
\begin{center}
\includegraphics[width=0.48\textwidth]{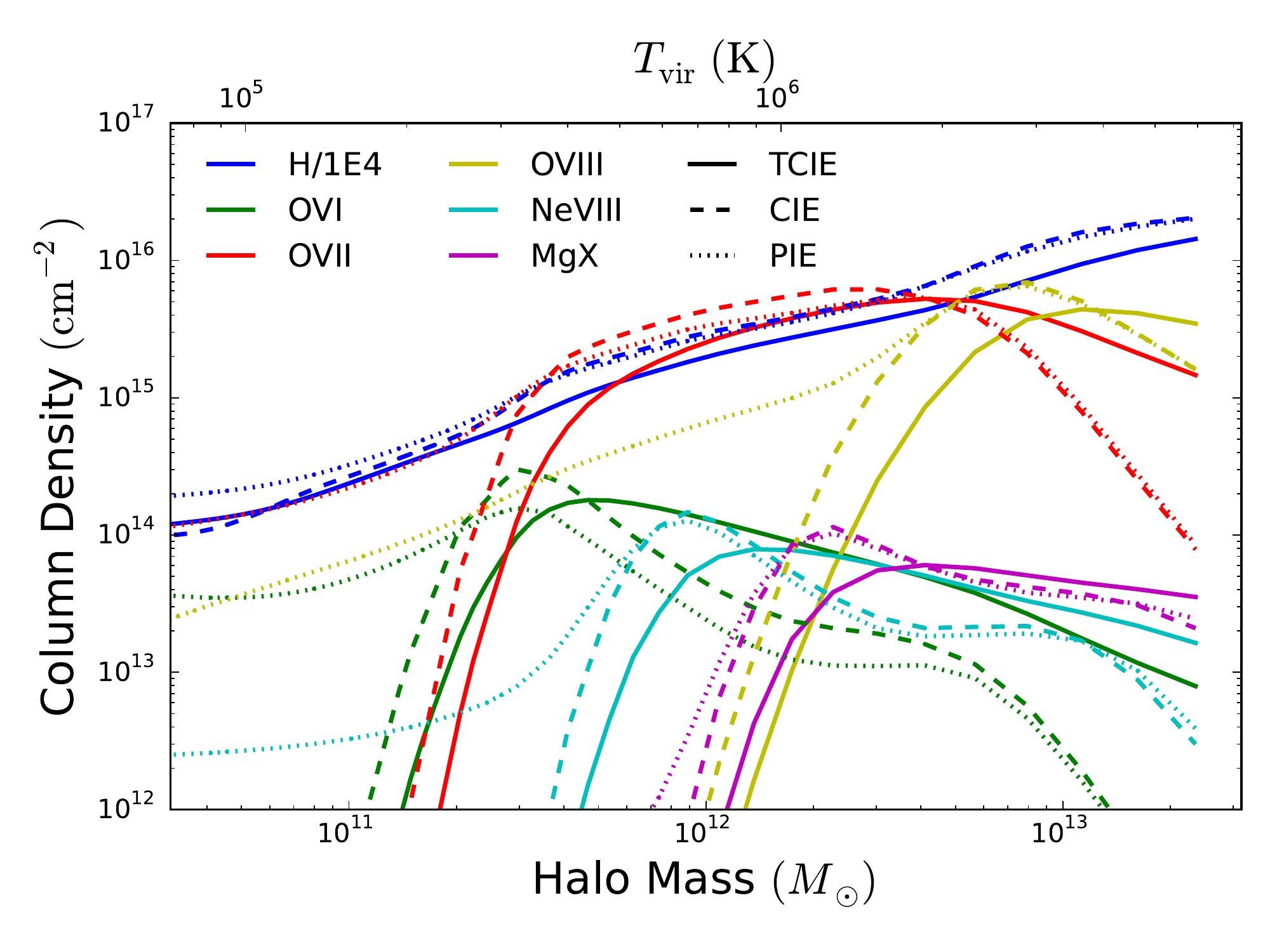}
\end{center}
\caption{The column density dependence on the halo mass of galaxies with fundamental plane SFR. The ion colors are same as Fig. \ref{ions_models}. Similar to the comparison between Fig. \ref{typical} and Fig. \ref{fiducial}, the changing of sSFR leads to higher column densities for low-mass galaxies, and lower column densities for massive galaxies, compared to Fig. \ref{ions_models}.}
\label{typical_ions}
\end{figure}

We also consider the SFR modification on the ion column density, shown in Fig. \ref{typical_ions}. Due to the more massive gaseous halo of low-mass galaxies, the total hydrogen column density is increased significantly, while high ionization state ions (e.g., {\OVII} and {\OVIII}) changes within a factor of $1.5$. The massive galaxy shows ions with column densities slightly smaller than Fig. \ref{ions_models}. Overall, the sSFR dependence on stellar mass does not change the phenomena illustrated by the fixed sSFR models, such as the convergence between models.

\section{Discussion}
Currently, modeling of gaseous components in halos often assume a single-temperature CIE model or a photon-heated PIE model \citep{Stocke:2013aa, Werk:2014aa, Miller:2015aa, Nicastro:2016aa, Faerman:2017aa}.
Improvements in such assumptions are warranted by observations that indicate the presence of multi-phase medium and temperature variations over different radii \citep{Anderson:2016aa, Tumlinson:2017aa}. However, both of these two temperature issues are not well constrained observationally -- there is not a universal temperature distribution for multi-phase gas or a universal radial-dependence of the temperature \citep{Anderson:2016aa, Bogdan:2017aa, Tumlinson:2017aa}.

Besides the gas temperature, the contribution by photoionization is controversial regarding the effect on high ionization state ions. The photoionization model is employed to explain the low and intermediate ionization state absorption system seen against the UV spectrum of background QSOs \citep{Werk:2014aa}. High ionization state ions (e.g., {\OVI} and {\NeVIII}) are normally explained in the collisional ionization model \citep{Savage:2005aa, Narayanan:2012aa, Meiring:2013aa, Pachat:2017aa}, while they are also possible to be modeled by photoionization \citep{Hussain:2015aa, Hussain:2017aa}.

In the last decade, the effect of the photoionization and the radiative cooling has been considered in theoretical calculations \citep{Wiersma:2009aa, Oppenheimer:2013aa, Gnat:2017aa}. Benefiting from these numerical calculations, we apply two improvements to the modeling of gaseous halos -- the photoionization modification and the radiative cooling multi-phase medium. These two improvements change the radiative cooling rate of the gaseous halo, subsequently the mass of the gaseous halo, and ions hosted by the halo. In this section, we compare our models with observations and theoretical simulations, and discuss implications and limitations.

\subsection{The Most Applicable Models}

In our gaseous halo model, we consider the modification of the photoionization and the effect of the radiatively cooling multi-phase medium, separately. In general, the photoionization mainly affects the low temperature or the low-density regions. In the context of a gaseous halo, the low temperature gas is most common in low-mass galaxies, while the low-density gas is on the outskirts of gaseous halos. Meanwhile, the radiative cooling mainly occurs within the cooling radius, where the density is high enough for the effective radiative cooling. However, these modifications have several limitations, which should be considered when one evaluates the applicability of these models.

For the photoionization of the gaseous halo, there are two main limitations -- the lack of ionizing photons from the host galaxy and the lack of the radiation transfer inside the halo. First, we only employ the UVB to supply ionizing photons, however, it is not the only source of the high-energy ionizing photons. The host galaxy also provides ionizing radiation from the star formation or soft X-rays from shock heated gases due to stellar winds and supernovae. Assuming an escape fraction of the unity of soft X-rays, \citet{Suresh:2017aa} show that the impact of the photoionization due to the host galaxy is limited to the inner $50\rm~kpc$ traced by the {\OVI}. Typically, the escape fraction is only several percent for low-redshift galaxies ($z<2$; \citealt{Grimes:2009aa, Khaire:2015ab}), which indicates that the galaxy escaping ionizing flux dominates the innermost region within the cooling radius. Second, ionizing UVB photons will be diluted by the absorption inside the gaseous halo, which leads to the suppression of the photoionization with the decreasing radius. The decreased photoionization means that the medium in the inner region is more likely to be collisionally ionized. Therefore, the UVB-only {\tt PIE} model is more likely the case for the outer region of massive halos.

We assume a time-independent cooling model for the multi-phase radiatively cooling medium, where the mass cooling rate is constant over all temperatures. This model has two assumptions -- the gases at different temperatures are ``well-mixed" and the hot medium can always be supplied to keep the cooling time independent. The first assumption means that the multi-phase ionic structure should have a physical connection (mixture) at various temperatures to keep the same mass cooling rate, and the cooling is only due to the radiative losses of the whole gaseous halo. However, some of the physical aspects surrounding the cooling are still uncertain, which is related to various processes besides the pure radiative cooling -- the galactic fountain, the accretion from IGM and even the tidal effect due to nearby galaxies \citep{Bregman:1980aa, Kwak:2010aa, Marinacci:2010aa, Gnat:2017aa}. These processes introduce perturbations to the gaseous halo, which leads to denser regions and possible thermal instabilities that might enhance the radiative cooling \citep{Armillotta:2016aa, Armillotta:2017aa}. These processes cannot be investigated in our analytic model, but might find solutions in the highest resolution cosmological simulations, which is beyond the scope of this paper. However, \citet{Thompson:2016aa} show that the cooling can follow the constant mass cooling rate model in a detailed radiative-cooling hot-wind model with mass-loading and energy transfer. In their calculation, the stable cooling model is evident in the large mass-loading factor $\dot{M}_{\rm hot}/{\rm SFR} \gtrsim 1$ case, which is related to another assumption on the boundary condition.

%the boundary condition.
This boundary condition -- an infinite ambient hot medium -- is required in a steady state model to balance the cooling rate of the gas. Without this boundary condition, the cooling will reduce the mass and the density of the hot gas, and subsequently reduce the cooling rate in the high temperature region, which violates the constant mass cooling rate assumption. In \citet{Thompson:2016aa}, the low mass-loading case shows the flattened ${\rm d} L/ {\rm d} \ln T$ in the high temperature region, indicating the lower cooling rate, which directly results from the weakening of the boundary condition. In the context of our gaseous halo model, this boundary condition could be satisfied automatically when the cooling radius is significantly smaller than the virial radius, which means that the gas beyond the cooling radius can be treated as the ambient hot phase gas to supply the cooling medium.

%summary
Based on above discussions, we suggest that gaseous halos should be divided into two categories based on their host galaxy masses. For the low-mass galaxy ($M_{\rm h} \lesssim 4\times 10^{11}~ M_\odot$), the {\tt PIE} model is a good assumption, since its halo size is small, and the virial temperature is low. For such a gaseous halo, the photoionization must be considered, since it provides additional heating to support a more massive gaseous halo, and changes the distribution of ionization states significantly. Also, low-mass galaxies normally have lower SFR ($\lesssim 1~M_\odot~\rm yr^{-1}$), which reduces the ionizing flux from the host galaxy, and it is approximately correct to assume the UVB-dominated photoionization. However, one potential issue is that the stellar feedback is stronger with the decreasing halo mass, which implies the feedback heating is higher and the $\gamma$ factor is smaller in Equation (2). Considering this effect, the low-mass galaxy may host a higher mass halo than the {\tt PIE} model predicts. Considering the radiative cooling, we could use the {\tt TPIE} model, which is more realistic, since the cooling radius is smaller than the virial radius as shown in Fig. \ref{typical}. In the next section, we will show that the photoionization modification on the {\tt TCIE} model is similar to its effect on the {\tt CIE} model.

For a high-mass galaxy ($M_{\rm h} \gtrsim 4\times 10^{11}~ M_\odot$), the virial temperature is high ($> 10^{5.5}\rm~ K$) and the cooling radius is smaller than the virial radius. Therefore, we suggest that the whole gaseous halo should be divided into two parts -- the inner high-density region within the cooling radius and the outskirts , which is a low-density region. In the inner region, the cooling produces a multi-phase medium, therefore, the {\tt TCIE} model is preferred. Also, in this case the boundary condition is satisfied to maintain the system in a steady-state. Beyond the cooling radius, the {\tt PIE} model is appropriated due to the density of $\lesssim 5\times 10^{-5}~\rm cm^{-3}$.

\begin{figure*}
\begin{center}
\subfigure{
\includegraphics[width=0.48\textwidth]{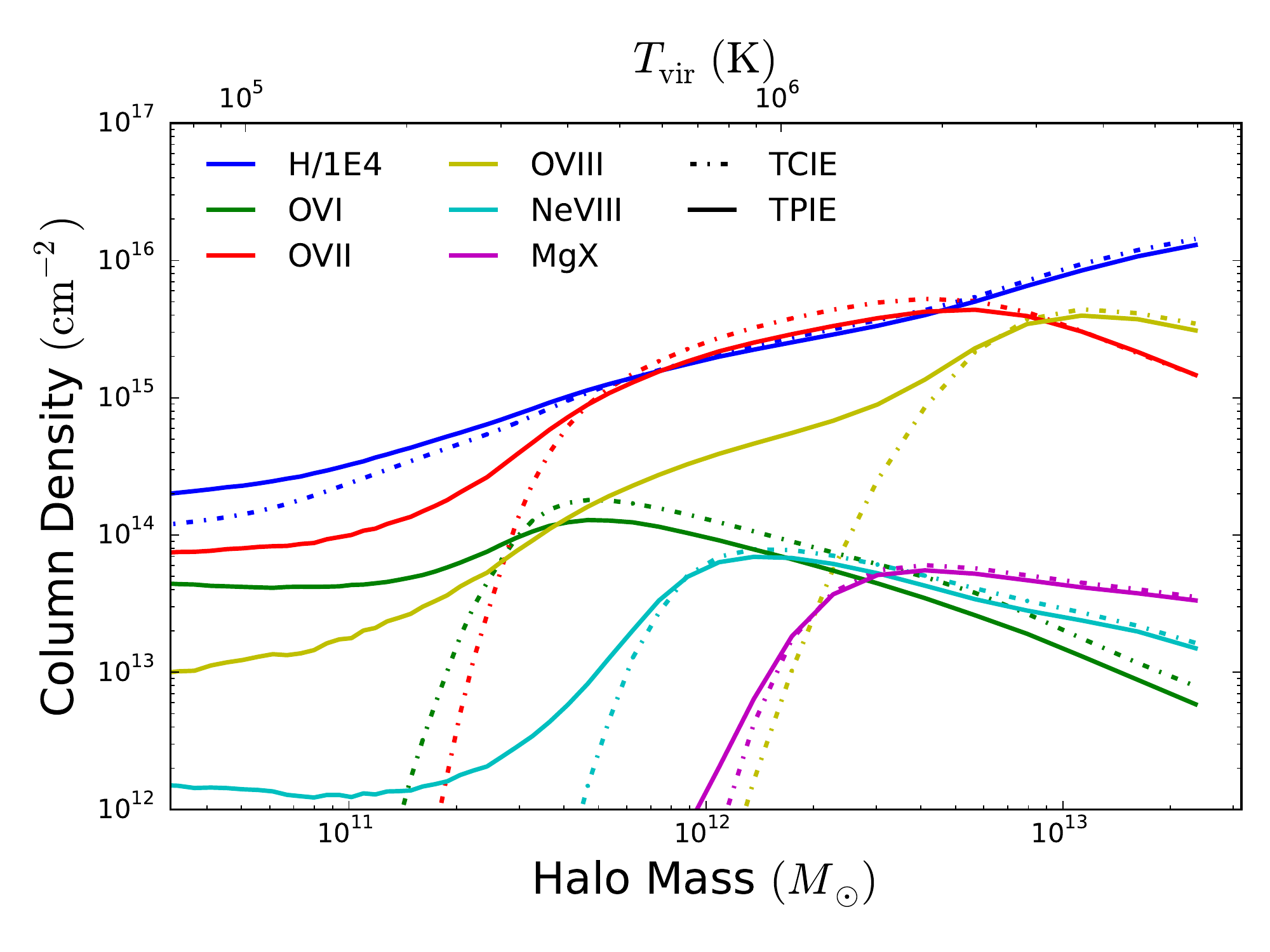}
\includegraphics[width=0.48\textwidth]{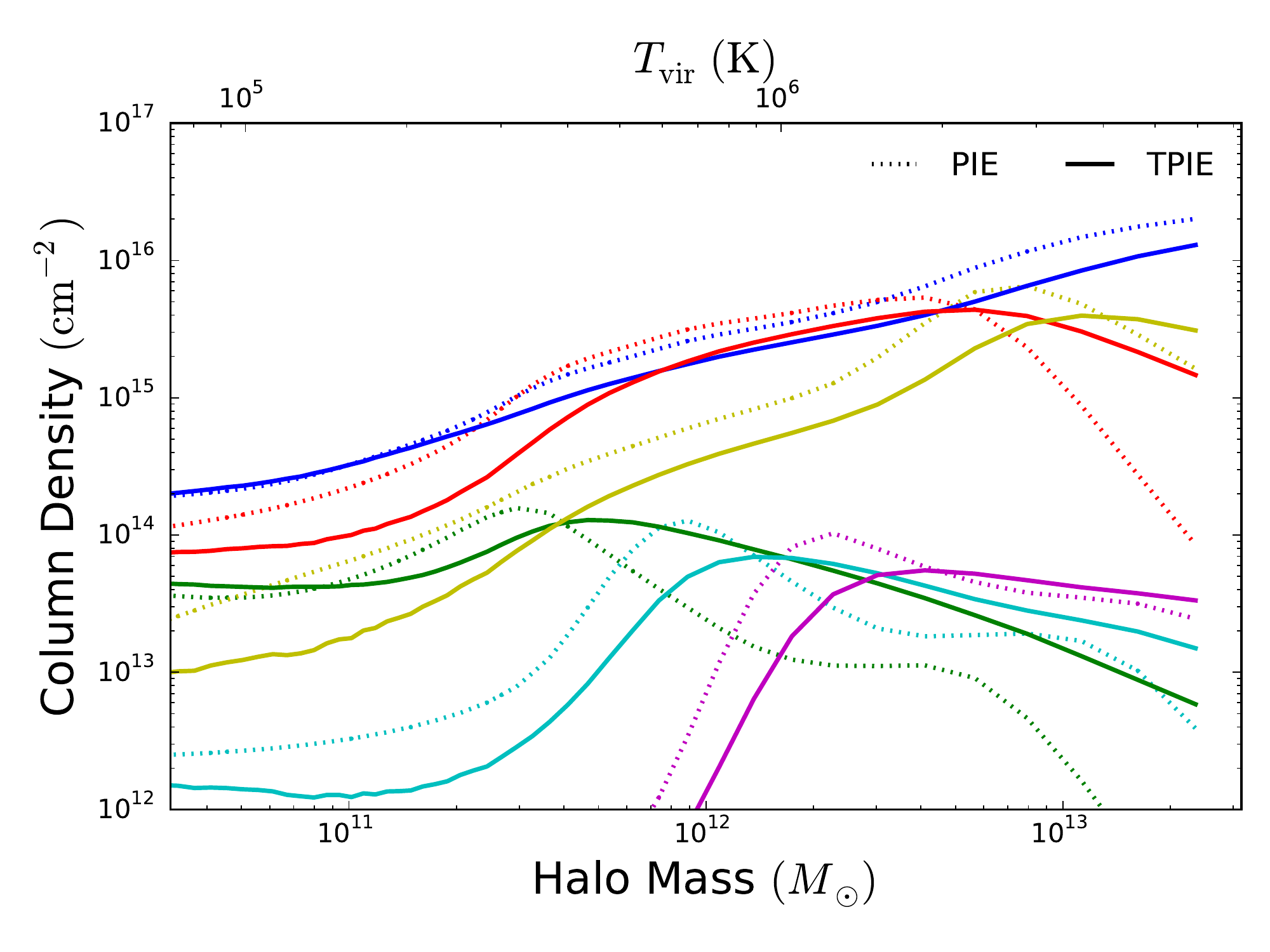}
}
\end{center}
\caption{The comparison between {\tt PIE}, {\tt TCIE}, and {\tt TPIE} models. Ions have the same colors as in Fig. \ref{ions_models}, while solid, dotted and dot-dashed lines are {\tt TCIE}, {\tt PIE}, {\tt TPIE} models, respectively. {\it Left panel}: We compare the {\tt TCIE} and {\tt TPIE} models, which are the most realistic. For $T_{\rm vir}$ values below the peak column densities of the {\tt TCIE} model, the column is considerably enhanced for the metal ions due to the photoionization modification. {\it Right panel}: We compare the {\tt PIE} and {\tt TPIE} models, where more relatively low ions for massive galaxies (e.g., {\OVI}) is produced due to the cooling of the high temperature medium.}
\label{pie_cd}
\end{figure*}

\subsection{The Multi-Phase Cooling Medium With Photoionization}
The {\tt TPIE} model -- the stable radiative cooling model with the photoionization -- is a more complex extension of the earlier models, and there are two potential issues with such a model. First, with the effect of photoionization increasing, the gas could be in a net heating phase, which could break our assumption on the stable radiative cooling model. This phenomenon is important where the photoionization might support gaseous components in special situations, such as at $10^4 \rm~ K$ and within $\sim 1\rm~ kpc$ of the plane (as occurs in the Milky Way -- the gaseous disk). However, this case is not the aim of our models. Second, the {\tt TPIE} model leads to a radial dependence in the temperature distribution, since {\tt PIE} cooling curves have a dependence on the density. This involves the modeling of the cooling flow, which is not included in our models. Meanwhile, as we will show below, the {\tt TCIE} and {\tt PIE} models can be a good approximation for the {\tt TPIE} model in different situations.

In the {\tt TPIE} model, there are two modifications compared to the previously defined model. First, the lower limit of temperature in {\tt TPIE} is no longer fixed at $10^{4.5}\rm~K$ due to the potential heating. We set a minimum emissivity of $10^{-26}\rm~erg~cm^3~s^{-1}$ (corresponding to a minimum temperature), which sets a dynamic range of more than three orders of magnitude for the emissivity. If this temperature is higher than $10^{4.5}\rm~K$, then the local minimum temperature is changed to the new temperature with the minimum emissivity. In practice, only few percent of gas has the new lower limit of temperature (about $5\times 10^4\rm~K$ to $10^5~\rm K$), which is not far from our fixed minimum temperature ($3\times 10^4\rm~K$). Second, for different temperatures, the density is also different due to the pressure balance, which leads to different cooling curves. However, this involves radiative transfer to obtain the photon spatial distribution in the gaseous halo. Therefore, we ignore such an effect and use the cooling curve of the total density to calculate the mass-temperature distribution for all different temperatures.

In the calculation of {\tt TPIE}, we use the galaxy sample with the typical SFR dependence on the stellar mass. Other parameters are fixed, including the metallicity of $Z = 0.3~Z_\odot$, redshift of $z=0$, $\beta = 0.5$, and the impact parameter of $0.3~R_{\rm vir}$. The results are compared to {\tt TCIE} and {\tt PIE} models, as shown in Fig. \ref{pie_cd}.

The gaseous halo mass is proportional to the normalization factor in the $\beta$-model, which is also indicated by the hydrogen column density. For the gaseous halo mass, the {\tt TPIE} model converges to the {\tt TCIE} model for massive galaxies ($\gtrsim 10^{12.5}~M_\odot$), and show similarities with the {\tt PIE} model (with the shift to the high mass galaxy) for low-mass galaxies ($\lesssim 10^{11.5}~M_\odot$) as expected. Overall, the {\tt TPIE} is roughly a direct summation of the effect of the photoionization and the cooling temperature distribution. For massive galaxies, the {\tt TCIE} model is a good approximation of the {\tt TPIE} model, which has enhanced ``low" ionization state ions (e.g., {\OVII} from the cooling of {\OVIII}). In the low-mass range, the gaseous halo is mainly dominated by the photoionization, which enriches the high ionization state abundance at low temperatures (e.g., {\OVI} or {\OVII}). Therefore, the {\tt TPIE} model can be approximated by a combination of {\tt PIE} and {\tt TCIE} models for all mass ranges.

\subsection{The $\gamma$ Factor}
In our model, we introduce a $\gamma$ factor in Equation (2) to account for the stellar feedback heating. In the previous calculation, this $\gamma$ factor is fixed at unity for simplicity. However, the $\gamma$ factor varies over different galaxies, which is determined by the detailed physics of the stellar feedback -- supernovae (SNe), stellar winds, photoionization due to the star light and the radiation pressure \citep{Hopkins:2017aa}. \citet{Hopkins:2017aa} shows that the SN feedback dominates the stellar feedback, although only SN feedback cannot produce observations. 

SNe can launch a galactic wind that ejects materials and energy into the gaseous halo or beyond the virial radius \citep{Fielding:2017aa}. The strength of the galactic wind can be modeled by the mass-loading factor $\eta$ as $\dot{M}_{\rm out} = \eta {\rm SFR}$. The mass-loading factor (at $0.25 ~R_{\rm vir}$) has a dependence on the galaxy halo mass \citep{Muratov:2015aa}:
\begin{eqnarray}
\eta &=& 2.9 (1+z)^{1.3} \left(\frac{V_c}{60~\rm km~ s^{-1}}\right)^{-3.2}, ~ V_c \leq 60~\rm km~s^{-1}, \notag\\
\eta &=& 2.9 (1+z)^{1.3} \left(\frac{V_c}{60~\rm km~ s^{-1}}\right)^{-1.0}, ~ \rm otherwise,
\end{eqnarray}
where $V_c$ is the circular velocity. This is a specific wind model obtained by parametrizing the galactic winds in the FIRE simulations \citep{Hopkins:2014aa}. For the energy carried by the galactic wind, we only consider the kinetic energy, ignoring the internal energy, since the temperature of the galactic wind is found to be much lower than the virial temperature in simulations \citep{Fielding:2017aa}. For the wind internal energy, \citet{Thompson:2016aa} showed a semi-analytic model for the cooling wind, and found a drop of the temperature at the radius of $5-10\rm~kpc$ for high mass-loading cases ($\eta > 0.8$), which agree with \citet{Muratov:2015aa}. The wind temperature is only about one tenth of the virial temperature, when it enters the innermost radius in our model. Therefore, the internal energy of the wind is negligible compared to the virial temperature of the halo.

\begin{figure}
\begin{center}
\includegraphics[width=0.48\textwidth]{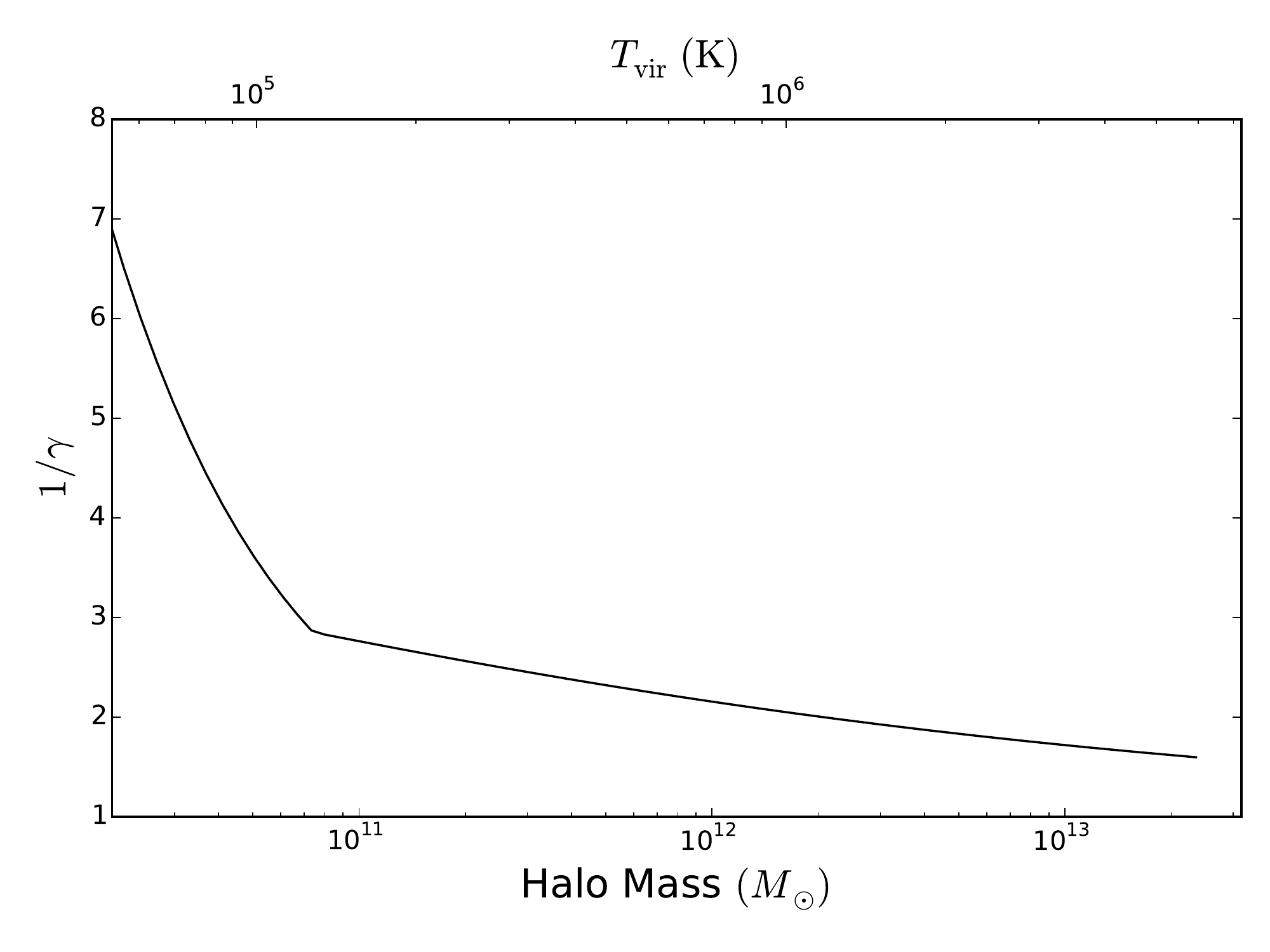}
\end{center}
\caption{The stellar feedback parameter $\gamma$ as a function of halo mass, where higher $1/\gamma$ indicates stronger stellar feedback heating.}
\label{gamma}
\end{figure}

The wind velocity is $V_{\rm wind} = 0.85V_c^{1.1}$, which is also measured at $0.25~R_{\rm vir}$ \citep{Muratov:2015aa}. Based on these relationships, we can set an upper limit for the stellar feedback, since not all of the kinetic energy can be converted into the internal energy of the gaseous halo. Some of the galactic wind will be recycled before it is well-mixed with the gaseous halo, and some will be ejected out of the galaxy halo. Therefore, the lower limit of $\gamma$ factor is calculated by:
\begin{equation}
\frac{1}{\gamma} = 1 + \eta(V_{\rm c})(\frac{V_{\rm wind}^2}{V_{\rm c}^2} - 1),
\end{equation}
Then, the $\gamma$ factor is around $0.14$ for the lowest mass galaxies, and around $0.5$ for galaxies with masses higher than $10^{11}~M_\odot$. We show the halo mass dependence of $\gamma$ in Fig. \ref{gamma}.

\begin{figure*}
\begin{center}
\subfigure{
\includegraphics[width=0.48\textwidth]{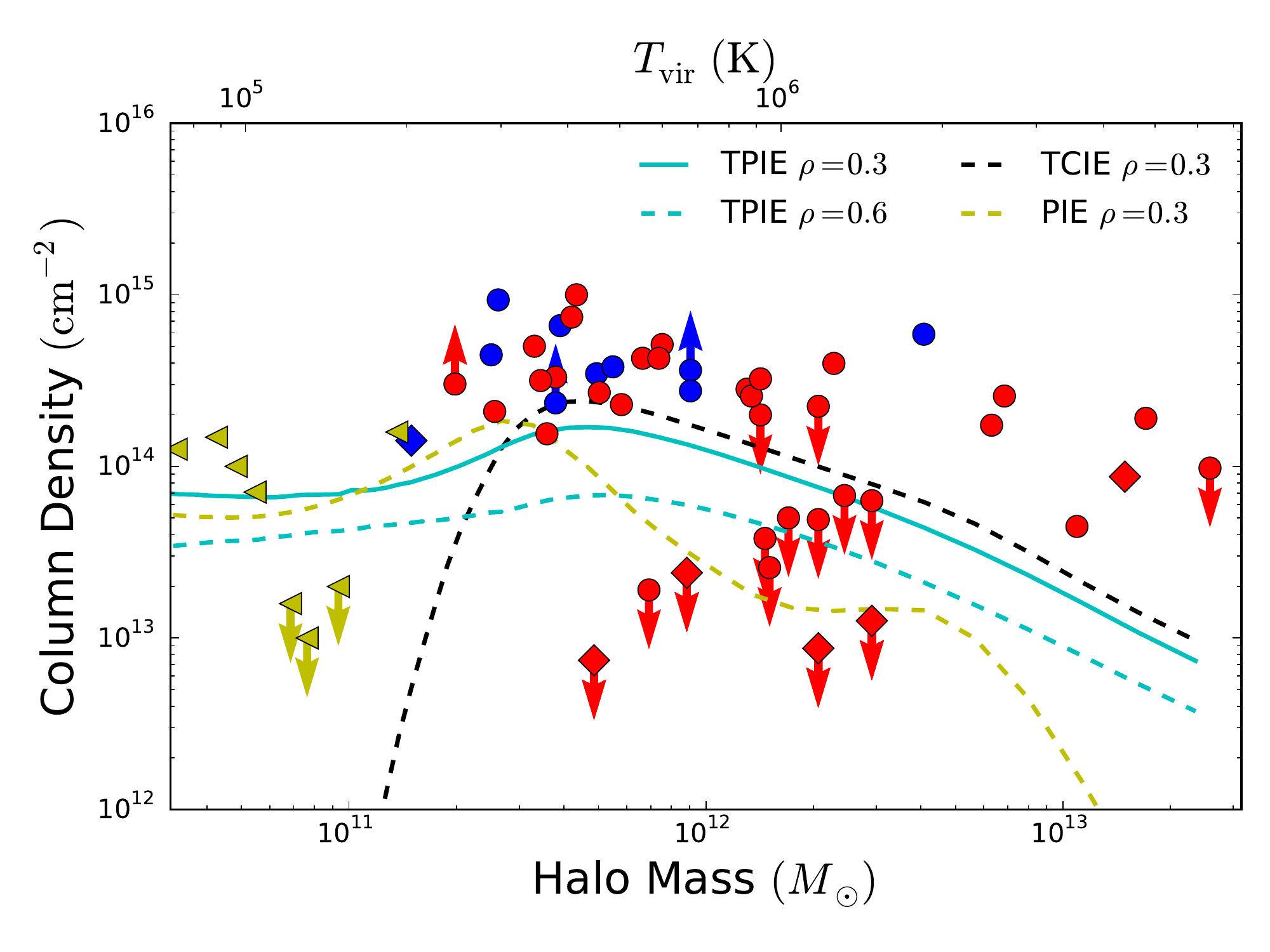}
\includegraphics[width=0.48\textwidth]{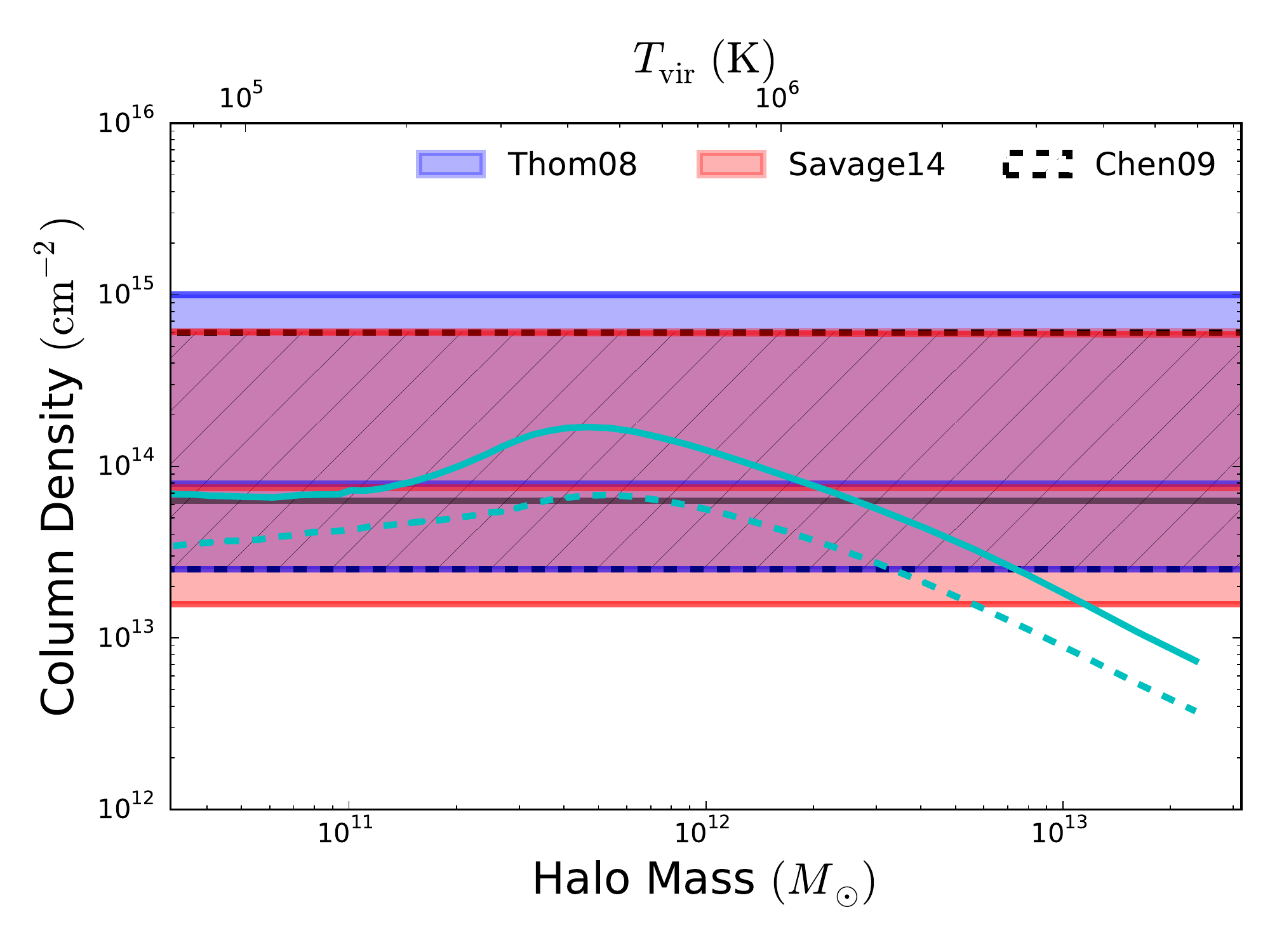}
}
\end{center}
\caption{Comparison of OVI columns in our models and observations. {\it Left panel}: Comparison with {\OVI}-galaxy pairs. The cyan lines are the {\tt TPIE} model, and the solid line has an impact parameter of $0.3~R_{\rm vir}$, while the dashed cyan line is $0.6~R_{\rm vir}$; most observations fall between these two impact parameters. The black dashed line is the {\tt TCIE} model with an impact parameter of $0.3~R_{\rm vir}$, while the yellow dashed line is {\tt PIE} model with the same impact parameter. The filled circle marks the {\OVI} column in \citet{Werk:2013aa}, and the red color indicates that the SFR is lower than the typical SFR, while the blue color indicates higher SFR values. The upper limit for column density is the detection threshold. The sample of \citet{Johnson:2015aa} is shown by diamond symbols, and the color indicates whether the galaxy is early-type (red) or late-type (blue). The sample of \citet{Johnson:2017aa} is shown in yellow left triangles. The COS-Halos sample lies about $0.5\rm dex$ over our models. {\it Right panel}: Comparison with blind {\OVI} surveys. Three samples are marked in blue, black slashed, and red slashed regions for \citet{Thom:2008aa}, \citet{Chen:2009aa}, and \citet{Savage:2014aa}, respectively. The median value for these samples are shown in corresponding solid lines, which are consistent with our models.}
\label{OVI}
\end{figure*}

\subsection{The {\OVI} Puzzle}
From our models, we find that the {\OVI} column lies in a moderately narrow range close to $10^{14}\rm~cm^{-2}$ for all masses of galaxies due to either the photoionization (in low-mass galaxies) or the low-temperature cooling medium (in massive galaxies; Fig. \ref{OVI}). Current observations show that the {\OVI} has a significant dependence on the star formation rather than the stellar mass \citep{Tumlinson:2011aa}, which seems to be reproduced by the {\tt TPIE} model. Therefore, we compared the prediction from our models with observations, and we adopt three samples -- COS-Halos \citep{Werk:2013aa}, \citet{Johnson:2015aa} and \citet{Johnson:2017aa}. Our models have the parameters $Z=0.3~Z_\odot$, $\beta = 0.5$ and the SFR is from the main-sequence relationship (also modified using redshift; ${\rm sSFR} \propto (1+z)^3$). The impact parameters of $0.3~R_{\rm vir}$ and $0.6~R_{\rm vir}$ are shown since the COS-Halos sample is limited to $< 0.55~R_{\rm vir}$. The redshift is set to $0.2$ since most of the {\OVI} samples have $\bar{z} \approx 0.2$. As shown in Fig. \ref{OVI}, the {\tt TPIE} model can be approximated by the combination of {\tt PIE} and {\tt TCIE} with a broad transition around $M_{\rm h} \approx 3 \times 10^{11} M_\odot$. Therefore, in the following discussion, the {\tt TPIE} designation represents the combination of {\tt PIE} and {\tt TCIE} models. For low-mass galaxies ($M_{\rm h} \lesssim 3 \times 10^{11} M_\odot$), the {\OVI} is mainly ionized through photoionization for the whole gaseous halo, while for the higher mass galaxies, the cooling medium corresponds to the majority of observed {\OVI}. The {\tt TPIE} model predicts a narrow range of column densities from $M_{\rm h} = 3\times 10^{10}~M_{\odot}$ to $2\times 10^{13}~M_{\odot}$. In most mass regions, the {\tt TPIE} model predicts an {\OVI} column density of $10^{13.5}\rm~cm^{-2}$ to $10^{14.1}\rm~cm^{-2}$.

For the COS-Halos sample, the line shape of {\OVI} usually shows multiple components, which can be separated by using the Voigt profile fitting. However, in addition to component separation, there is the issue of the host galaxy. \citet{Werk:2013aa} assigned one galaxy for each multi-component absorption system and calculated the SFR. Therefore, we adopt their measurements based on the apparent optical depth method, which does not separate different components. One caveat is that this treatment of different components might give higher {\OVI} column densities than the isolated gaseous halo. Several galaxies in the COS-Halos sample have other lower mass galaxies at the same redshift in the same field or lie in galaxy groups, which may introduce contamination from the other galaxies or from the intragroup medium. Since the COS-Halos sample also provides the SFR for each galaxy, we mark with blue a galaxy that has a SFR higher than the SFR calculated from the SFR main sequence, while a lower one is red. For galaxies with non-detections of {\OVI}, we set the upper limit as the detection limit if it is available. 

The Johnson 2015 sample considered galaxies in groups or clusters with isolated galaxies, so we used only isolated galaxies with impact parameters smaller than the virial radius. \citet{Johnson:2015aa} do not have information on the SFR, but they report the galaxy type. Therefore, we assign early-type galaxies with red colors and late-type with blue colors. This color encoding is different from the COS-Halos sample, but will not affect the general tendency. The Johnson 2017 sample focuses on dwarf galaxies with stellar masses in $\log M_\star = 7.7-9.2$ \citep{Johnson:2017aa}. For these galaxies, no color is assigned, since no SFR information is available.

In Fig. \ref{OVI}, most of the detected {\OVI} have column densities of $\gtrsim 10^{14}\rm~cm^{-2}$, no matter whether the SFR is above or below the typical SFR. Overall, the difference is about $0.3-0.5\rm~dex$, which may be accounted for by two explanations. One is the heating from the stellar feedback, which might produce a smaller $\gamma$ factor in Equation (2). A factor of $4$ to $10$ can raise the normalization factor in the $\beta$-model by a factor of $2$ to $3$, which also raises the {\OVI} column density by the same ratio. Similarly, \citet{McQuinn:2018aa} showed that a cooling flow with $100~M_\odot\rm~yr^{-1}$ can account for the observed {\OVI} column density, and that most of the cooling flow might be destroyed by stellar feedback. However, the galactic wind only feedback model is unlikely to be sufficiently energetic based on our calculation in Section 4.3. Another possibility is that the observed {\OVI} is overestimated due to the intragroup medium contamination or the overlap of multiple gaseous halos in the sightline \citep{Stocke:2014aa}.

To address this possibility further, we do not limit the sample to {\OVI}-galaxy pairs, but also consider all intervening {\OVI} absorption systems  from blind surveys. We consider three samples -- \citet{Thom:2008aa}, \citet{Chen:2009aa}, and \citet{Savage:2014aa}. These three samples have some overlap with each other, but since the detection methods are not the same, they are also complementary to each other. In \citet{Thom:2008aa}, the median {\OVI} column density is $\log N({\rm OVI}) = 13.9$ with a standard deviation of $0.4\rm~dex$, while in \citet{Chen:2009aa}, the {\OVI} column density has a median value of $\log N({\rm OVI}) = 13.8$ and a scatter of $0.4\rm~dex$. In \citet{Savage:2014aa}, the components are reported separately. The median column density of single {\OVI} components is $\log N({\rm OVI}) = 13.68$ with a range of $13.00$ to $14.59$. The average number of components per {\OVI} absorption system is around $1.6$, which leads to the average {\OVI} column density for each {\OVI} system of $\log N({\rm OVI}) = 13.87$. These three surveys are consistent with each other in terms of the median and range of $N({\rm OVI})$.

Although these intervening {\OVI} systems currently do not have detected host galaxies, it is possible that they are also the gaseous halo of galaxies, whose luminosities are below $0.1~L^*$ ($M_{h} < 3\times 10^{11}~M_\odot$; the   detection limit of the COS-Halos galaxy sample). If this is the case, these {\OVI} observations show a significant difference from the COS-Halos sample, whose median is $\log N({\rm OVI}) = 14.5$ with a scatter of $0.26\rm~dex$ (only accounting for the detected {\OVI}).

Since the COS-Halos sample provides the SFR for each {\OVI}-galaxy pair, we built a specific model (with $\gamma$) for individual systems based on its stellar mass, SFR, and impact parameter from \citet{Werk:2014aa}. We only use the physical impact parameter from the COS-Halos sample, and we recalculate the relative impact parameter using $R_{\rm vir}$ calculated from Equation (3) with a mean halo density of $200\rho_{\rm crit}$ rather than $200 \rho_{\rm matter}$. These recalculated virial radii are smaller than those presented in \citet{Werk:2014aa} by a factor of $\approx 30 - 40\%$ within the redshift range of $z = 0.14 - 0.36$. This modification leads to larger relative impact parameters. For the QSO-galaxy pair J1437+5045 and 317\_38 (here we use the same notation of COS-Halos sample -- position angle and angular separation), our calculation leads to the virial radius of $140.2\rm~kpc$, which indicates that the absorption system is beyond the virial radius at $z=0.246$ with the impact parameter of $143 \rm~kpc$. Although different stellar mass-halo mass relationships and different cosmological constants can lead to different results on the virial radius, the systems with reported $\rho/R_{\rm vir} > 0.5$ are actually further out in the halo ($\rho/R_{\rm vir} > 0.8$).

\begin{figure*}
\begin{center}
\subfigure{
\includegraphics[width=0.48\textwidth]{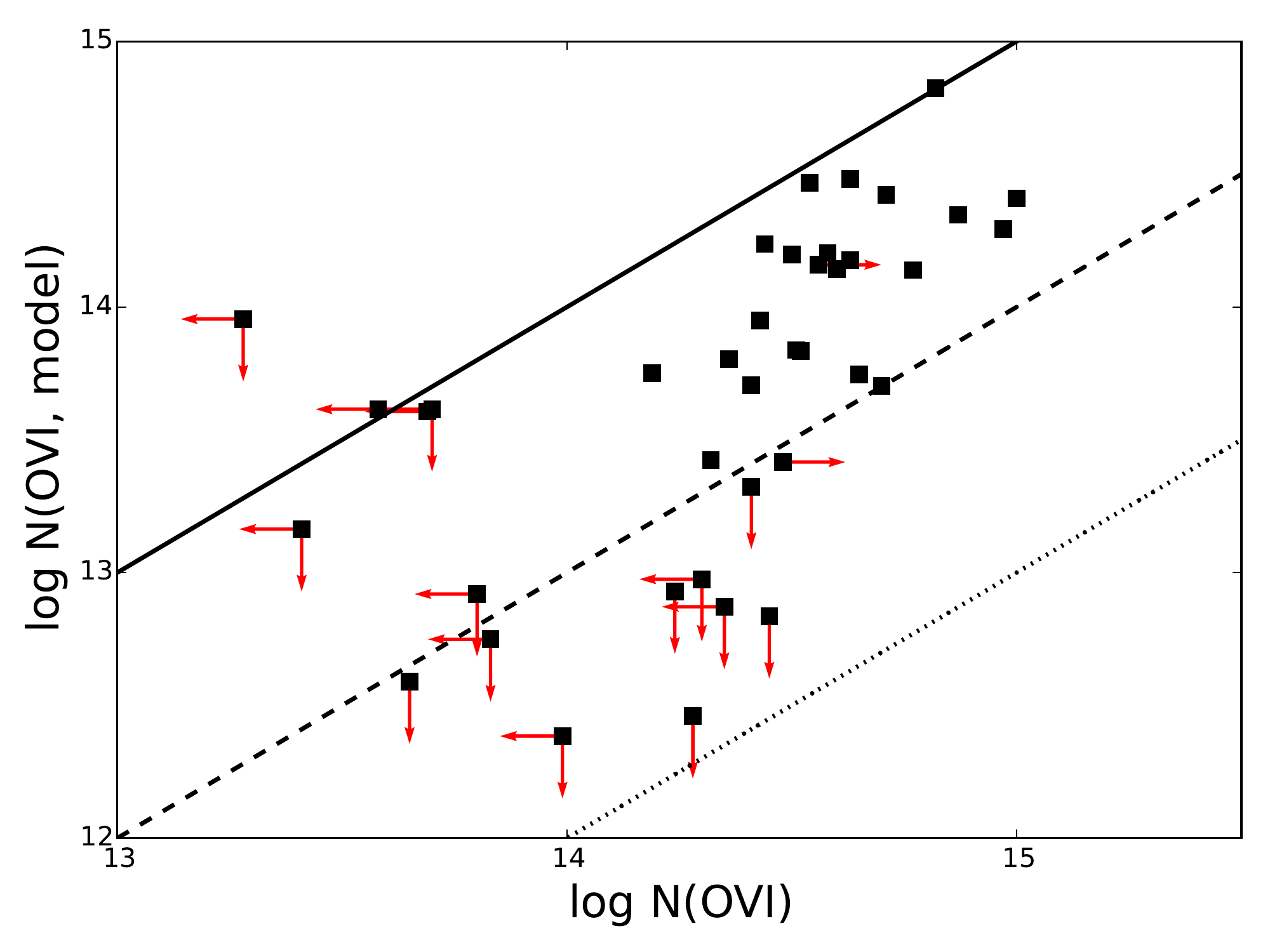}
\includegraphics[width=0.48\textwidth]{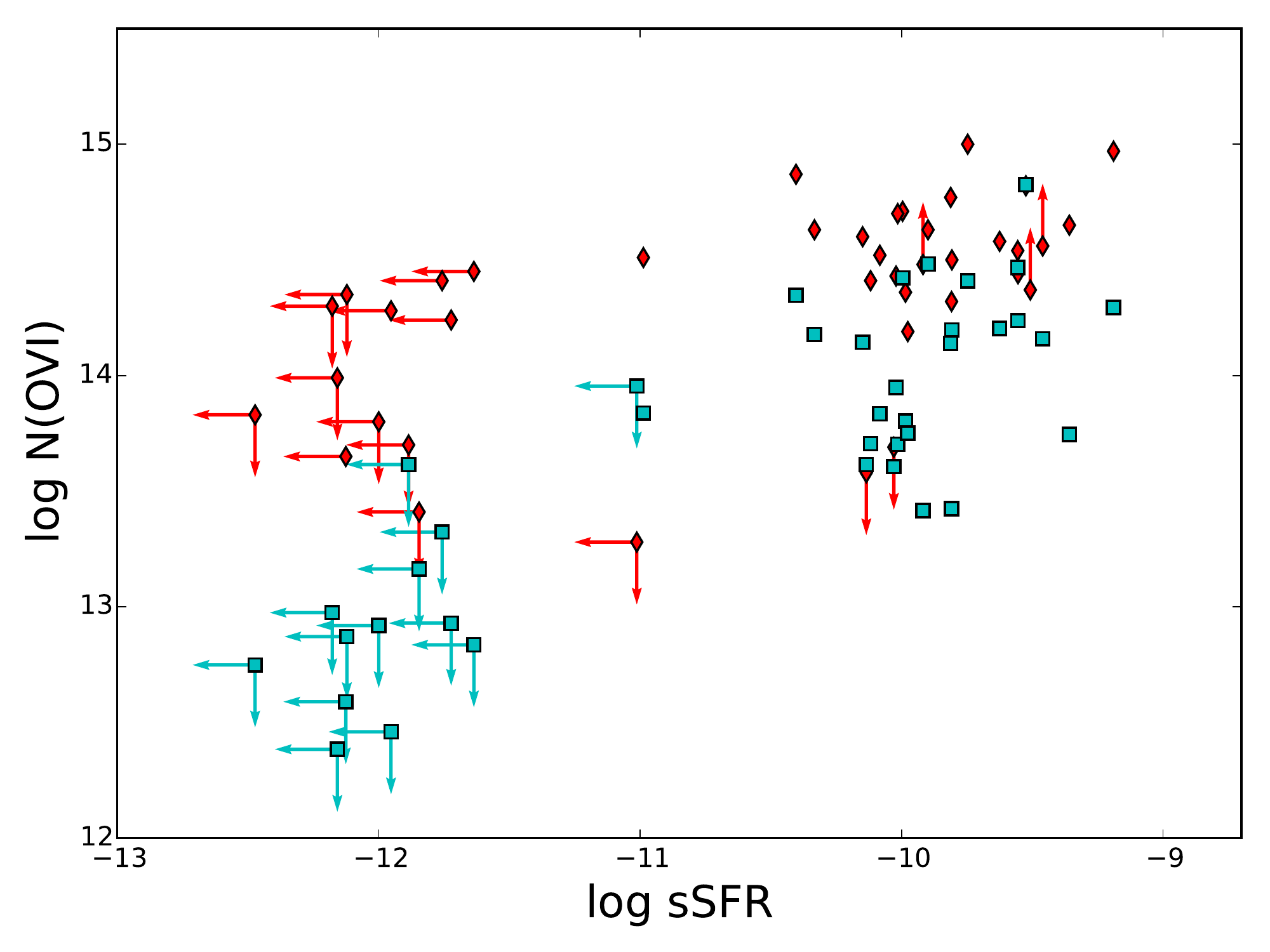}
}
\end{center}
\caption{Specific models for the COS-Halos {\OVI} sample. {\it Left panel}: For each system, the predicted {\OVI} column density is calculated with the reported stellar mass, the SFR, and the impact parameter. If the SFR is only an upper limit, then our models predict an upper limit. The solid, dashed, and dotted lines indicate models are equal to observations, $10\%$, and $1\%$ of observations. For detected objects, the model are typically a factor of $3-5$ lower than the observations, which is consistent with Fig. \ref{OVI}. {\it Right panel}: The sSFR dependence of the {\OVI} column density. The red diamonds are the observations, while the cyan squares are our models.}
\label{coshalo_OVI}
\end{figure*}

Our calculations for the {\OVI} column density are shown in Fig. \ref{coshalo_OVI}. If the SFR of a galaxy is the upper limit, we can only derive the upper limit of the {\OVI} column density. Among 30 systems with detectable {\OVI}, there are five that do not have measurable SFRs. It is clear that some of these five systems are in galaxy groups -- the galaxy 211\_33 of QSO J1133+0327 has a similar redshift ($\Delta z<0.0002$) to the galaxy 110\_5 adopted in the COS-Halos study, and the galaxy 35\_14 of QSO J0910+1014 has the similar redshift of the galaxy 242\_34. For these {\OVI} absorption systems, it is possible that they are due to the intragroup medium or smaller galaxies that are closer, since we have shown that the {\OVI} column density can be detected for low-mass galaxies ($M_{\rm h} < 3\times 10^{11}~M_\odot$), which is consistent with observations \citep{Johnson:2017aa}. Such contamination may also explain some of the difference between the model and the observation. As shown in \citet{Bregman:2018aa}, there are two small galaxies surrounding QSO J1009+0713 with a redshift of 0.3556 similar to the galaxy 170\_9 (0.3557). For this system, the {\OVI} has four components, which have velocities of $-95\rm~km~s^{-1}$, $25\rm~km~s^{-1}$, $117.2\rm~km~s^{-1}$, and $200.7\rm~km~s^{-1}$. For the galaxy 170\_9, our model predicts $\log N({\rm OVI}) = 14.41$, while the two components of two high velocities are $14.21\pm 0.16$ and $14.34\pm 0.05$, respectively. The two components at low velocities are larger contributors ($14.61\pm 0.05$ and $14.52\pm 0.12$) to the total {\OVI} column density of $15.00\pm 0.03$, and we suggest that they may be associated with the two small galaxies with slightly lower redshifts. Overall, our specific models show a difference of 0.5 dex smaller than the detected {\OVI} from COS-Halos, which is similar to our general comparison shown in Fig. \ref{OVI}.

In Fig. \ref{coshalo_OVI}, we also show the relationship between the sSFR and the {\OVI} column density. Our models show results similar to the Evolution and Assembly of GaLaxies and their Environments (EAGLE) simulation \citep{Oppenheimer:2016aa}. For star-forming galaxies, the predicted {\OVI} column densities are about 0.5 dex lower than the observation, while we predict lower {\OVI} column densities for passive galaxies (our upper limit is lower than those simulated in EAGLE; \citealt{Oppenheimer:2016aa}). The difference for passive galaxies may be due to the lack of AGN heating in our models. The AGN feedback is also proposed to explain the strong {\OVI} in star forming galaxies \citep{Oppenheimer:2017aa}, but as discussed above, this difference is possibly due to contamination in COS-Halos sample.

\begin{figure*}
\begin{center}
\subfigure{
\includegraphics[width=0.48\textwidth]{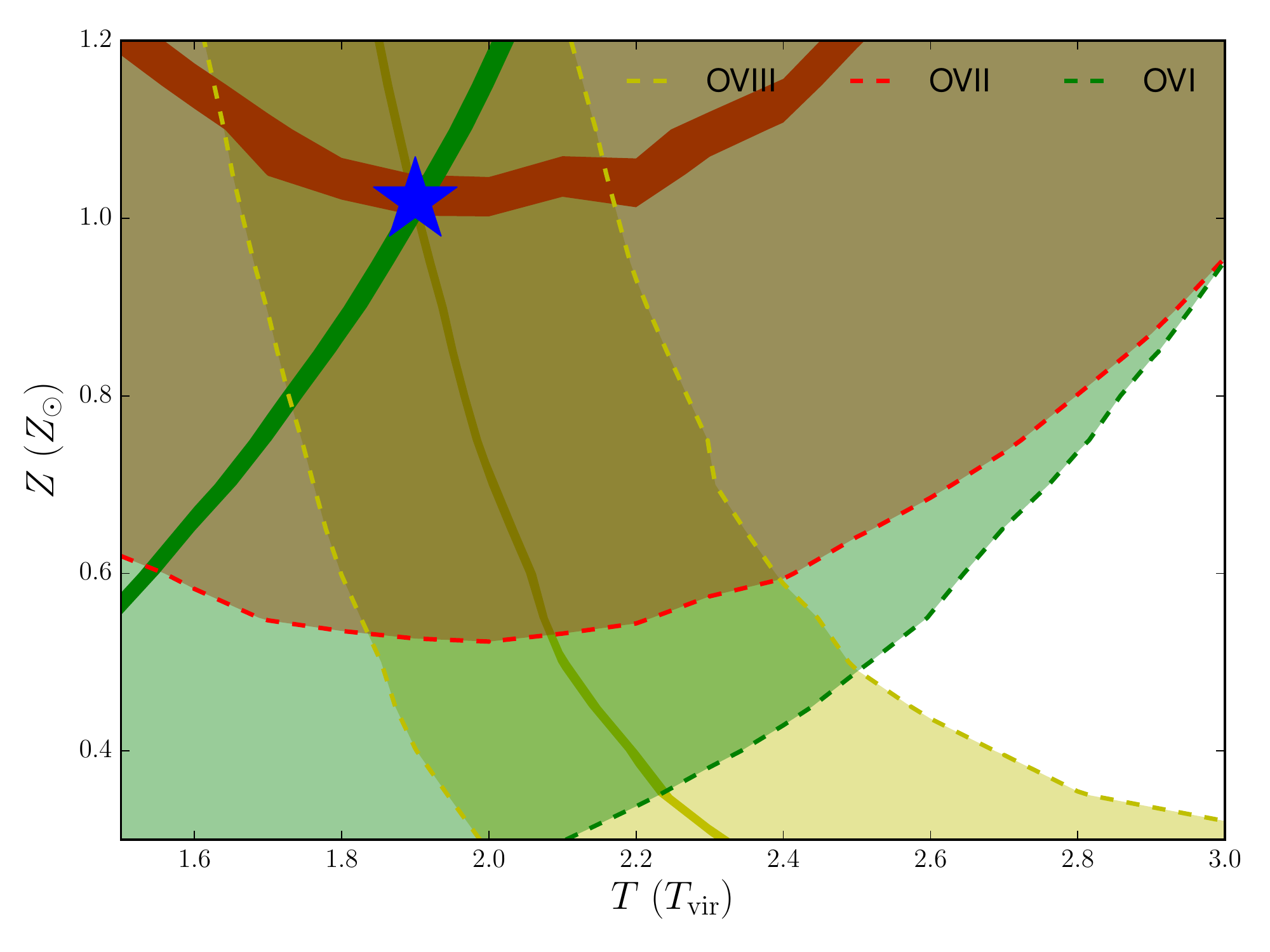}
\includegraphics[width=0.48\textwidth]{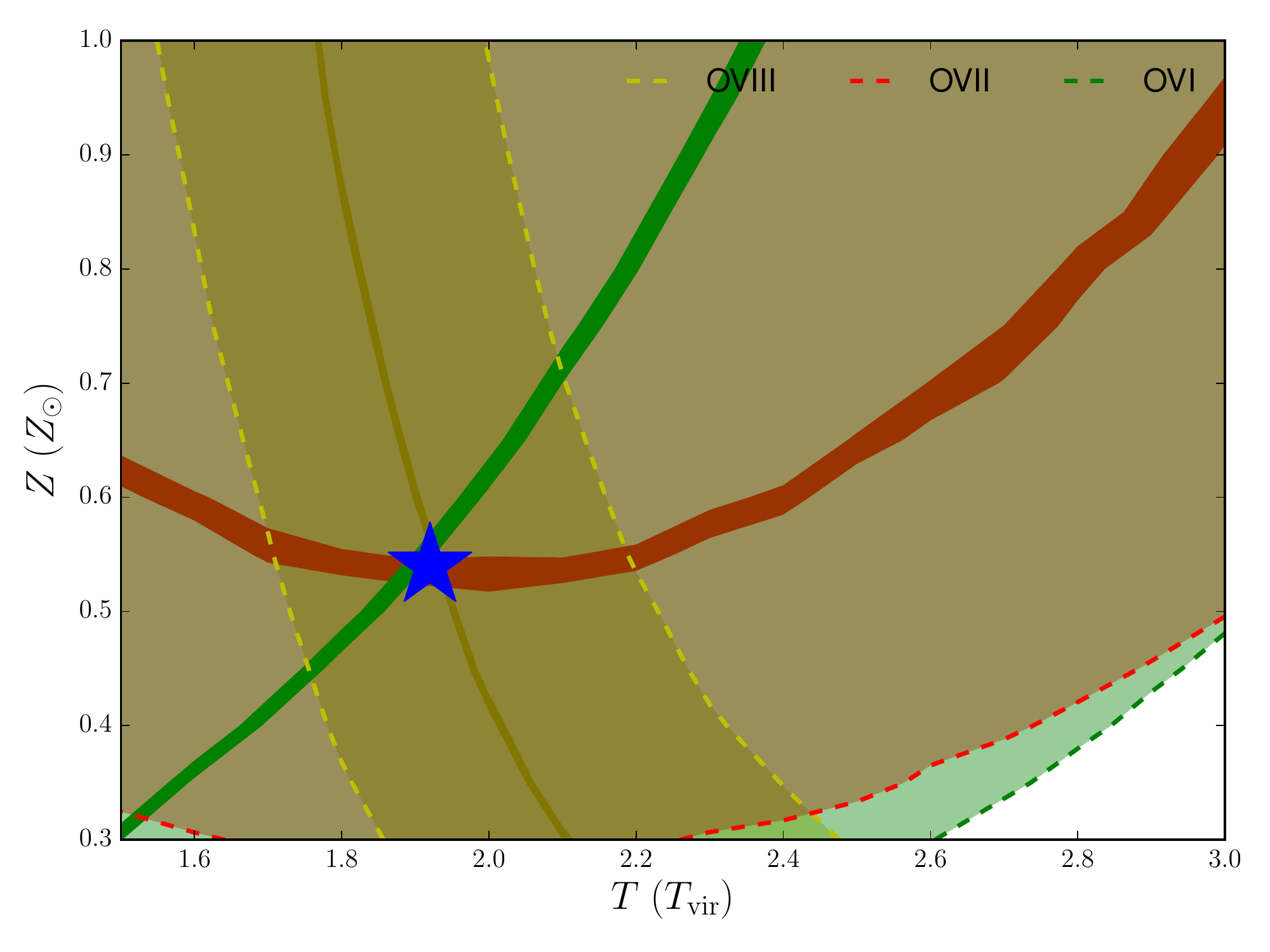}
}
\end{center}
\caption{Modified model for Galactic {\OVII} and {\OVIII}, constrained by observations, as a function of metallicity and temperature. {\it Left panel}: Models with the nominal value for $\gamma$ ($=1$). The dashed lines indicate the acceptable region for each ion within $1\sigma$. {\OVIII}, {\OVII}, and {\OVI} are in yellow, red, and green, respectively. The thick colored lines indicates the median value of different ions. The blue star indicates the preferred solution with $T\approx 1.90\times 10^6~\rm K$ and $Z = 1.02~Z_\odot$. {\it Right panel}: Models with enhanced stellar feedback ($\gamma = 0.5$), as given in Fig. \ref{gamma}. The symbols are the same as the left panel, but the preferred solution is $T\approx 1.93\times 10^6~\rm K$ and $Z = 0.55~Z_\odot$.}
\label{OVII}
\end{figure*}

\subsection{The Galactic {\OVII}/{\OVIII}}
The {\OVII} and {\OVIII} ions have resonant lines in the X-ray band at $21.60~\rm \AA$ and $18.97~\rm \AA$. However, the detection of these two ions is limited by the current X-ray observatory sensitivity, and only {\OVII} and {\OVIII} in the Milky Way have been confirmed in both absorption and emission \citep{Nicastro:2002aa, Wang:2005aa, Henley:2012aa}. The modeling of the {\OVII} and {\OVIII} emission lines shows that the gaseous halo of the MW has a normalization parameter of  $1.35\pm0.24\times 10^{-2}/Z~\rm cm^{-3}~kpc^{-1.5}$ \citep{Miller:2015aa}. For the MW model, we adopt $M_{\star} = 7\times 10^{10}~M_\odot$, $M_{\rm h} = 1.7\times 10^{12}~M_\odot$, $Z = 0.3~Z_\odot$ and a SFR of $1 ~M_\odot~\rm yr^{-1}$. This MW model leads to a normalization parameter of $1.75\times 10^{-2}~\rm cm^{-3}~kpc^{-1.5}$, $1.62\times 10^{-2}~\rm cm^{-3}~kpc^{-1.5}$, $1.24\times 10^{-2}~\rm cm^{-3}~kpc^{-1.5}$ and $1.15\times 10^{-2}~\rm cm^{-3}~kpc^{-1.5}$ for the {\tt CIE}, {\tt PIE}, {\tt TCIE} and {\tt TPIE} models, respectively. For a distance to the galactic center of $8\rm~ kpc$, the sightline with the galactic latitude of $90^\circ$ has a column density that is half of the sightline with an impact parameter of $0.03~R_{\rm vir}$ ($8\rm~kpc$). Then, the predicted column density is $\log N({\rm OVII}) = 15.6$, $15.7$, $15.8$, $15.6$ for {\tt CIE}, {\tt PIE}, {\tt TCIE} and {\tt TPIE}, respectively. These column densities lead to an EW of $12\rm~m\AA$ ($\log N({\rm OVII})=15.7$), which is less than the most of the observations as summarized in \citet{Hodges-Kluck:2016aa}, where the mean in this direction ($b>60^\circ$) is about $25\rm~m\AA$. The corresponding {\OVIII} column densities are $13.9$, $14.7$, $13.0$ and $14.4$ for our four models, which is about one order of magnitude lower than observations \citep{Gupta:2012aa}. These modelings show two issues -- the ratio of $N({\rm OVIII})/N({\rm OVII})$ is too low and the total amount of oxygen is less than the observation.

For the $N({\rm OVIII})/N({\rm OVII})$ ratio, there are two ways of improving the agreement with Galactic observations -- raising the maximum temperature or extending the gaseous halo beyond the virial radius. First, for a $M_{\rm h} = 1.7\times 10^{12}~M_\odot$ halo, the virial temperature is around $10^6\rm~K$, while the measured hot gas temperature is around $1.5-2\times 10^6\rm~K$ \citep{Henley:2012aa, Miller:2015aa, Nevalainen:2017aa}. Also, it is evident that the hot gas temperature is higher than the virial temperature for most elliptical galaxies \citep{Davis:1996aa, Brown:1998aa, Goulding:2016aa}. This higher temperature can increase the $N({\rm OVIII})/N({\rm OVII})$ ratio significantly, since the {\OVIII} ion traces the higher temperature gas. Second, an extended gaseous halo also changes this ratio involving the photoionization modification as shown in the Section 3.3. Therefore, increasing the maximum radius helps to increase the $N({\rm OVIII})/N({\rm OVII})$ ratio.

To increase the total amount of the oxygen, there are also two approaches -- having an extended gaseous halo and increasing the metallicity. For the MW, the cooling radius is smaller than the virial radius, which means that the larger maximum radius will not reduce the normalization factor in the $\beta$-model, so this modification only increases the gaseous component surrounding the galaxy, hence the metal mass. As stated in Section 3.2, for a given galaxy-gaseous halo pair (fixed the SFR and the halo mass), higher metallicity leads to a higher total metal mass, although the gaseous halo mass is reduced. Therefore, a higher solar metallicity halo can solve the problem of the small {\OVII} column density in our model.

To illustrate these possibilities, we construct a {\tt TPIE} model to match the galactic {\OVII} and {\OVIII} observations. In the modified model, we vary two parameters -- the maximum temperature $T_{\rm max} = \alpha T_{\rm vir}$ and the metallicity. We use the observation summarized in \citet{Faerman:2017aa}, which has  $N({\rm OVII})= 1.4 (1.0-2.0) \times 10^{16}\rm~cm^{-2}$ and $N({\rm OVIII}) = 0.36 (0.22 - 0.57)\times 10^{16}\rm~cm^{-2}$ (also see \citealt{Gupta:2012aa, Fang:2015aa}). The {\OVI} is also considered to show whether it can be reproduced in the same model, and the column density of halo {\OVI} is $\log N({\rm OVI}) = 13.95\pm 0.34$ \citep{Sembach:2003aa}. For the {\OVI} column density, the contribution from the disk is excluded based on the velocity criterion \citep{Savage:2003aa}.

In Fig. \ref{OVII}, we explore the parameter space of $\alpha = 1.5-3$ and $Z = 0.3-1.2~Z_\odot$, which is determined by the $N({\rm OVIII})/N({\rm OVII})$ ratio and column densities. The acceptable region for each ion is constrained by the observational limits, therefore, the overlap region indicates the preferred parameter space, while the cross of lines indicates the preferred model. Our modified model suggests a super-solar metallicity of $1.02~Z_\odot$ and a maximum temperature of $1.9\times 10^6\rm~K$. This high metallicity solution is not favored since it is very unlikely for the gaseous halo to have higher metallicity than the galaxy disk. This high metallicity is caused by the high oxygen column density in the observation, which can be solved when the $\gamma$ factor is considered.

Involving the $\gamma$ factor in Equation (2) can lead to a higher mass gaseous halo, since a $\gamma < 1$ leads to a higher radiative cooling rate. The typical $\gamma$ factor for the MW is around 0.5. Applying this modification to our model, we will obtain lower metallicity solutions, since the small $\gamma$ factor leads to a more massive gaseous halo and more metals. Then it is not necessary to have high metallicities to account for observed oxygen. We obtain the best model of the MW gaseous halo with the metallicity of $0.55~Z_\odot$ and the maximum temperature of $1.9\times 10^6\rm~ K$, showing column densities of $\log N = 13.95$, $16.15$ and $15.53$ for {\OVI}, {\OVII}, and {\OVIII} ions, respectively. The normalization factor of the $\beta$-model is $1.91\times 10^{-2}\rm~cm^{-3}~kpc^{3\beta}$, and the gaseous halo mass is $1.94\times 10^{10}~M_\odot$, which contributes $7 \%$ to the total baryon mass. The emission line studies of {\OVII} show a normalization factor of $3.39_{-0.55}^{+0.67} \times 10^{-2}\rm~cm^{-3}~kpc^{3\beta}$ with $0.3~Z_\odot$ \citep{Li:2017ab}, which is equivalent to $1.85\times 10^{-2}\rm~cm^{-3}~kpc^{3\beta}$ with $0.55~Z_\odot$. Therefore, our MW gaseous halo solution also matches with the emission line study.

We now consider the magnitude of column density increases when the gaseous halo is extended beyond the virial radius. Assuming the maximum radius is twice the virial radius, the change in the column density is about $5\%$ to $8\%$, which is not enough to account for the observed {\OVII} and {\OVIII} column density. Therefore, compared to the heating due to the stellar feedback, extending of the maximum radius is a secondary effect for the column density, although it leads to a significant increase in the mass, which will be discussed in Section 4.7.

\begin{figure}
\begin{center}
\includegraphics[width=0.48\textwidth]{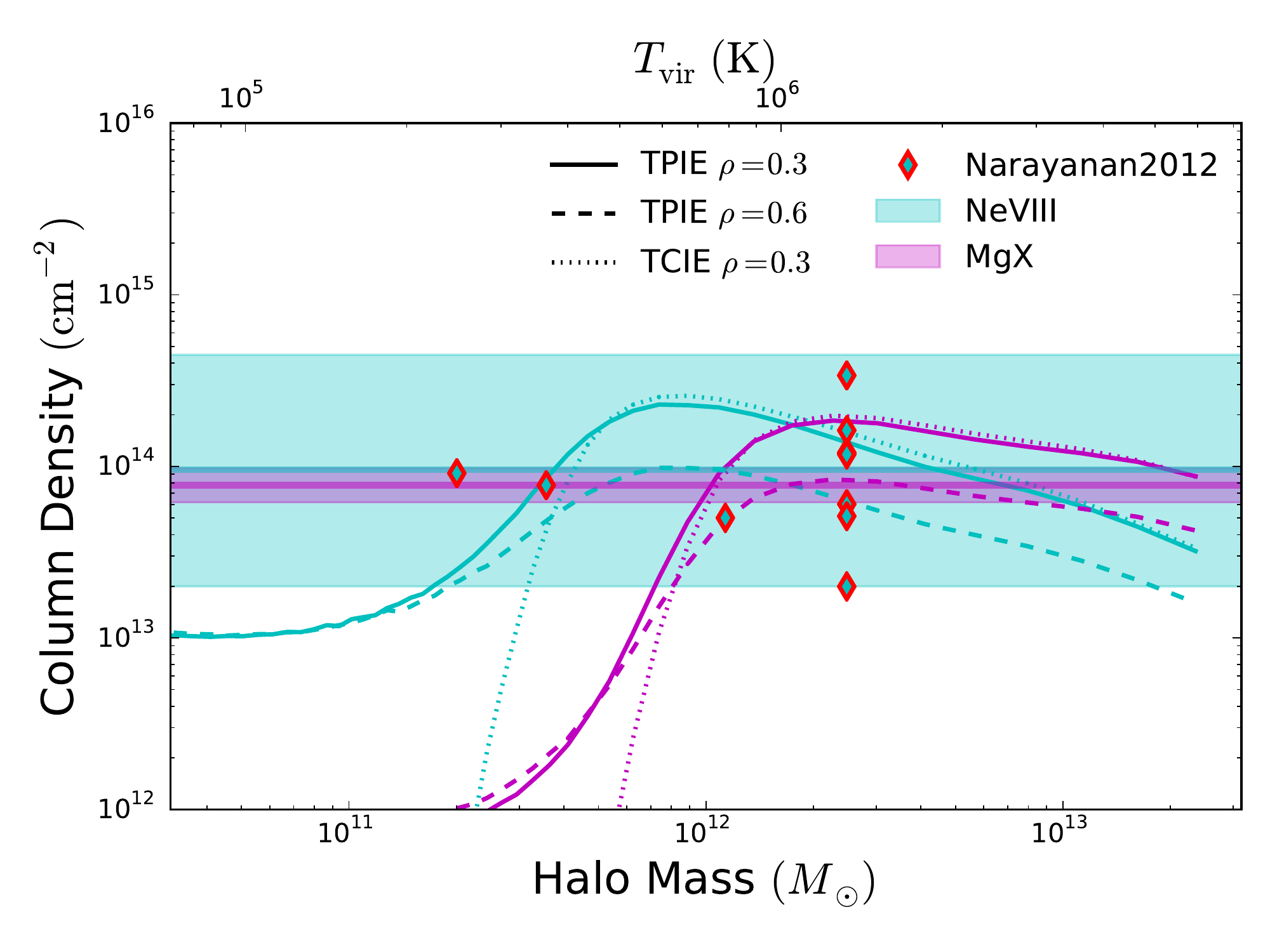}
\end{center}
\caption{Comparison of {\NeVIII} (cyan) and {\MgX} (magenta) in our models and observations. The solid, dashed, and dotted lines are the {\tt TPIE} model with impact parameters of $0.3~R_{\rm vir}$, $0.6~R_{\rm vir}$, and the {\tt TCIE} model with $0.3~R_{\rm vir}$. The {\NeVIII}-galaxy pair data is from \citet{Narayanan:2012aa}, while the {\NeVIII} absorption in PG 1206+459 is broken into seven components, and it is not clear which one corresponds  to the reported galaxy \citep{Tripp:2011aa}.}
\label{NeVIII}
\end{figure}

\subsection{Intervening {\NeVIII}/{\MgX} Systems}
The {\NeVIII} and {\MgX} occurs in the extreme UV band ($770\rm~\AA$ and $610\rm~\AA$ respectively), which can only be detected for extragalactic galaxies due to the wavelength limit of Galactic absorption ($912\rm~ \AA$) and due to the UV observing band of {\it HST}/COS ($1150-1750\rm~\AA$). The {\NeVIII} is detectable in the redshift of $0.5$ to $1.3$, while the {\MgX} is detectable between $z=0.9$ and $1.8$. At higher redshift, galaxies have a higher mean sSFR, which results in a more massive gaseous halo. In the  redshift range of $z=0.7$ to $z=1.2$, the SFR is raised by a factor of $5-10$, which makes the gaseous halo $2-3$ times more massive, with a similar increase in the column density of {\NeVIII} and {\MgX}. In Fig. \ref{NeVIII}, we show our models of the {\NeVIII} and {\MgX} columns at the redshift of 1.

As summarized in \citet{Pachat:2017aa}, the median of the detected {\NeVIII} is $\log N({\rm NeVIII}) = 13.98\pm0.31$, varying in the range of $\log N({\rm NeVIII}) = 13.30$ to $14.65$. The observed {\NeVIII} column density shows consistency with our model in a wide galaxy mass range. For the host galaxy, current observations show it varies from $0.08~L^*$ to $\approx 2~L^*$(\citealt{Narayanan:2012aa} and reference herein), which is also well matched with our models. For MgX, there is only one detection towards LBQS 1435-0134, with the column density of $\log N({\rm MgX}) = 13.89\pm0.10$ \citep{Qu:2016aa}, which is also consistent with the model for a (sub-)$L^*$ galaxy.

\begin{figure}
\begin{center}
\includegraphics[width=0.48\textwidth]{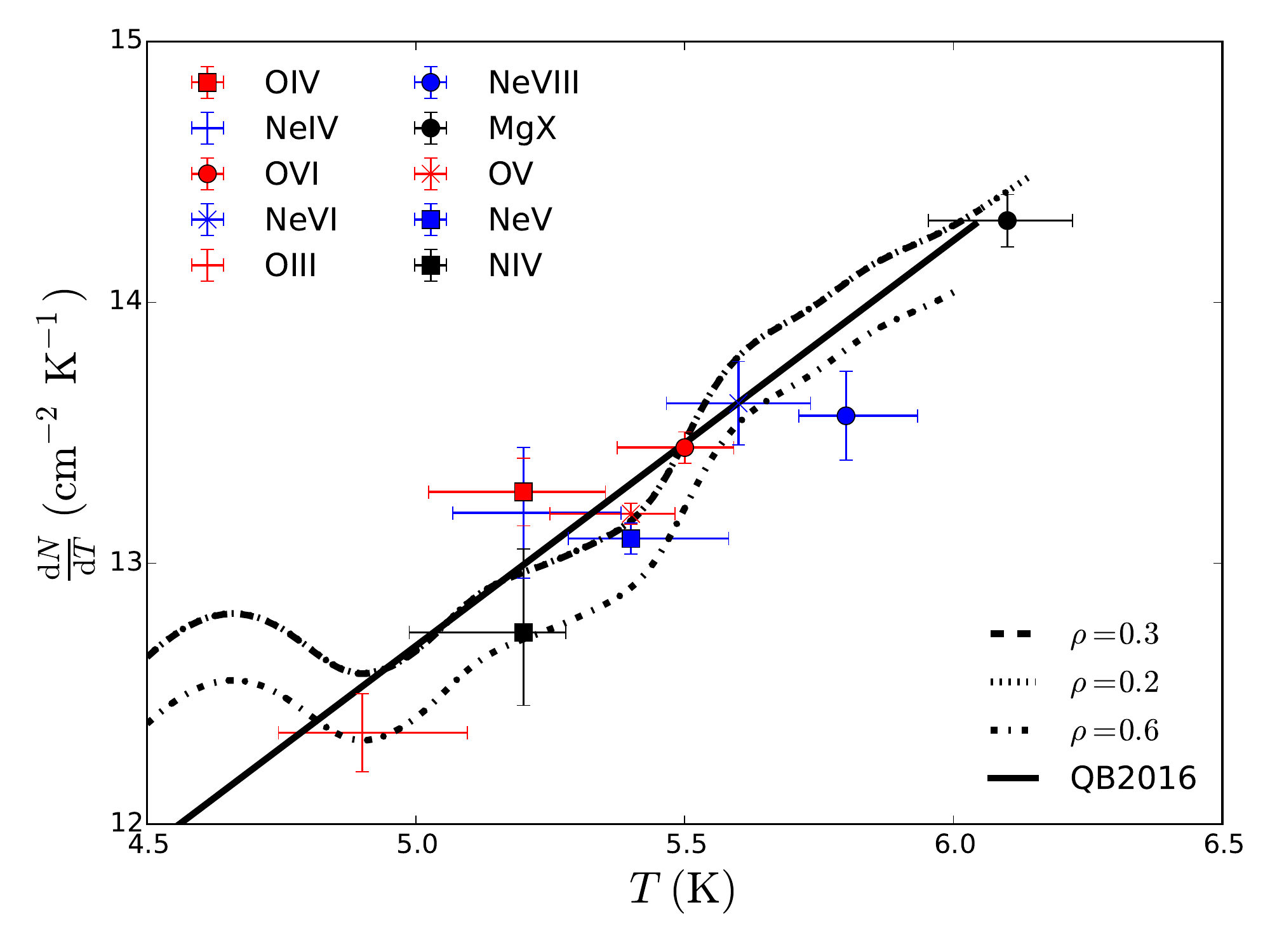}
\end{center}
\caption{The fitting result of {\tt TCIE} for the {\MgX} system in the sightline towards LBQS 1435-0134 \citep{Qu:2016aa}. For each ion, the temperature is the peak temperature of the ionization fraction, and the error bar is the full width of the half maximum, while the y axis value is the normalized column density gradient, which is given by $\frac{{\rm d}N_{\rm H, ion}}{{\rm d}T} = \frac{{\rm d}N_{\rm H, model}}{{\rm d}T} \times N_{\rm ion, observed} / N_{\rm ion, model}$. The solid line is the power law model in \citet{Qu:2016aa}, while dashed, dotted, dash-dotted lines are stable cooling models with impact parameters of $0.3~R_{\rm vir}$, $0.2~R_{\rm vir}$, and $0.6~R_{\rm vir}$, respectively. Note that the $\rho=0.2~R_{\rm vir}$ line overlaps with the $\rho=0.3~R_{\rm vir}$ line. Therefore, it is clear that the observed power law column density distribution is actually a result of the cooling medium.}
\label{lbqs1435}
\end{figure}

The {\MgX} system in LBQS 1435-0134 is a good example to study the multi-phase medium in the gaseous halo since it has a wide ionization state coverage (i.e., from {\OIII} to {\MgX}), and most of them are high ionization state ions (higher than {\OIV}). In \citet{Qu:2016aa}, it is modeled by a three-temperature {\tt CIE} model or a power law model of the column density-temperature distribution with an index of 1.55. This power law model has a total $\chi^2$ of 3.3 with 7 degrees of freedom. We notice the power law index is approximately the slope of the mass-temperature distribution $M(T)$ in the cooling model, therefore we fit this {\MgX} system using the {\tt TCIE} model. The only one variable is the stellar mass, and the SFR is calculated based on the stellar mass and modified by the redshift of 1.2. We use three different impact parameters of 0.2, 0.3 and 0.6. The fitting results are $\log M_{\star} = 10.76$ ($\log M_{\rm h} = 12.1$) with total $\chi^2 = 65.7~(dof = 10)$, $10.64$ ($12.0$) with $\chi^2 = 30.0~(10)$ and $10.56$ ($11.9$) with $\chi^2 = 131.1~(10)$, respectively. Therefore, the {\MgX} system is likely to be a (sub-)$L^*$ galaxy at the redshift of 1.2. For the best model with $\rho = 0.3~R_{\rm vir}$, the most of $\chi^2$ (22.0/30.0) is from three ions {\HI}, {\OIV} and {\NeVIII}. These deviations may be caused by the uncertainty in the lower and the upper limits of the temperature distribution. Extending the lower limit can increase the low and intermediate ionization state ions, while extending the upper limit can decrease the {\NeVIII} column density as shown in Fig. \ref{NeVIII} regarding the $\log M_{\rm h} = 12.0$. Therefore, although the best reduced $\chi^2$ is around 3, it is a valuable step in modeling such a complex object (the gaseous halo) with a simple physical model.

\begin{figure*}
\begin{center}
\subfigure{
\includegraphics[width=0.48\textwidth]{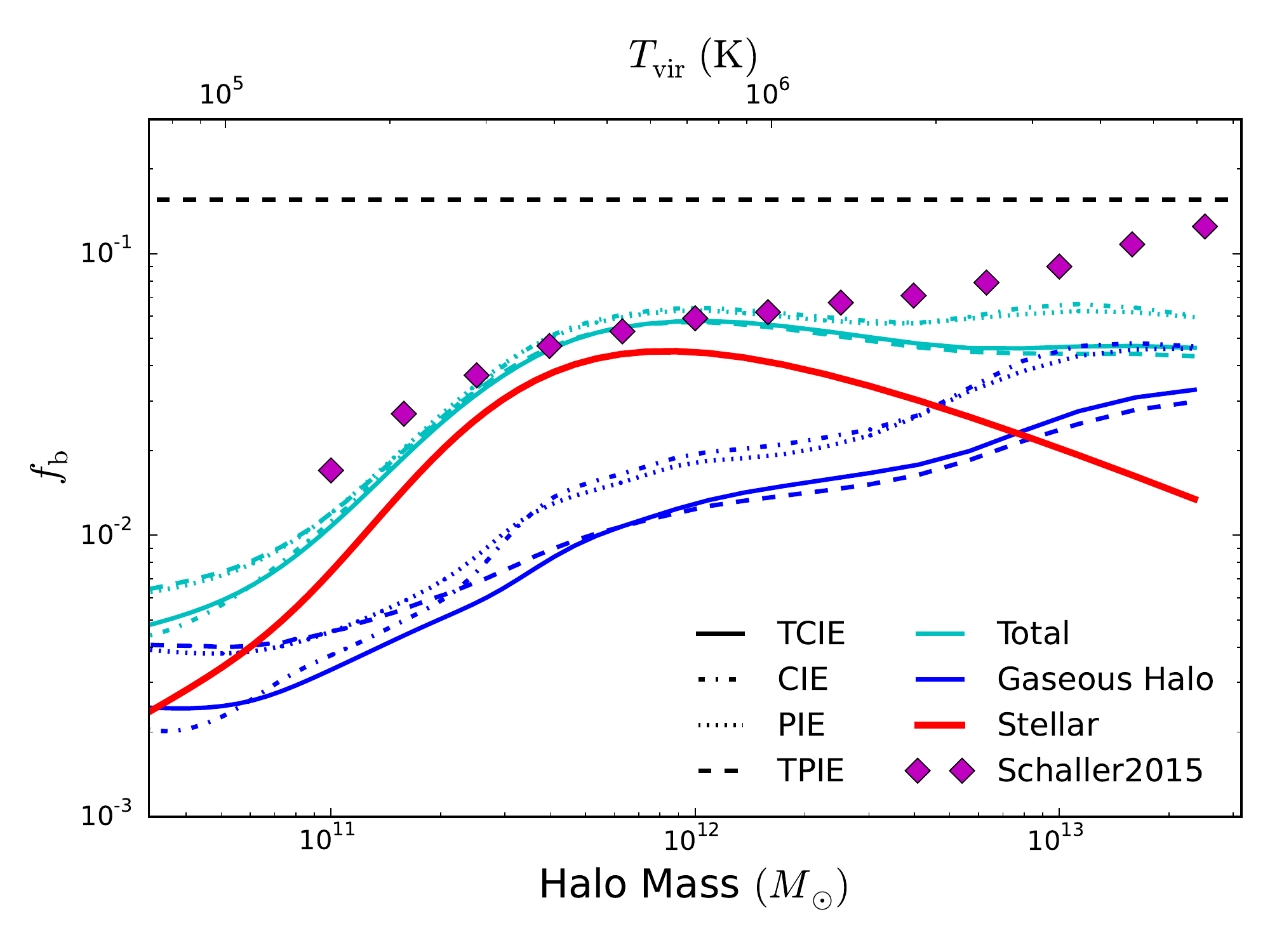}
\includegraphics[width=0.48\textwidth]{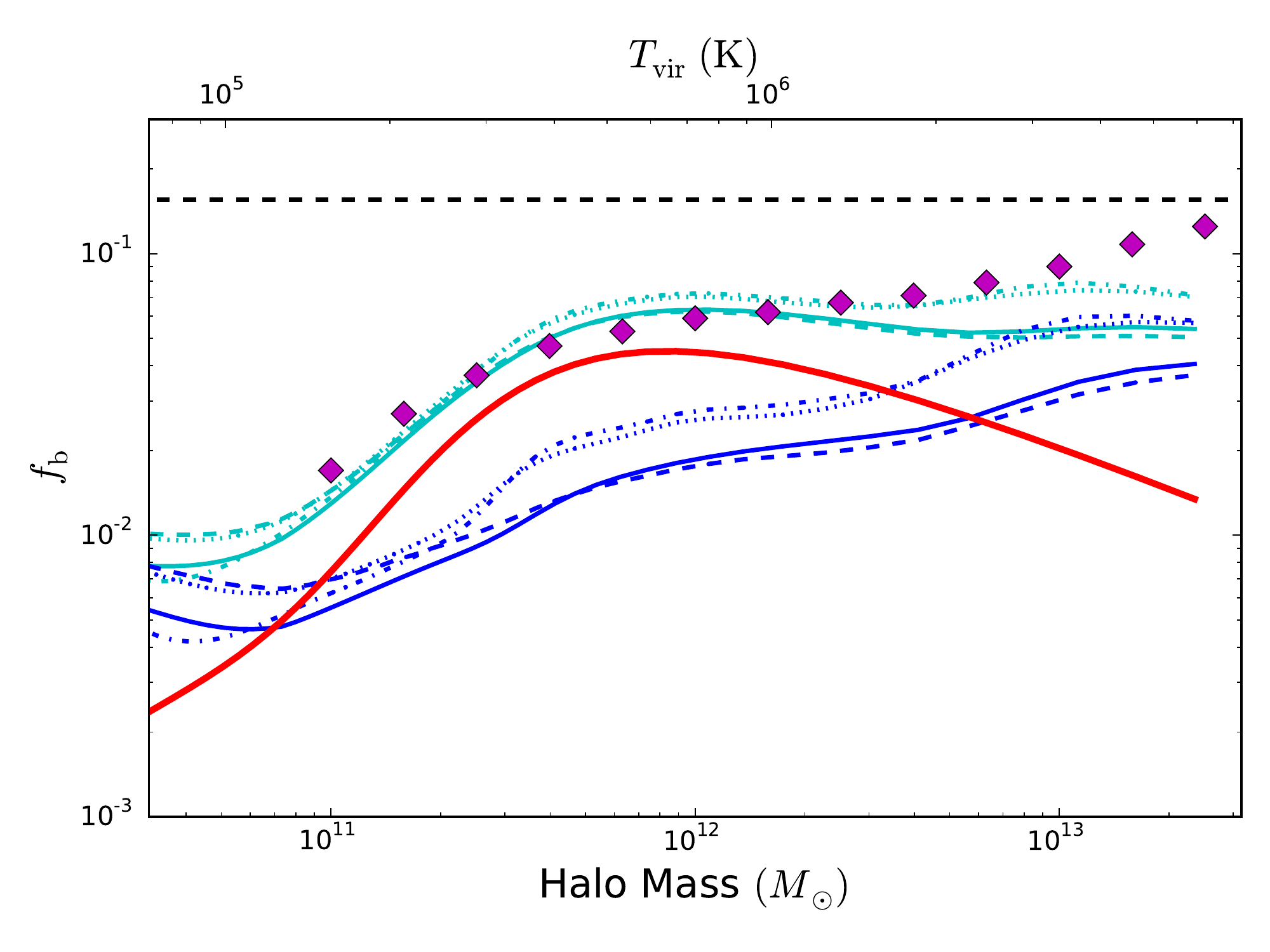}
}
\end{center}
\caption{The baryonic fraction dependence on the halo mass. {\it Left panel}: The {\tt CIE}, {\tt PIE}, {\tt TCIE} and {\tt TPIE} models are shown in dot-dashed, dotted, solid and dashed lines, while the black dashed line is the cosmic baryonic fraction, and the red solid line is the stellar content baryonic contribution. Blue lines are the gaseous-halo-only baryonic fraction, while cyan lines are the total baryonic fraction enclosed in the virial radius. The magenta diamonds are the baryonic fraction from the EAGLE simulation \citep{Schaller:2015aa}. {\it Right panel}: The baryonic fraction due to the gaseous halo is increased for the enhanced stellar feedback given in Fig. \ref{gamma}. The largest increases occur at lower masses and points at which the stellar and gaseous halos are equal occur at $M_{\rm h} = 6\times 10^{10}~M_\odot$ for the {\tt TPIE} model.}
\label{fb}
\end{figure*}

\subsection{Mass Budget}
The galaxy missing baryon problem is a crucial aspect of both observation and theory for  galaxy formation and evolution, so we also check whether our gaseous halo model can address this issue. In Fig. \ref{fb}, we show the baryonic fraction for models considered for typical star-forming galaxies. In all mass regions, the cosmic baryonic fraction is significantly higher than the total baryonic fraction of galaxies, which indicates that the gaseous halo within the virial radius cannot account for missing baryons of galaxies in the mass region  considered. The overall tendency shows that the low-mass galaxy is more baryon-poor than the massive galaxy, showing a sharp rise between $M_{\rm h} \sim 10^{11}~M_\odot$ and $4\times 10^{11}~M_\odot$. The baryonic fraction is almost a constant $f_{\rm b}\approx 0.05-0.06$ for galaxies with higher masses than $5\times10^{11}~M_\odot$. 

These baryonic fractions are the median values for galaxies at different masses, while it is possible to have significant scatter due to the SFR scatter. With the SFR scatter of $0.4\rm~dex$ \citep{Renzini:2015aa}, the baryonic fraction could have a scatter of $0.2\rm~dex$ based on the square root relationship between the sSFR and the gaseous halo mass. This scatter is consistent with cosmological simulations, which shows the baryonic fraction can vary between $20\%$ to $100\%$ of the cosmic baryonic fraction \citep{Marinacci:2014aa, Muratov:2015aa, Schaller:2015aa, Suresh:2017aa}.

That low-mass galaxies have a low $f_{\rm b}$ is the direct result of both low stellar mass and the high cooling rate. Although they have a relatively high sSFR, the gaseous halo is still low mass and comparable to the stellar mass within a factor of 2. This tendency is consistent with the simulation effort (EAGLE) when the halo mass is smaller than $\approx 3\times 10^{12}~M_\odot$ \citep{Schaye:2015aa, Schaller:2015aa}. EAGLE has prescriptions for the star formation, stellar evolution, stellar feedback and AGN feedback. The discrepancy between the EAGLE simulations and our models in the high mass region is due to the lack of heating from the AGN feedback in our model, which is positively related to the halo mass rather than the stellar mass. In the low-mass region, the baryonic fraction in our model is slightly less than  \citet{Schaller:2015aa} by a factor of $\lesssim 2$ at the halo mass of $10^{11}~M_\odot$. This might emphasize the importance of the stellar feedback for the low-mass galaxies, and a $\gamma$ factor of $0.1-0.2$ can account for such a difference. Involving the $\gamma$ factor described in Section 4.3, we set the upper limits for the baryonic fraction in our models, which is also shown in Fig. \ref{fb}. The modification on the $\gamma$ factor leads to higher baryonic fraction in low-mass galaxies, which is consistent with our hypothesis that low-mass galaxies have higher stellar feedback heating.

However, the trend of increasing baryonic fraction with the halo mass is significantly different from the Illustris simulation, which also has full stellar physics and AGN feedback \citep{Vogelsberger:2014aa}. The Illustris simulation shows the opposite tendency with the low-mass galaxy having more baryonic material (even higher than the cosmic baryonic fraction) enclosed in the virial radius \citep{Suresh:2017aa}. This result may be a result of the photoionization, which is included in Illustris but not EAGLE. Our models show that the photoionization modification is important for low-mass galaxies, since it can support a more massive gaseous halo. Nevertheless, the divergence between Illustris and EALGE is very unlikely to be caused by the photoionization modification, since it has also been shown that the photoionization can only raise the gaseous halo mass by a factor of about 2, down to the stellar mass of $8\times 10^{7} ~M_\odot$ (see Section 3.1). This difference is more likely to be caused by the weak stellar feedback employed in Illustris, which is set to keep gas inside the halo \citep{Suresh:2017aa}.

Another consideration that might moderate the missing baryon problem is having a gaseous halo extending beyond the virial radius. When we change the outermost radius for the gaseous halo from one virial radius to twice the virial radius, the gaseous halo mass is increased by a factor of the $2-3$. For an $L^*$ galaxy ($M_h=1.7\times 10^{12}~M_\odot$, ${\rm SFR} = 5~M_\odot~\rm yr^{-1}$), the cooling radius is $173\rm~kpc$ (less than the virial radius of $253\rm~kpc$), which indicates that the increasing of outermost radius will not change the normalization factor in the $\beta$-model. Modifying the outermost radius raises the mass from $2.6\times 10^{10}~M_\odot$ to $7.4\times 10^{10}~M_\odot$, raising the baryon fraction from $0.055$ to $0.083$. Therefore, in the case that the cooling radius is smaller than the virial radius, the factor is fixed to $2.83$, otherwise, the factor is slightly smaller but still around $2$. This would raise the baryonic fraction, but still not enough to account for all of the missing baryons for $L^*$ galaxies. For the high mass and the low-mass end of the galaxy distribution, the total baryonic fraction is raised by a factor of $\approx 2$, since most of the mass is in the gaseous halo rather than the stellar content.

\subsection{Future Observations}
An issue highlighted by this work is that one needs measurements for ions that are the dominant volume filling ions, which traces the gas that is near hydrostatic equilibrium and is at the temperature of most of the gaseous mass.  In practice, this requires that we obtain {\OVII} and {\OVIII} absorption line data for galaxies with $M_{\rm h} > 3\times 10^{11}~ M_\odot$.  Absorption in {\OVII} is available for the MW for about two dozen sight lines and in {\OVIII} for a handful of objects \citep{Fang:2015aa, Hodges-Kluck:2016aa}. A significant advance can be realized through improved $S/N$ for {\OVII} and especially {\OVIII} as well as for a larger number of sightlines. This will not happen with existing instruments ({\it XMM-Newton} and {\it Chandra}), which have already devoted about 20 Msec of observing time toward bright objects, so improvements would require several times this amount. 

For external galaxies, no {\OVII} or {\OVIII} absorption lines have been detected \citep{Nicastro:2016ab}. Sight lines through the halos of external galaxies ($0.3-1~ R_{\rm vir}$) are expected to be nearly an order of magnitude weaker than those from the MW.  The failure to see these lines is consistent with model predictions, given the sensitivity of current instruments and the amount of redshift space that has been probed.

Detecting {\OVII} and {\OVIII} through a sample of external galaxy halos will require a new instrument with capabilities that offer at least an order of magnitude improvement. Such an instrument would also offer a breakthrough in the study of these lines in the MW. This level of improvement is possible through {\it Arcus} \citep{Smith:2016aa}, an Explorer class mission that will have nearly an order of magnitude improvement in both spectral resolution and in collecting area, relative to the {\it XMM-Newton}/RGS (and a larger improvement relative to the {\it Chandra}/LETG).  The spectral resolution will be about 3000 ($100~\rm km~s^{-1}$), providing kinematic information as well as insights into turbulence. The {\it Athena} mission will also add to our understanding of these absorption systems, but its spectral resolution is poorer than that of {\it XMM-Newton} ($1300~\rm km~s^{-1}$), so kinematic information will be limited \citep{Barcons:2017aa}. The {\it Lynx} mission concept will offer another order of magnitude increase in collecting area, relative to {\it Arcus} and with double the resolution ($50~\rm km~s^{-1}$), which approaches the thermal width of gas at $2 \times 10^6\rm~ K$ \citep{Gaskin:2016aa}.  It will be sensitive to much weaker lines and will provide excellent kinematic information.

\section{Summary}
We report upon a gaseous halo model connecting the SFR and the radiative cooling rate, including photoionization and a multi-phase medium. This model predicts a comparable gaseous halo mass to the stellar mass, and can be employed to understand observations of high ionization state ions (i.e., {\OVI}, {\OVII}, {\NeVIII}, {\MgX}, and {\OVIII}). We summarize our major results:
\begin{enumerate}
\item Photoionization is the most important physical process in determining the relative ion distribution in the entire extended  gaseous halo of low-mass galaxies and the outskirts of massive galaxies. For low-mass galaxies ($M_{\rm h}< 3\times 10^{11}~M_\odot$), photoionization supports a more massive gaseous halo, and generates high ionization state ions (e.g., {\OVI} and {\OVII}). For more massive galaxies, photoionization leads to more high ionization state ions in the outskirts (i.e., the {\OVIII} of the MW).
\item The multi-phase medium within the cooling radius can be modeled by the distribution of $M(T)\propto T/\Lambda(T)$. This multi-phase medium leads to a flattened dependence of high ionization state ion column densities with galaxy halo mass. More relatively low ionization state ions (compared to the virial temperature) are generated because of the cooling from the high temperature medium.
\item Overall, our models predict the mass of the gaseous halo is comparable to the stellar mass (within one order of magnitude) for star forming galaxies over all halo masses. The cooling radius is expected to vary between $50-200~\rm kpc$, which is a small variation when compared to the two order of magnitude range in the halo mass.
\item {\OVI} has a narrow range ($\log N({\rm OVI}) = 13.5-14.3$) for galaxies with $M_{\rm h}<10^{13}~M_\odot$. Above $M_{\rm h}=3\times 10^{11}~M_\odot$, the {\OVI} is mainly from the collisional ionization, while below this mass, photons from the UVB ionizes most of {\OVI} ions. The predicted {\OVI} column density range is consistent with blind {\OVI} surveys.
\item A modified model is constructed for the Galactic {\OVII} and {\OVIII}, with changes in the standard metallicity of $0.55~Z_\odot $ and a maximum temperature of $1.93\times 10^6~\rm K$, which is above the virial temperature but similar to that derived from emission ratios. Such a gaseous halo leads to a hot halo mass of $1.9\times 10^{10}~M_\odot$ within the virial radius, which contributes to $7\%$ of the total baryonic mass of the MW.
\item For intervening {\NeVIII} and {\MgX} at $z=0.5-1.3$, our models predict column densities of $\approx 10^{14}~\rm cm^{-2}$, which is consistent with observations and informs the detection limit for future observations.
\item Such a gaseous halo cannot close the census of the galaxy missing baryons within $R_{\rm vir}$. Where it is possible to compare, our models results are similar to those of the EAGLE simulations, and about the half of baryons are still missing for $L^*$ galaxies within the virial radius.
\end{enumerate}

\acknowledgments
The authors would like to thank the referee for detailed and helpful comments, and Ben Oppenheimer, Edmund Hodges-Kluck, Oleg Gnedin and Hui Li for thoughtful comments and assistance. We also acknowledge the software {\tt Astropy} and {\tt Chianti}, and their developers \citep{Astropy-Collaboration:2013aa, Del-Zanna:2015aa}. We gratefully acknowledge support from NASA ADAP grant NNX16AF23G and from the Department of Astronomy at the University of Michigan.

\bibliographystyle{apj}
\bibliography{MissingBaryon}

\begin{thebibliography}{}
\expandafter\ifx\csname natexlab\endcsname\relax\def\natexlab#1{#1}\fi

\bibitem[{{Anderson} \& {Bregman}(2010)}]{Anderson:2010aa}
{Anderson}, M.~E., \& {Bregman}, J.~N. 2010, \apj, 714, 320

\bibitem[{{Anderson} {et~al.}(2013){Anderson}, {Bregman}, \&
  {Dai}}]{Anderson:2013aa}
{Anderson}, M.~E., {Bregman}, J.~N., \& {Dai}, X. 2013, \apj, 762, 106

\bibitem[{{Anderson} {et~al.}(2016){Anderson}, {Churazov}, \&
  {Bregman}}]{Anderson:2016aa}
{Anderson}, M.~E., {Churazov}, E., \& {Bregman}, J.~N. 2016, \mnras, 455, 227

\bibitem[{{Armillotta} {et~al.}(2016){Armillotta}, {Fraternali}, \&
  {Marinacci}}]{Armillotta:2016aa}
{Armillotta}, L., {Fraternali}, F., \& {Marinacci}, F. 2016, \mnras, 462, 4157

\bibitem[{{Armillotta} {et~al.}(2017){Armillotta}, {Fraternali}, {Werk},
  {Prochaska}, \& {Marinacci}}]{Armillotta:2017aa}
{Armillotta}, L., {Fraternali}, F., {Werk}, J.~K., {Prochaska}, J.~X., \&
  {Marinacci}, F. 2017, \mnras, 470, 114

\bibitem[{{Asplund} {et~al.}(2009){Asplund}, {Grevesse}, {Sauval}, \&
  {Scott}}]{Asplund:2009aa}
{Asplund}, M., {Grevesse}, N., {Sauval}, A.~J., \& {Scott}, P. 2009, \araa, 47,
  481

\bibitem[{{Astropy Collaboration} {et~al.}(2013){Astropy Collaboration},
  {Robitaille}, {Tollerud}, {Greenfield}, {Droettboom}, {Bray}, {Aldcroft},
  {Davis}, {Ginsburg}, {Price-Whelan}, {Kerzendorf}, {Conley}, {Crighton},
  {Barbary}, {Muna}, {Ferguson}, {Grollier}, {Parikh}, {Nair}, {Unther},
  {Deil}, {Woillez}, {Conseil}, {Kramer}, {Turner}, {Singer}, {Fox}, {Weaver},
  {Zabalza}, {Edwards}, {Azalee Bostroem}, {Burke}, {Casey}, {Crawford},
  {Dencheva}, {Ely}, {Jenness}, {Labrie}, {Lim}, {Pierfederici}, {Pontzen},
  {Ptak}, {Refsdal}, {Servillat}, \&
  {Streicher}}]{Astropy-Collaboration:2013aa}
{Astropy Collaboration}, {Robitaille}, T.~P., {Tollerud}, E.~J., {et~al.} 2013,
  \aap, 558, A33

\bibitem[{{Baldi} {et~al.}(2012){Baldi}, {Ettori}, {Molendi}, \&
  {Gastaldello}}]{Baldi:2012aa}
{Baldi}, A., {Ettori}, S., {Molendi}, S., \& {Gastaldello}, F. 2012, \aap, 545,
  A41

\bibitem[{{Balogh} {et~al.}(1998){Balogh}, {Schade}, {Morris}, {Yee},
  {Carlberg}, \& {Ellingson}}]{Balogh:1998aa}
{Balogh}, M.~L., {Schade}, D., {Morris}, S.~L., {et~al.} 1998, \apjl, 504, L75

\bibitem[{{Barcons} {et~al.}(2017){Barcons}, {Barret}, {Decourchelle}, {den
  Herder}, {Fabian}, {Matsumoto}, {Lumb}, {Nandra}, {Piro}, {Smith}, \&
  {Willingale}}]{Barcons:2017aa}
{Barcons}, X., {Barret}, D., {Decourchelle}, A., {et~al.} 2017, Astronomische
  Nachrichten, 338, 153

\bibitem[{{Behroozi} {et~al.}(2013){Behroozi}, {Wechsler}, \&
  {Conroy}}]{Behroozi:2013aa}
{Behroozi}, P.~S., {Wechsler}, R.~H., \& {Conroy}, C. 2013, \apj, 770, 57

\bibitem[{{Bogd{\'a}n} {et~al.}(2017){Bogd{\'a}n}, {Bourdin}, {Forman},
  {Kraft}, {Vogelsberger}, {Hernquist}, \& {Springel}}]{Bogdan:2017aa}
{Bogd{\'a}n}, {\'A}., {Bourdin}, H., {Forman}, W.~R., {et~al.} 2017, \apj, 850,
  98

\bibitem[{{Bogd{\'a}n} {et~al.}(2013){Bogd{\'a}n}, {Forman}, {Kraft}, \&
  {Jones}}]{Bogdan:2013aa}
{Bogd{\'a}n}, {\'A}., {Forman}, W.~R., {Kraft}, R.~P., \& {Jones}, C. 2013,
  \apj, 772, 98

\bibitem[{{Bregman}(1980)}]{Bregman:1980aa}
{Bregman}, J.~N. 1980, \apj, 236, 577

\bibitem[{{Bregman} {et~al.}(2018){Bregman}, {Anderson}, {Miller},
  {Hodges-Kluck}, {Dai}, {Li}, \& {Li}}]{Bregman:2018aa}
{Bregman}, J.~N., {Anderson}, M.~E., {Miller}, M.~J., {et~al.} 2018, submitted

\bibitem[{{Bregman} \& {Lloyd-Davies}(2007)}]{Bregman:2007ab}
{Bregman}, J.~N., \& {Lloyd-Davies}, E.~J. 2007, \apj, 669, 990

\bibitem[{{Brown} \& {Bregman}(1998)}]{Brown:1998aa}
{Brown}, B.~A., \& {Bregman}, J.~N. 1998, \apjl, 495, L75

\bibitem[{{Bryans} {et~al.}(2006){Bryans}, {Badnell}, {Gorczyca}, {Laming},
  {Mitthumsiri}, \& {Savin}}]{Bryans:2006aa}
{Bryans}, P., {Badnell}, N.~R., {Gorczyca}, T.~W., {et~al.} 2006, \apjs, 167,
  343

\bibitem[{{Cen} \& {Ostriker}(1999)}]{Cen:1999aa}
{Cen}, R., \& {Ostriker}, J.~P. 1999, \apj, 514, 1

\bibitem[{{Chen} \& {Mulchaey}(2009)}]{Chen:2009aa}
{Chen}, H.-W., \& {Mulchaey}, J.~S. 2009, \apj, 701, 1219

\bibitem[{{Dai} {et~al.}(2010){Dai}, {Bregman}, {Kochanek}, \&
  {Rasia}}]{Dai:2010aa}
{Dai}, X., {Bregman}, J.~N., {Kochanek}, C.~S., \& {Rasia}, E. 2010, \apj, 719,
  119

\bibitem[{{Danforth} \& {Shull}(2008)}]{Danforth:2008aa}
{Danforth}, C.~W., \& {Shull}, J.~M. 2008, \apj, 679, 194

\bibitem[{{Davis} \& {White}(1996)}]{Davis:1996aa}
{Davis}, D.~S., \& {White}, III, R.~E. 1996, \apjl, 470, L35

\bibitem[{{Del Zanna} {et~al.}(2015){Del Zanna}, {Dere}, {Young}, {Landi}, \&
  {Mason}}]{Del-Zanna:2015aa}
{Del Zanna}, G., {Dere}, K.~P., {Young}, P.~R., {Landi}, E., \& {Mason}, H.~E.
  2015, \aap, 582, A56

\bibitem[{{Fabian}(2012)}]{Fabian:2012aa}
{Fabian}, A.~C. 2012, \araa, 50, 455

\bibitem[{{Faerman} {et~al.}(2017){Faerman}, {Sternberg}, \&
  {McKee}}]{Faerman:2017aa}
{Faerman}, Y., {Sternberg}, A., \& {McKee}, C.~F. 2017, \apj, 835, 52

\bibitem[{{Fang} {et~al.}(2015){Fang}, {Buote}, {Bullock}, \&
  {Ma}}]{Fang:2015aa}
{Fang}, T., {Buote}, D., {Bullock}, J., \& {Ma}, R. 2015, \apjs, 217, 21

\bibitem[{{Fielding} {et~al.}(2017){Fielding}, {Quataert}, {Martizzi}, \&
  {Faucher-Gigu{\`e}re}}]{Fielding:2017aa}
{Fielding}, D., {Quataert}, E., {Martizzi}, D., \& {Faucher-Gigu{\`e}re}, C.-A.
  2017, \mnras, 470, L39

\bibitem[{{Fukugita} \& {Peebles}(2006)}]{Fukugita:2006aa}
{Fukugita}, M., \& {Peebles}, P.~J.~E. 2006, \apj, 639, 590

\bibitem[{{Gaskin} {et~al.}(2016){Gaskin}, {{\"O}zel}, \&
  {Vikhlinin}}]{Gaskin:2016aa}
{Gaskin}, J., {{\"O}zel}, F., \& {Vikhlinin}, A. 2016, in \procspie, Vol. 9904,
  Space Telescopes and Instrumentation 2016: Optical, Infrared, and Millimeter
  Wave, 99040N

\bibitem[{{Gnat}(2017)}]{Gnat:2017aa}
{Gnat}, O. 2017, \apjs, 228, 11

\bibitem[{{Goulding} {et~al.}(2016){Goulding}, {Greene}, {Ma}, {Veale},
  {Bogdan}, {Nyland}, {Blakeslee}, {McConnell}, \& {Thomas}}]{Goulding:2016aa}
{Goulding}, A.~D., {Greene}, J.~E., {Ma}, C.-P., {et~al.} 2016, \apj, 826, 167

\bibitem[{{Grimes} {et~al.}(2009){Grimes}, {Heckman}, {Aloisi}, {Calzetti},
  {Leitherer}, {Martin}, {Meurer}, {Sembach}, \& {Strickland}}]{Grimes:2009aa}
{Grimes}, J.~P., {Heckman}, T., {Aloisi}, A., {et~al.} 2009, \apjs, 181, 272

\bibitem[{{Gupta} {et~al.}(2012){Gupta}, {Mathur}, {Krongold}, {Nicastro}, \&
  {Galeazzi}}]{Gupta:2012aa}
{Gupta}, A., {Mathur}, S., {Krongold}, Y., {Nicastro}, F., \& {Galeazzi}, M.
  2012, \apjl, 756, L8

\bibitem[{{Haardt} \& {Madau}(2012)}]{Haardt:2012aa}
{Haardt}, F., \& {Madau}, P. 2012, \apj, 746, 125

\bibitem[{{Henley} \& {Shelton}(2012)}]{Henley:2012aa}
{Henley}, D.~B., \& {Shelton}, R.~L. 2012, \apjs, 202, 14

\bibitem[{{Hodges-Kluck} {et~al.}(2016){Hodges-Kluck}, {Miller}, \&
  {Bregman}}]{Hodges-Kluck:2016aa}
{Hodges-Kluck}, E.~J., {Miller}, M.~J., \& {Bregman}, J.~N. 2016, \apj, 822, 21

\bibitem[{{Hopkins} {et~al.}(2014){Hopkins}, {Kere{\v s}}, {O{\~n}orbe},
  {Faucher-Gigu{\`e}re}, {Quataert}, {Murray}, \& {Bullock}}]{Hopkins:2014aa}
{Hopkins}, P.~F., {Kere{\v s}}, D., {O{\~n}orbe}, J., {et~al.} 2014, \mnras,
  445, 581

\bibitem[{{Hopkins} {et~al.}(2017){Hopkins}, {Wetzel}, {Keres},
  {Faucher-Giguere}, {Quataert}, {Boylan-Kolchin}, {Murray}, {Hayward},
  {Garrison-Kimmel}, {Hummels}, {Feldmann}, {Torrey}, {Ma}, {Angles-Alcazar},
  {Su}, {Orr}, {Schmitz}, {Escala}, {Sanderson}, {Grudic}, {Hafen}, {Kim},
  {Fitts}, {Bullock}, {Wheeler}, {Chan}, {Elbert}, \&
  {Narananan}}]{Hopkins:2017aa}
{Hopkins}, P.~F., {Wetzel}, A., {Keres}, D., {et~al.} 2017, ArXiv e-prints,
  arXiv:1702.06148

\bibitem[{{Hussain} {et~al.}(2017){Hussain}, {Khaire}, {Srianand}, {Muzahid},
  \& {Pathak}}]{Hussain:2017aa}
{Hussain}, T., {Khaire}, V., {Srianand}, R., {Muzahid}, S., \& {Pathak}, A.
  2017, \mnras, 466, 3133

\bibitem[{{Hussain} {et~al.}(2015){Hussain}, {Muzahid}, {Narayanan},
  {Srianand}, {Wakker}, {Charlton}, \& {Pathak}}]{Hussain:2015aa}
{Hussain}, T., {Muzahid}, S., {Narayanan}, A., {et~al.} 2015, \mnras, 446, 2444

\bibitem[{{Johnson} {et~al.}(2015){Johnson}, {Chen}, \&
  {Mulchaey}}]{Johnson:2015aa}
{Johnson}, S.~D., {Chen}, H.-W., \& {Mulchaey}, J.~S. 2015, \mnras, 449, 3263

\bibitem[{{Johnson} {et~al.}(2017){Johnson}, {Chen}, {Mulchaey}, {Schaye}, \&
  {Straka}}]{Johnson:2017aa}
{Johnson}, S.~D., {Chen}, H.-W., {Mulchaey}, J.~S., {Schaye}, J., \& {Straka},
  L.~A. 2017, ArXiv e-prints, arXiv:1710.06441

\bibitem[{{Kere{\v s}} {et~al.}(2009){Kere{\v s}}, {Katz}, {Dav{\'e}},
  {Fardal}, \& {Weinberg}}]{Keres:2009aa}
{Kere{\v s}}, D., {Katz}, N., {Dav{\'e}}, R., {Fardal}, M., \& {Weinberg},
  D.~H. 2009, \mnras, 396, 2332

\bibitem[{{Kere{\v s}} {et~al.}(2005){Kere{\v s}}, {Katz}, {Weinberg}, \&
  {Dav{\'e}}}]{Keres:2005aa}
{Kere{\v s}}, D., {Katz}, N., {Weinberg}, D.~H., \& {Dav{\'e}}, R. 2005,
  \mnras, 363, 2

\bibitem[{{Khaire} \& {Srianand}(2015)}]{Khaire:2015ab}
{Khaire}, V., \& {Srianand}, R. 2015, \mnras, 451, L30

\bibitem[{{Kravtsov} {et~al.}(2014){Kravtsov}, {Vikhlinin}, \&
  {Meshscheryakov}}]{Kravtsov:2014aa}
{Kravtsov}, A., {Vikhlinin}, A., \& {Meshscheryakov}, A. 2014, ArXiv e-prints,
  arXiv:1401.7329

\bibitem[{{Kwak} \& {Shelton}(2010)}]{Kwak:2010aa}
{Kwak}, K., \& {Shelton}, R.~L. 2010, \apj, 719, 523

\bibitem[{{Larson}(1981)}]{Larson:1981aa}
{Larson}, R.~B. 1981, \mnras, 194, 809

\bibitem[{{Larson}(1994)}]{Larson:1994aa}
{Larson}, R.~B. 1994, in Lecture Notes in Physics, Berlin Springer Verlag, Vol.
  439, The Structure and Content of Molecular Clouds, ed. T.~L. {Wilson} \&
  K.~J. {Johnston}, 13

\bibitem[{{Leroy} {et~al.}(2008){Leroy}, {Walter}, {Brinks}, {Bigiel}, {de
  Blok}, {Madore}, \& {Thornley}}]{Leroy:2008aa}
{Leroy}, A.~K., {Walter}, F., {Brinks}, E., {et~al.} 2008, \aj, 136, 2782

\bibitem[{{Li} {et~al.}(2016){Li}, {Bregman}, {Wang}, {Crain}, \&
  {Anderson}}]{Li:2016aa}
{Li}, J.-T., {Bregman}, J.~N., {Wang}, Q.~D., {Crain}, R.~A., \& {Anderson},
  M.~E. 2016, \apj, 830, 134

\bibitem[{{Li} {et~al.}(2017){Li}, {Bregman}, {Wang}, {Crain}, \&
  {Anderson}}]{Li:2017aa}
---. 2017, Submitted

\bibitem[{{Li} {et~al.}(2014){Li}, {Crain}, \& {Wang}}]{Li:2014aa}
{Li}, J.-T., {Crain}, R.~A., \& {Wang}, Q.~D. 2014, \mnras, 440, 859

\bibitem[{{Li} \& {Bregman}(2017)}]{Li:2017ab}
{Li}, Y., \& {Bregman}, J. 2017, \apj, 849, 105

\bibitem[{{Li} \& {Bregman}(2018)}]{Li:2018aa}
{Li}, Y., \& {Bregman}, J. 2018, in American Astronomical Society Meeting
  Abstracts, Vol. 231, American Astronomical Society Meeting Abstracts,
  \#451.02

\bibitem[{{Li} {et~al.}(2015){Li}, {Bryan}, {Ruszkowski}, {Voit}, {O'Shea}, \&
  {Donahue}}]{Li:2015aa}
{Li}, Y., {Bryan}, G.~L., {Ruszkowski}, M., {et~al.} 2015, \apj, 811, 73

\bibitem[{{Lilly} {et~al.}(2013){Lilly}, {Carollo}, {Pipino}, {Renzini}, \&
  {Peng}}]{Lilly:2013aa}
{Lilly}, S.~J., {Carollo}, C.~M., {Pipino}, A., {Renzini}, A., \& {Peng}, Y.
  2013, \apj, 772, 119

\bibitem[{{Madau} \& {Dickinson}(2014)}]{Madau:2014aa}
{Madau}, P., \& {Dickinson}, M. 2014, \araa, 52, 415

\bibitem[{{Marinacci} {et~al.}(2010){Marinacci}, {Binney}, {Fraternali},
  {Nipoti}, {Ciotti}, \& {Londrillo}}]{Marinacci:2010aa}
{Marinacci}, F., {Binney}, J., {Fraternali}, F., {et~al.} 2010, \mnras, 404,
  1464

\bibitem[{{Marinacci} {et~al.}(2014){Marinacci}, {Pakmor}, {Springel}, \&
  {Simpson}}]{Marinacci:2014aa}
{Marinacci}, F., {Pakmor}, R., {Springel}, V., \& {Simpson}, C.~M. 2014,
  \mnras, 442, 3745

\bibitem[{{McGaugh} \& {Schombert}(2015)}]{McGaugh:2015aa}
{McGaugh}, S.~S., \& {Schombert}, J.~M. 2015, \apj, 802, 18

\bibitem[{{McQuinn} \& {Werk}(2018)}]{McQuinn:2018aa}
{McQuinn}, M., \& {Werk}, J.~K. 2018, \apj, 852, 33

\bibitem[{{Meiring} {et~al.}(2013){Meiring}, {Tripp}, {Werk}, {Howk},
  {Jenkins}, {Prochaska}, {Lehner}, \& {Sembach}}]{Meiring:2013aa}
{Meiring}, J.~D., {Tripp}, T.~M., {Werk}, J.~K., {et~al.} 2013, \apj, 767, 49

\bibitem[{{M{\'e}nard} {et~al.}(2010){M{\'e}nard}, {Scranton}, {Fukugita}, \&
  {Richards}}]{Menard:2010aa}
{M{\'e}nard}, B., {Scranton}, R., {Fukugita}, M., \& {Richards}, G. 2010,
  \mnras, 405, 1025

\bibitem[{{Miller} \& {Bregman}(2015)}]{Miller:2015aa}
{Miller}, M.~J., \& {Bregman}, J.~N. 2015, \apj, 800, 14

\bibitem[{{Mitra} {et~al.}(2013){Mitra}, {Ferrara}, \&
  {Choudhury}}]{Mitra:2013aa}
{Mitra}, S., {Ferrara}, A., \& {Choudhury}, T.~R. 2013, \mnras, 428, L1

\bibitem[{{Mo} {et~al.}(2010){Mo}, {van den Bosch}, \& {White}}]{Mo:2010aa}
{Mo}, H., {van den Bosch}, F.~C., \& {White}, S. 2010, {Galaxy Formation and
  Evolution}

\bibitem[{{Morselli} {et~al.}(2016){Morselli}, {Renzini}, {Popesso}, \&
  {Erfanianfar}}]{Morselli:2016aa}
{Morselli}, L., {Renzini}, A., {Popesso}, P., \& {Erfanianfar}, G. 2016,
  \mnras, 462, 2355

\bibitem[{{Muratov} {et~al.}(2015){Muratov}, {Kere{\v s}},
  {Faucher-Gigu{\`e}re}, {Hopkins}, {Quataert}, \& {Murray}}]{Muratov:2015aa}
{Muratov}, A.~L., {Kere{\v s}}, D., {Faucher-Gigu{\`e}re}, C.-A., {et~al.}
  2015, \mnras, 454, 2691

\bibitem[{{Myers} {et~al.}(1986){Myers}, {Dame}, {Thaddeus}, {Cohen},
  {Silverberg}, {Dwek}, \& {Hauser}}]{Myers:1986aa}
{Myers}, P.~C., {Dame}, T.~M., {Thaddeus}, P., {et~al.} 1986, \apj, 301, 398

\bibitem[{{Nakanishi} \& {Sofue}(2016)}]{Nakanishi:2016aa}
{Nakanishi}, H., \& {Sofue}, Y. 2016, \pasj, 68, 5

\bibitem[{{Narayanan} {et~al.}(2012){Narayanan}, {Savage}, \&
  {Wakker}}]{Narayanan:2012aa}
{Narayanan}, A., {Savage}, B.~D., \& {Wakker}, B.~P. 2012, \apj, 752, 65

\bibitem[{{Nelson} {et~al.}(2013){Nelson}, {Vogelsberger}, {Genel}, {Sijacki},
  {Kere{\v s}}, {Springel}, \& {Hernquist}}]{Nelson:2013aa}
{Nelson}, D., {Vogelsberger}, M., {Genel}, S., {et~al.} 2013, \mnras, 429, 3353

\bibitem[{{Nelson} {et~al.}(2017){Nelson}, {Kauffmann}, {Pillepich}, {Genel},
  {Springel}, {Pakmor}, {Hernquist}, {Weinberger}, {Torrey}, {Vogelsberger}, \&
  {Marinacci}}]{Nelson:2017aa}
{Nelson}, D., {Kauffmann}, G., {Pillepich}, A., {et~al.} 2017, ArXiv e-prints,
  arXiv:1712.00016

\bibitem[{{Nevalainen} {et~al.}(2017){Nevalainen}, {Wakker}, {Kaastra},
  {Bonamente}, {Snowden}, {Paerels}, \& {de Vries}}]{Nevalainen:2017aa}
{Nevalainen}, J., {Wakker}, B., {Kaastra}, J., {et~al.} 2017, \aap, 605, A47

\bibitem[{{Nicastro} {et~al.}(2016{\natexlab{a}}){Nicastro}, {Senatore},
  {Gupta}, {Guainazzi}, {Mathur}, {Krongold}, {Elvis}, \&
  {Piro}}]{Nicastro:2016aa}
{Nicastro}, F., {Senatore}, F., {Gupta}, A., {et~al.} 2016{\natexlab{a}},
  \mnras, 457, 676

\bibitem[{{Nicastro} {et~al.}(2016{\natexlab{b}}){Nicastro}, {Senatore},
  {Gupta}, {Mathur}, {Krongold}, {Elvis}, \& {Piro}}]{Nicastro:2016ab}
---. 2016{\natexlab{b}}, \mnras, 458, L123

\bibitem[{{Nicastro} {et~al.}(2002){Nicastro}, {Zezas}, {Drake}, {Elvis},
  {Fiore}, {Fruscione}, {Marengo}, {Mathur}, \& {Bianchi}}]{Nicastro:2002aa}
{Nicastro}, F., {Zezas}, A., {Drake}, J., {et~al.} 2002, \apj, 573, 157

\bibitem[{{Oosterloo} {et~al.}(2007){Oosterloo}, {Fraternali}, \&
  {Sancisi}}]{Oosterloo:2007aa}
{Oosterloo}, T., {Fraternali}, F., \& {Sancisi}, R. 2007, \aj, 134, 1019

\bibitem[{{Oppenheimer} \& {Schaye}(2013)}]{Oppenheimer:2013aa}
{Oppenheimer}, B.~D., \& {Schaye}, J. 2013, \mnras, 434, 1043

\bibitem[{{Oppenheimer} {et~al.}(2017){Oppenheimer}, {Segers}, {Schaye},
  {Richings}, \& {Crain}}]{Oppenheimer:2017aa}
{Oppenheimer}, B.~D., {Segers}, M., {Schaye}, J., {Richings}, A.~J., \&
  {Crain}, R.~A. 2017, ArXiv e-prints, arXiv:1705.07897

\bibitem[{{Oppenheimer} {et~al.}(2016){Oppenheimer}, {Crain}, {Schaye},
  {Rahmati}, {Richings}, {Trayford}, {Tumlinson}, {Bower}, {Schaller}, \&
  {Theuns}}]{Oppenheimer:2016aa}
{Oppenheimer}, B.~D., {Crain}, R.~A., {Schaye}, J., {et~al.} 2016, \mnras, 460,
  2157

\bibitem[{{Pachat} {et~al.}(2017){Pachat}, {Narayanan}, {Khaire}, {Savage},
  {Muzahid}, \& {Wakker}}]{Pachat:2017aa}
{Pachat}, S., {Narayanan}, A., {Khaire}, V., {et~al.} 2017, \mnras, 471, 792

\bibitem[{{Pannella} {et~al.}(2009){Pannella}, {Carilli}, {Daddi}, {McCracken},
  {Owen}, {Renzini}, {Strazzullo}, {Civano}, {Koekemoer}, {Schinnerer},
  {Scoville}, {Smol{\v c}i{\'c}}, {Taniguchi}, {Aussel}, {Kneib}, {Ilbert},
  {Mellier}, {Salvato}, {Thompson}, \& {Willott}}]{Pannella:2009aa}
{Pannella}, M., {Carilli}, C.~L., {Daddi}, E., {et~al.} 2009, \apjl, 698, L116

\bibitem[{{Planck Collaboration} {et~al.}(2016){Planck Collaboration}, {Ade},
  {Aghanim}, {Arnaud}, {Ashdown}, {Aumont}, {Baccigalupi}, {Banday},
  {Barreiro}, {Bartlett}, \& et~al.}]{Planck-Collaboration:2016aa}
{Planck Collaboration}, {Ade}, P.~A.~R., {Aghanim}, N., {et~al.} 2016, \aap,
  594, A13

\bibitem[{{Qu} \& {Bregman}(2016)}]{Qu:2016aa}
{Qu}, Z., \& {Bregman}, J.~N. 2016, \apj, 832, 189

\bibitem[{{Renzini} \& {Peng}(2015)}]{Renzini:2015aa}
{Renzini}, A., \& {Peng}, Y.-j. 2015, \apjl, 801, L29

\bibitem[{{Robitaille} \& {Whitney}(2010)}]{Robitaille:2010aa}
{Robitaille}, T.~P., \& {Whitney}, B.~A. 2010, \apjl, 710, L11

\bibitem[{{Sancisi} {et~al.}(2008){Sancisi}, {Fraternali}, {Oosterloo}, \& {van
  der Hulst}}]{Sancisi:2008aa}
{Sancisi}, R., {Fraternali}, F., {Oosterloo}, T., \& {van der Hulst}, T. 2008,
  \aapr, 15, 189

\bibitem[{{Savage} {et~al.}(2014){Savage}, {Kim}, {Wakker}, {Keeney}, {Shull},
  {Stocke}, \& {Green}}]{Savage:2014aa}
{Savage}, B.~D., {Kim}, T.-S., {Wakker}, B.~P., {et~al.} 2014, \apjs, 212, 8

\bibitem[{{Savage} {et~al.}(2005){Savage}, {Lehner}, {Wakker}, {Sembach}, \&
  {Tripp}}]{Savage:2005aa}
{Savage}, B.~D., {Lehner}, N., {Wakker}, B.~P., {Sembach}, K.~R., \& {Tripp},
  T.~M. 2005, \apj, 626, 776

\bibitem[{{Savage} {et~al.}(2003){Savage}, {Sembach}, {Wakker}, {Richter},
  {Meade}, {Jenkins}, {Shull}, {Moos}, \& {Sonneborn}}]{Savage:2003aa}
{Savage}, B.~D., {Sembach}, K.~R., {Wakker}, B.~P., {et~al.} 2003, \apjs, 146,
  125

\bibitem[{{Scannapieco} {et~al.}(2008){Scannapieco}, {Tissera}, {White}, \&
  {Springel}}]{Scannapieco:2008aa}
{Scannapieco}, C., {Tissera}, P.~B., {White}, S.~D.~M., \& {Springel}, V. 2008,
  \mnras, 389, 1137

\bibitem[{{Scannapieco} \& {Br{\"u}ggen}(2015)}]{Scannapieco:2015aa}
{Scannapieco}, E., \& {Br{\"u}ggen}, M. 2015, \apj, 805, 158

\bibitem[{{Schaller} {et~al.}(2015){Schaller}, {Frenk}, {Bower}, {Theuns},
  {Jenkins}, {Schaye}, {Crain}, {Furlong}, {Dalla Vecchia}, \&
  {McCarthy}}]{Schaller:2015aa}
{Schaller}, M., {Frenk}, C.~S., {Bower}, R.~G., {et~al.} 2015, \mnras, 451,
  1247

\bibitem[{{Schaye} {et~al.}(2015){Schaye}, {Crain}, {Bower}, {Furlong},
  {Schaller}, {Theuns}, {Dalla Vecchia}, {Frenk}, {McCarthy}, {Helly},
  {Jenkins}, {Rosas-Guevara}, {White}, {Baes}, {Booth}, {Camps}, {Navarro},
  {Qu}, {Rahmati}, {Sawala}, {Thomas}, \& {Trayford}}]{Schaye:2015aa}
{Schaye}, J., {Crain}, R.~A., {Bower}, R.~G., {et~al.} 2015, \mnras, 446, 521

\bibitem[{{Sembach} {et~al.}(2003){Sembach}, {Wakker}, {Savage}, {Richter},
  {Meade}, {Shull}, {Jenkins}, {Sonneborn}, \& {Moos}}]{Sembach:2003aa}
{Sembach}, K.~R., {Wakker}, B.~P., {Savage}, B.~D., {et~al.} 2003, \apjs, 146,
  165

\bibitem[{{Smith} {et~al.}(2016){Smith}, {Abraham}, {Allured}, {Bautz},
  {Bookbinder}, {Bregman}, {Brenneman}, {Brickhouse}, {Burrows}, {Burwitz},
  {Carvalho}, {Cheimets}, {Costantini}, {Dawson}, {DeRoo}, {Falcone}, {Foster},
  {Grant}, {Heilmann}, {Hertz}, {Hine}, {Huenemoerder}, {Kaastra}, {Madsen},
  {McEntaffer}, {Miller}, {Miller}, {Morse}, {Mushotzky}, {Nandra}, {Nowak},
  {Paerels}, {Petre}, {Plice}, {Poppenhaeger}, {Ptak}, {Reid}, {Sanders},
  {Schattenburg}, {Schulz}, {Smale}, {Temi}, {Valencic}, {Walker},
  {Willingale}, {Wilms}, \& {Wolk}}]{Smith:2016aa}
{Smith}, R.~K., {Abraham}, M.~H., {Allured}, R., {et~al.} 2016, in \procspie,
  Vol. 9905, Space Telescopes and Instrumentation 2016: Ultraviolet to Gamma
  Ray, 99054M

\bibitem[{{Stern} {et~al.}(2016){Stern}, {Hennawi}, {Prochaska}, \&
  {Werk}}]{Stern:2016aa}
{Stern}, J., {Hennawi}, J.~F., {Prochaska}, J.~X., \& {Werk}, J.~K. 2016, \apj,
  830, 87

\bibitem[{{Stocke} {et~al.}(2013){Stocke}, {Keeney}, {Danforth}, {Shull},
  {Froning}, {Green}, {Penton}, \& {Savage}}]{Stocke:2013aa}
{Stocke}, J.~T., {Keeney}, B.~A., {Danforth}, C.~W., {et~al.} 2013, \apj, 763,
  148

\bibitem[{{Stocke} {et~al.}(2014){Stocke}, {Keeney}, {Danforth}, {Syphers},
  {Yamamoto}, {Shull}, {Green}, {Froning}, {Savage}, {Wakker}, {Kim},
  {Ryan-Weber}, \& {Kacprzak}}]{Stocke:2014aa}
---. 2014, \apj, 791, 128

\bibitem[{{Suresh} {et~al.}(2017){Suresh}, {Rubin}, {Kannan}, {Werk},
  {Hernquist}, \& {Vogelsberger}}]{Suresh:2017aa}
{Suresh}, J., {Rubin}, K.~H.~R., {Kannan}, R., {et~al.} 2017, \mnras, 465, 2966

\bibitem[{{Thom} \& {Chen}(2008)}]{Thom:2008aa}
{Thom}, C., \& {Chen}, H.-W. 2008, \apjs, 179, 37

\bibitem[{{Thompson} {et~al.}(2016){Thompson}, {Quataert}, {Zhang}, \&
  {Weinberg}}]{Thompson:2016aa}
{Thompson}, T.~A., {Quataert}, E., {Zhang}, D., \& {Weinberg}, D.~H. 2016,
  \mnras, 455, 1830

\bibitem[{{Tripp} {et~al.}(2011){Tripp}, {Meiring}, {Prochaska}, {Willmer},
  {Howk}, {Werk}, {Jenkins}, {Bowen}, {Lehner}, {Sembach}, {Thom}, \&
  {Tumlinson}}]{Tripp:2011aa}
{Tripp}, T.~M., {Meiring}, J.~D., {Prochaska}, J.~X., {et~al.} 2011, Science,
  334, 952

\bibitem[{{Tumlinson} {et~al.}(2017){Tumlinson}, {Peeples}, \&
  {Werk}}]{Tumlinson:2017aa}
{Tumlinson}, J., {Peeples}, M.~S., \& {Werk}, J.~K. 2017, \araa, 55, 389

\bibitem[{{Tumlinson} {et~al.}(2011){Tumlinson}, {Thom}, {Werk}, {Prochaska},
  {Tripp}, {Weinberg}, {Peeples}, {O'Meara}, {Oppenheimer}, {Meiring}, {Katz},
  {Dav{\'e}}, {Ford}, \& {Sembach}}]{Tumlinson:2011aa}
{Tumlinson}, J., {Thom}, C., {Werk}, J.~K., {et~al.} 2011, Science, 334, 948

\bibitem[{{Vogelsberger} {et~al.}(2014){Vogelsberger}, {Genel}, {Springel},
  {Torrey}, {Sijacki}, {Xu}, {Snyder}, {Nelson}, \&
  {Hernquist}}]{Vogelsberger:2014aa}
{Vogelsberger}, M., {Genel}, S., {Springel}, V., {et~al.} 2014, \mnras, 444,
  1518

\bibitem[{{Wang} {et~al.}(2005){Wang}, {Yao}, {Tripp}, {Fang}, {Cui},
  {Nicastro}, {Mathur}, {Williams}, {Song}, \& {Croft}}]{Wang:2005aa}
{Wang}, Q.~D., {Yao}, Y., {Tripp}, T.~M., {et~al.} 2005, \apj, 635, 386

\bibitem[{{Weinberg} {et~al.}(1997){Weinberg}, {Miralda-Escud{\'e}},
  {Hernquist}, \& {Katz}}]{Weinberg:1997aa}
{Weinberg}, D.~H., {Miralda-Escud{\'e}}, J., {Hernquist}, L., \& {Katz}, N.
  1997, \apj, 490, 564

\bibitem[{{Werk} {et~al.}(2013){Werk}, {Prochaska}, {Thom}, {Tumlinson},
  {Tripp}, {O'Meara}, \& {Peeples}}]{Werk:2013aa}
{Werk}, J.~K., {Prochaska}, J.~X., {Thom}, C., {et~al.} 2013, \apjs, 204, 17

\bibitem[{{Werk} {et~al.}(2014){Werk}, {Prochaska}, {Tumlinson}, {Peeples},
  {Tripp}, {Fox}, {Lehner}, {Thom}, {O'Meara}, {Ford}, {Bordoloi}, {Katz},
  {Tejos}, {Oppenheimer}, {Dav{\'e}}, \& {Weinberg}}]{Werk:2014aa}
{Werk}, J.~K., {Prochaska}, J.~X., {Tumlinson}, J., {et~al.} 2014, \apj, 792, 8

\bibitem[{{White} \& {Frenk}(1991)}]{White:1991aa}
{White}, S.~D.~M., \& {Frenk}, C.~S. 1991, \apj, 379, 52

\bibitem[{{Wiersma} {et~al.}(2009){Wiersma}, {Schaye}, \&
  {Smith}}]{Wiersma:2009aa}
{Wiersma}, R.~P.~C., {Schaye}, J., \& {Smith}, B.~D. 2009, \mnras, 393, 99

\bibitem[{{Willson}(2000)}]{Willson:2000aa}
{Willson}, L.~A. 2000, \araa, 38, 573

\bibitem[{{Zaritsky} {et~al.}(1994){Zaritsky}, {Kennicutt}, \&
  {Huchra}}]{Zaritsky:1994aa}
{Zaritsky}, D., {Kennicutt}, Jr., R.~C., \& {Huchra}, J.~P. 1994, \apj, 420, 87

\bibitem[{{Zheng} {et~al.}(2017){Zheng}, {Peek}, {Werk}, \&
  {Putman}}]{Zheng:2017aa}
{Zheng}, Y., {Peek}, J.~E.~G., {Werk}, J.~K., \& {Putman}, M.~E. 2017, \apj,
  834, 179

\end{thebibliography}

\end{document}